\documentclass[screen,acmsmall,authorversion,nonacm]{acmart}

\setcopyright{cc}

\usepackage{hyperref}
\usepackage{enumitem}
\usepackage{subcaption}
\usepackage{float}
\usepackage{longtable}

\begin{document}

\title[Understanding Self-Regulated Learning Behavior]
{Understanding Self-Regulated Learning Behavior Among High and Low Dropout Risk Students During CS1: Combining Trace Logs, Dropout Prediction and Self-Reports}

\author{Denis Zhidkikh}
\orcid{0000-0003-1335-8346}
\affiliation{%
  \institution{University of Jyväskylä}
  \department{Faculty of Information Technology}
  \city{Jyväskylä}
  \country{Finland}
}
\email{denis.d.zhidkikh@jyu.fi}

\author{Ville Isomöttönen}
\orcid{0000-0002-5274-236X}
\affiliation{%
  \institution{University of Jyväskylä}
  \department{Faculty of Information Technology}
  \city{Jyväskylä}
  \country{Finland}
}

\author{Toni Taipalus}
\orcid{0000-0003-4060-3431}
\affiliation{%
  \institution{Tampere University}
  \department{Faculty of Information Technology and Communication Sciences}
  \city{Tampere}
  \country{Finland}
}

\authorsaddresses{Corresponding author: Denis Zhidkikh,
\href{mailto:denis.d.zhidkikh@jyu.fi}{denis.d.zhidkikh@jyu.fi}; University of Jyväskylä, Faculty of Information Technology, P.O. Box 35, FI-40014 Jyväskylä, Finland}

\begin{abstract}
  The introductory programming course (CS1) at the university level is often perceived as particularly challenging, contributing to high dropout rates among Computer Science students. Identifying when and how students encounter difficulties in this course is critical for providing targeted support. This study explores the behavioral patterns of CS1 students at varying dropout risks using self-regulated learning (SRL) as the theoretical framework. Using learning analytics, we analyzed trace logs and task performance data from a virtual learning environment to map resource usage patterns and used student dropout prediction to distinguish between low and high dropout risk behaviors. Data from 47 consenting students were used to carry out the analysis. Additionally, self-report questionnaires from 29 participants enriched the interpretation of observed patterns.
  The findings reveal distinct weekly learning strategy types and categorize course behavior.
  Among low dropout risk students, three learning strategies were identified that differed in how students prioritized completing tasks and reading course materials. High dropout risk students exhibited nine different strategies, some representing temporary unsuccessful strategies that can be recovered from, while others indicated behaviors of students on the verge of dropping out.
  This study highlights the value of combining student behavior profiling with predictive learning analytics to explain dropout predictions and devise targeted interventions. Practical findings of the study can in turn be used to help teachers, teaching assistants and other practitioners to better recognize and address students at the verge of dropping out.
\end{abstract}

\ccsdesc[500]{Social and professional topics~CS1}

\keywords{self-regulation, CS1, trace logs, student dropout prediction}

\maketitle

\section{Introduction}

Developing solid learning and working skills is paramount when transitioning to higher education.
When students begin their university studies, they may face numerous challenges, such as increased workload, higher expectations, and greater emphasis on independent learning \citep{Leese2010BridgingGapSupporting}.
This notion is especially true in Computer Science, where the introductory programming course (CS1) is often perceived as laborious and fast-paced by first-year students \citep[e.g.,][]{Hamalainen2019}.
Students who struggle with the learning process may face hits to their self-efficacy beliefs, affecting later studies and potentially leading to burnout \citep[e.g.,][]{Jarvinen2024AcademicExperiencesInformation}.
Lack of motivation, perceived course difficulty, and low interest can, at worst, result in dropping out \citep{Petersen2016RevisitingWhyStudents, Barr2022WhyStudentsDrop}.
To mitigate dropout rates and foster effective learning habits, universities employ various support mechanisms, such as introductory courses, peer mentoring programs and course-specific interventions \citep[e.g.,][]{Turner2017EasingTransitionFirsta, Kangas2017HowFacilitateFreshmen, Pon-Barry2017ExpandingCapacityPromotinga, Borrella2022TakingActionReduce}.

Identifying struggling students and those at risk of dropping out is crucial to ensuring these interventions are effective.
Academic performance prediction and student dropout detection have been extensively researched \citep{Hellas2018PredictingAcademicPerformance, Prenkaj2020-af}.
Many studies apply learning analytics techniques to various student data, including demographics \citep[e.g.,][]{Naseem2019UsingEnsembleDecision}, basic activity indicators \citep[e.g.,][]{Waheed2023EarlyPredictionLearnersa}, outcomes of student assignments \citep[e.g.,][]{Zhidkikh2024ReproducingPredictiveLearning} and self-reports \citep[e.g.,][]{Tempelaar2018StudentProfilingDispositional}.
Notably, \citet{Moreno-Marcos2020TemporalAnalysisDropout} evaluated an effective prediction approach using students' self-regulation behavior data collected from self-reports and trace logs.
Although most student dropout prediction methods demonstrate high accuracy, easy explainability of the predictions remains a challenge \citep{Zhidkikh2024ReproducingPredictiveLearning}.

To increase the effectiveness of the predictions and interventions, it is therefore essential to understand how students study and how they utilize available learning resources.
Recent research has increasingly focused on identifying student learning behaviors, emphasizing the self-regulation, metacognition, and motivation aspects \citep{Loras2022StudyBehaviorComputinga}.
For instance, common activity indicators can be used to model students' learning approaches and to identify student behavior profiles \citep[e.g.,][]{Kim2020ExploringStudentTeacher, Sharma2015IdentifyingStylesPaths}.
Similarly, trace logs from virtual learning environments can reveal patterns in students' learning resource usage and detect common learning strategies used in a classroom \citep[e.g.,][]{Jovanovic2017, Zhidkikh2023MeasuringSelfRegulated}.
Additionally, self-reports, such as questionnaires, learning journals, and interviews may provide insights into students' use of learning strategies and cognitive processes \citep[e.g.,][]{Loksa2016RoleSelfregulationProgramminga, Loksa2020InvestigatingNovicesSitu, Silva2024WhatLearningStrategies}.
However, while these methods help map learning behaviors, they usually do not directly link behavior to early dropout \citep{Zhidkikh2023MeasuringSelfRegulated}.
At most, some studies have compared behaviors of high and low-performing students, often in a post-course setting.
For instance, \citet{Liao2019BehaviorsHigherLowera} interviewed 19 CS1 students who took the final CS1 exam and identified general differences in learning paths between higher-performing and lower-performing students.

Motivated by the examples above, this paper aims to deepen the understanding of learning behaviors among high and low-risk students during CS1.
The present paper complements prior key studies \citep[][]{Liao2019BehaviorsHigherLowera, Moreno-Marcos2020TemporalAnalysisDropout}.
Specifically, we map students' learning paths, extending previous work \citep[cf.
][]{Liao2019BehaviorsHigherLowera} by tracking weekly learning behaviors using trace logs collected during the course.
Further, we employ dropout prediction to identify at-risk students for every course week \citep[cf.
][]{Moreno-Marcos2020TemporalAnalysisDropout} and use this information to discern weekly learning behavior by the associated dropout risk.
To this end, the paper presents a learning analytics-based method to link learning resource usage patterns with dropout predictions, combining tools developed in prior studies \citep{Zhidkikh2023MeasuringSelfRegulated, Zhidkikh2024ReproducingPredictiveLearning}.
A post-course self-report questionnaire was also administered to support the interpretation of the behavioral patterns and to evaluate the potential of the learning analytics used.
Both qualitative and quantitative data were inductively explored to gain insights into students' behaviors at different dropout risk levels.
We use self-regulated learning as the foundational theoretical framework widely applied in CSEd research to understand the learning process \citep{Prather2020WhatWeThinka}.
The study ultimately seeks to answer the following research questions:
\begin{enumerate}[label=\textbf{RQ\arabic*}]
  \item \label{rq1} What learning behaviors emerge in students when identified as low dropout risk?
  \item \label{rq2}
    What learning behaviors emerge in students when identified as high dropout risk?
  \item \label{rq3} How did students' self-reported views on their learning inform the interpretation of behaviors from trace log and dropout prediction data?
\end{enumerate}

\section{Theoretical framework and related work}

In this paper, we employ self-regulated learning (SRL) theory as the primary framework for mapping and understanding student behavior in CS1.
As such, we first provide a brief overview into SRL and how it is measured in academic contexts.
To contextualize the research and aid interpretation of the results, we also present recent related work on measuring SRL in CSEd.
Finally, we present related works on student dropout prediction to provide further context on the methods applied in this study.

\subsection{Self-regulated learning and learning strategies}\label{subsec:srl_theory}

Self-regulation generally refers to the ability to monitor and manage one's effort, time, motivation and behavior when learning or problem-solving \citep{Pintrich2000, Schunk2005SelfRegulatedLearningEducational}. While metacognition describes student's knowledge to self-evaluate learning outcomes and approaches, self-regulation describes the ability to control them \citep{Prather2020WhatWeThinka}.
Therefore, SRL is a strategic learning process in which a student self-regulates by actively setting goals,
engaging with the learning resources using various strategies, monitoring progress and refining their approach based on the outcomes and external learning context \citep{Zimmerman2008InvestigatingSelfRegulationMotivation,Winne1998, Winne2019LearningStrategiesSelfRegulateda}. Although multiple models detailing the SRL process exist \citep{panadero2017review,Loksa2022MetacognitionSelfRegulationProgrammingb}, SRL is often described as a cyclical process consisting of forethought and goal-setting, enacting on the goals and self-controlling, and self-reflection on the outcomes and the process \citep{Zimmerman2009}.
While the general SRL process can be modelled, the specific ways a student regulates their behavior can depend highly on the context, such as the subject, student's prior experiences, the task at hand and even other students \citep{Ben-Eliyahu2015}.

Various measures of SRL have been applied in academics. Common measures include self-report tools that assess student's aptitudes for self-regulation (e.g., questionnaires, think-aloud protocols, learning journals), and event measures that map SRL processes via observable, externally collected data \citep{Roth2016}.
While some researchers have noted issues related to the reliability of self-reports compared to event-based measures \citep[e.g.,][]{Choi2023LogsSelfReportsMisalignment, Zimmerman2008InvestigatingSelfRegulationMotivation}, a growing number of studies has argued for the combined use of both measures \citep[e.g.,][]{Fan2023FullerPictureTriangulation, Lim2021ImpactLearningAnalytics,Tempelaar2020SubjectiveDataObjective}. As \citet{Roth2016} summarize, the triangulation of new and established measures enables researchers to gain new insights into SRL.

A common aspect used to evaluate and study SRL is the use of learning strategies. In a classic conceptualization outlined by \citet{Zimmerman1986}, a learning strategy is a set of planned actions, such as rehearsing, seeking assistance, and reviewing course materials, aimed at acquiring information and skills relevant to the goal or task at hand.
This definition has been commonly used in SRL research based on self-reports or theoretical frameworks that build on Zimmerman's work \citep[e.g.,][]{falkner2014identifying,Silva2024WhatLearningStrategies,Moreno-Marcos2020TemporalAnalysisDropout,Garcia2018SystematicLiteratureReview}.

For event-based measures, \citet{Winne1998} proposed a more fine-grained distinction of student actions into learning tactics and learning strategies.
The conceptualization was later expanded by \citet{Winne2001} and \citet{Winne2019LearningStrategiesSelfRegulateda}.
In this conceptualization, a \emph{learning tactic} is a sequence of discrete, atomic actions that a student performs to achieve a sub-goal or a concrete task.
A \emph{learning strategy}, in turn, is a higher-level sequence of tactics that a student employs to reach a larger learning goal and that is guided by the student's metacognitive monitoring and control.
For example, a student may first aim to gain a basic understanding of a new programming topic by perusing parts of the coursebook and reading and completing examples within the book --- all as part of a single coursebook-oriented tactic.
To complete programming exercises for a week, in contrast, a student's strategy may be to first use a coursebook-oriented tactic to build content knowledge and then, once that understanding is in place, shift to a more task-oriented tactic that involves completing tasks while using course materials as a reference.
To reach the learning goals, students may adjust their tactics over time based on newly set sub-goals, self-evaluations, and changing learning context.
In terms of the overall SRL process, \citet{Winne2001} positions learning strategy selection as part of the forethought phase, while learning tactics are dynamically adjusted during the enacting phase in response to the state of the task.
\citet{Winne2019LearningStrategiesSelfRegulateda} note that tactics are specifically selected to enact on the plans of a learning strategy.

Compared to the classic conceptualization of learning strategies, the tactic--strategy distinction provides a hierarchical view of student actions, where tactics are seen as building blocks of strategies \citep{Winne2019LearningStrategiesSelfRegulateda}.
Such an event-based characterization allows describing the study process as an enactment of learning actions based on the current conditions. Consequently, this definition allows the graphical representation of student behavior as sequences of actions, tactics, and strategies over time \citep{Winne2019LearningStrategiesSelfRegulateda}.

The tactic--strategy conceptualization of student behavior has been widely applied in learning analytics research over the past decade. The most prominent operationalization builds on the seminal work of \citet{Matcha2019AnalyticsLearningStrategies}, in which learning tactics and strategies are identified using process mining and cluster analysis on student trace logs obtained from an online learning environment.
Learning tactics are operationalized as patterns in student action sequences within an uninterrupted period of study, referred to as a learning session.
These patterns are extracted from study sessions using cluster analysis on the sequence of actions performed by students.
Learning strategies are, in turn, defined as students' tactic usage patterns across the studied period and are identified through further cluster analysis.
This approach has since been replicated and verified in various study contexts, such as flipped classrooms \citep{Uzir2020}, online courses \citep{Fan2021LearningAnalyticsReveal}, medical education \citep{Piotrkowicz2021DatadrivenExplorationEngagement}, and secondary education \citep{Zhidkikh2023MeasuringSelfRegulated}.

While the process mining and cluster analysis approach has been widely adopted, other operationalizations exist. For example, \citet{Fincham2019StudyTacticsLearning} analyzed the learning tactics and strategies of first-year engineering students by modeling tactics as a Hidden Markov Model rather than applying cluster analysis to the sequence of actions. Some subsequent studies have also used Hidden Markov Models \citep{Yang2025SelfRegulatedLearningProcesses,Villalobos2022SupportingSelfregulatedLearning}, noting their ability to model the sequential nature of learning tactics.
Other recent studies have explored network analysis approaches. \citet{Fan2023DissectingLearningTactics} applied ordered network analysis to identify and interpret learning tactics in an online course, with results suggesting that network-based methods may provide deeper characterizations of learning tactics than process mining. Further network-based approaches to tactic--strategy identification include epistemic network analysis \citep{Rakovic2023NetworkAnalyticsUnveil}
and psychological networks \citep{Saqr2020UsingPsychologicalNetworks}.

Overall, there are multiple approaches to implementing event-based measures for SRL when applying the tactic--strategy characterization of \citet{Winne1998}.
However, it is important to recognize that no single approach yields a definitive description of self-regulated learning in a given context. In a comparative study by \citet{Matcha2019DetectionLearningStrategies}, process mining, sequence analysis, and network analysis approaches were evaluated for their ability to detect learning tactics and strategies. The study found that while the approaches produced slightly different sets of learning tactics and strategies, the differences were mainly attributable to the varying emphases of the approaches. As such, \citet{Matcha2019DetectionLearningStrategies} highlight that what matters more than the exact operationalization of student behavior is how the results are interpreted and grounded in the studied context. In other words, domain knowledge plays a critical role in implementing event-based measures of self-regulation.

\subsection{Self-regulated learning in computing education}\label{subsec:srl_comp_ed}

There have been several prior studies on self-regulation in computing education.
Collectively, many studies highlight the importance of SRL in computing education, and especially the act of supporting SRL for novice learners.
The findings across many of such studies underscore the role of SRL in academic success, particularly in computer science and programming education.
For example, \citet{schraw2006promoting} identified cognition, metacognition, and motivation as the core components of SRL and recommended strategies to promote self-regulation in science education.
\citet{falkner2014identifying} identified discipline-specific SRL strategies that are especially useful in software development courses, emphasizing the need for cognitive and metacognitive scaffolding. \citet{cigdem2015does} also shows that self-regulation positively influences programming achievement, particularly when coupled with high self-efficacy in blended learning environments.

In programming education, studies like \citet{Loksa2016RoleSelfregulationProgramminga} and \citet{Silva2024WhatLearningStrategies} have shown that self-regulation strategies such as planning, monitoring, and reflection are linked to problem-solving success, although their effectiveness depends on programming skill. Along similar lines, and with those of \citet{schraw2006promoting}, \citet{bergin2005examining} found that high-performing students use more metacognitive and resource management strategies alongside intrinsic motivation to enhance their performance, supporting the need for targeted SRL training as advocated by \citet{bielaczyc1995training} and \citet{Silva2024WhatLearningStrategies}.
In contrast, \citet{flanigan2023relationship} found that poor SRL experiences in undergraduate computer science students can reinforce a fixed mindset. That is, fostering effective SRL strategies is essential for encouraging a growth mindset, but a failed attempt to support SRL is not detrimental only on applying SRL, but also on course performance. Finally, \citet{Liao2019BehaviorsHigherLowera} reported on multiple strategies applied by high-performing programming students such as encouraging consistent study habits, active help-seeking, reflective practices, timely feedback, and creating a supportive learning environment.

\citet{Silva2024WhatLearningStrategies} also discussed the challenges students face in maintaining motivation and managing time, which reinforces the need for educational interventions to help students better apply SRL strategies. \citet{Jovanovic2017} found that students in flipped classrooms often rely on less effective, passive strategies, such as video watching, rather than more active approaches like formative assessments.
Similarly, \citet{Matcha2019DetectionLearningStrategies} and \citet{Liao2019BehaviorsHigherLowera} discussed how different analytic approaches can identify varying learning strategies in online courses, and argued for selecting appropriate methods such as tailoring and scaffolding to better support learners.
Similar insights were also discussed by \citet{panadero2017review} who reviewed six SRL models, concluding that these frameworks should be tailored to students' developmental stages.
Focusing their review of SRL models applied in computing education, \citet{Loksa2022MetacognitionSelfRegulationProgrammingb} further suggested that researchers should select a framework and concepts that best fit their research context.
\citet{greene2007theoretical} discussed the complexity of SRL, advocating for research that considers the interactions between cognitive and task conditions across learning phases.

\citet{ko2016applying} demonstrated that successful students tend to spend more time on challenging concepts, indicating that persistence and focus are the key components of effective self-regulation.
This aligns with \citet{maldonado2018predicting}, who emphasize the importance of consistent planning, monitoring, and reflection in the context of self-paced MOOCs, further reinforcing the value of SRL behaviors in online learning success.
On a broader level, \citet{ergen2017effect} confirmed the large positive effect of SRL strategies on academic achievement across various disciplines and education levels, while \citet{emagnaw2019self} cautioned about the importance of self-regulation and time management for high achievers.
Collectively, these studies argue for supporting SRL through targeted feedback, as suggested by \citet{butler1995feedback}, and addressing challenges in real-time \citep{arakawa2021situ} can potentially enhance student outcomes.

\subsection{At-risk student detection}

Current attempts at identifying at-risk students in CS primarily rely on predictive learning analytics to detect low-performing individuals \citep{Hellas2018PredictingAcademicPerformance}. Such analytics are usually based on attempting to classify students as passing or dropout using some measurable features about the student \citep{Prenkaj2020-af}.
\citet{Hellas2018PredictingAcademicPerformance} outlined various predictive features that has been used, including demographic, academic, and behavioral factors. For instance, \citet{Naseem2019UsingEnsembleDecision} addressed high attrition rates in CS using 23 attributes encompassing students' demographics, academic performance, financial information, and prior educational history.
\citet{Quille2019-ta} in turn demonstrated that student success in CS1 can be predicted with high accuracy using self-efficacy and mathematical ability metrics as key predictors. Further examining at-risk indicators, \citet{Ifenthaler2020-di} identified crucial factors such as GPA, learning history, and clickstream data. Similarly, \citet{Ifenthaler2022-hb} found that increased engagement with self-assessments correlates with better exam performance, underscoring the importance of integrating assessment data into performance predictions.

More recently, \citet{Sghir2022-ei} noted the increasing use of machine and deep learning models in predicting academic outcomes, paving way for using more performance data in predictions. \citet{Waheed2023EarlyPredictionLearnersa} demonstrated high effectiveness of deep learning techniques in predicting students at risk of failing in self-paced online education. Focusing on privacy concerns, \citet{VanPetegem2022} presented a privacy-friendly framework for early detection of students at risk of failing introductory programming courses. Utilizing interpretable machine learning techniques, they achieved high predictive accuracy early in the semester. A further study by \citet{Zhidkikh2024ReproducingPredictiveLearning} reproduced these results using the same framework in a different CS1 course, highlighting the generalizability and explainability of at-risk student predictions through submission metadata and noting the importance of student self-evaluations in improving early dropout predictions.
Finally, \citet{Moreno-Marcos2020TemporalAnalysisDropout} investigated dropout prediction in self-paced MOOCs incorporating self-regulated learning strategies as predictors. They found that event-based SRL strategies, derived from learners' interactions with course materials, had high predictive power, outperforming traditional demographic and self-reported measures. Their result emphasizes the importance of considering complex, activity-based features over conventional data in predictive models.

Collectively, these studies underscore the critical role of predictive analytics in identifying at-risk students in CS education. They highlight the shift toward incorporating complex, activity-based features and advanced machine learning techniques to enhance predictive accuracy.

\section{Research context and data}

The present study was conducted at the university of the first author which provides education and conducts research in the natural and social sciences.
Students begin their studies with a major subject and have the ability to personalize their study plans by pick minor subject courses from other disciplines at the university.
Computer Science (CS) is one of the subjects available as a major or a minor for the university's students.
First-year CS students usually complete the introductory CS1 and CS2 programming courses along with more theoretical courses on algorithms, data management, and data networks.

\subsection{Course context}\label{subsec:course_context}

This study focused on the 6-credit (ECTS) CS1 course taught during fall 2023.
The course is aimed primarily for first-year CS majors, but the course is open to all students at the university.
In fall 2023, 321 students enrolled, with a pass rate of 63.5\%.
The 11-week course included multiple weekly activities, diverse learning resources, and a rigorous grading scheme.
To pass, students were required to complete at least 40\% of the weekly tasks, complete a course assignment (a mini-game project) and attend the final exam.
The final grade was determined based on the number of completed weekly tasks and performance in the final exam. The course design allows getting the best grade even if either the final exam or weekly tasks were passed with minimal marks by compensating one with another.

\begin{table}[t]
  \centering
  \small
  \begin{tabular}{@{}rll@{}}
    \toprule
    Week & Topic                                                           & Course assignment schedule \\ \midrule
    1    & Introduction to command line, simple C\# program                &                            \\ \addlinespace[0.5em]
    2    & Functions, documentation, variables                             &                            \\ \addlinespace[0.5em]
    3    & IDE usage, more functions and variables                         &                            \\ \addlinespace[0.5em]
    4    & Parameters, functions, introduction to unit tests               &                            \\ \addlinespace[0.5em]
    5    & Conditionals, strings                                           & Milestone 1 (Planning)     \\ \addlinespace[0.5em]
    6    & Loops, arrays                                                   &                            \\ \addlinespace[0.5em]
    7    & Multi-dimensional arrays, string splitting                      &                            \\ \addlinespace[0.5em]
    8    & Dynamic data structures, intro into object oriented programming & Milestone 2 (Prototype)    \\ \addlinespace[0.5em]
    9    & Game development, file handling                                 &                            \\ \addlinespace[0.5em]
    10   & Recursion, exceptions, function references, lambda expressions  &                            \\ \addlinespace[0.5em]
    11   & Floating point numbers, questions from students                 & Milestone 3 (Final)        \\
    \bottomrule
  \end{tabular}
  \caption{
    The syllabus of the studied CS1 course. The course comprises 11 weekly topics that each includes two lectures and exercises. Additionally, the course includes a mini-project assignment that is split into three milestones that students have to present to the teaching assistants.
  }\label{tab:cs1_syllabus}
\end{table}

\begin{figure}[t]
  \centering
  \includegraphics[width=\linewidth]{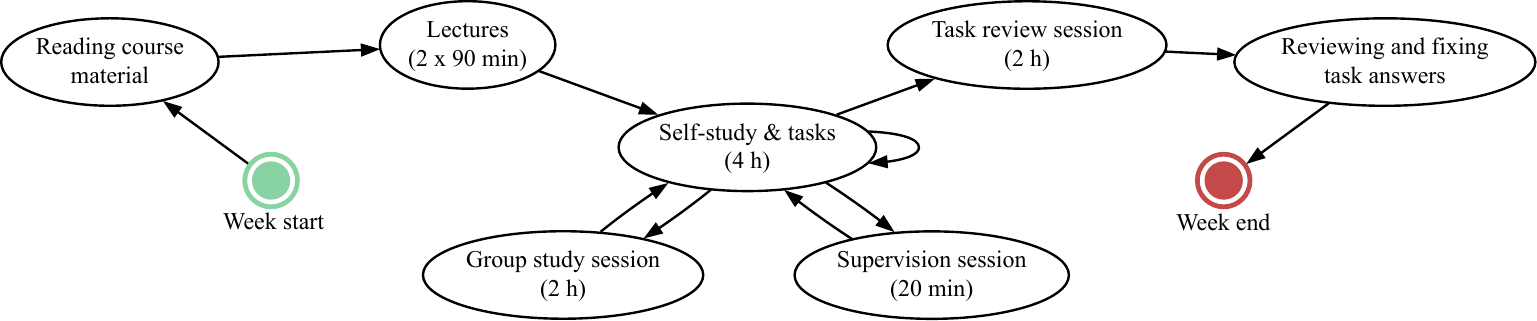}
  \caption{Course week design of the studied CS1 course.
    The course design follows a weekly pattern, in which students first learn the week's content, then apply the new content knowledge in homework, and finally, review their answers before proceeding to new content.
  } \label{fig:cs1_weekly_schedule}
\end{figure}

The course syllabus is outlined in Table \ref{tab:cs1_syllabus}.
The course followed a weekly schedule, with a learning strategy suggested by the course teacher depicted in Figure \ref{fig:cs1_weekly_schedule}.
Each week began with two lectures for which the students were encouraged to prepare by reviewing the week's materials.
The lectures were held in-person and online, during which the lecturer and students interacted through quizzes, interactive examples and discussions on the week's topic.

Following the lectures, students completed a weekly task set consisting of 10 to 20 graded exercises of varying difficulty levels. All exercises were divided into four main types.
\emph{Introductory tasks} included quizzes and tasks based on visual simulators with the goal to scaffold students into the week's topic.
\emph{Basic tasks} refer to programming tasks relevant to the week's topic. For example, basic tasks in week 2 revolved around defining and using functions to draw staircases within the game engine used in the course, and basic tasks in week 6 focused on solving the rainfall problem using arrays and loops.
\emph{Bonus tasks} contained any extra tasks related to the week's topic and were mainly provided for those interested in additional practice or extra points for a better course grade. For instance, bonus tasks in week 1 included practicing programming with Alice\footnote{\url{https://alice.org}}, a logic quiz and answering the course's preliminary survey.
Finally, \emph{guru tasks} required students to seek extra knowledge beyond the week's material and were mainly provided for more skilled students. Guru tasks included, for instance, programming a Conway's Game of Life simulator and solving Project Euler problems or prior Advent of Code problems.
For brevity, we collectively refer to bonus and guru tasks as \emph{extra tasks} in this paper.

Students generally had the freedom to choose which exercises to submit. However, a small selection of introductory and basic tasks were designated as mandatory \emph{core tasks}.
Submitting answers to the core tasks was a requirement for passing the course to ensure that all students completed at least some tasks from every week.
The number of core tasks varied weekly from two to three and they were selected by the course teacher beforehand.

In addition to the weekly tasks, students completed a course assignment throughout the course. The assignment comprised a self-made mini-project, which was usually a game made with the game engine used in the course or a small command-line program. Students were allocated 30 hours of course work for the assignment, which was split into three milestones: a plan document, a working prototype and the finished project.
During the course, students were required to participate in three one-on-one supervision sessions with a teaching assistant to assess the progress of the project throughout the milestones (see Table \ref{tab:cs1_syllabus}).
The assignment was evaluated as a whole in the final milestone and was graded as pass/fail. Those whose assignment did not match the passing criteria as evaluated by the TA were required to revise their work and participate in a new one-on-one supervision session.

Throughout the course, the students had access to several learning resources, such as a coursebook, lectures, supplementary examples and ungraded exercises, lecture recordings, and documentation.
Students were encouraged to use only the course materials and not search the internet, as all necessary and verified materials are available within the course. In addition, students could attend group study sessions in which they could study and get live help from teaching assistants.
Additionally, each course week ended with a task review session held after the task deadline, allowing students to compare each other's answers.
After the session, the students were encouraged to review and fix their answers using model answers provided by the course teacher.
Similar to course materials, students were encouraged to attend lectures, group study sessions and task review sessions, but there was no required attendance.
In fall 2023, $62.6\%$ of the course students self-reported attending some of the lectures, with an average self-reported attendance of $19.4\%$ per lecture; the lowest reported attendance was $2.8\%$ (week 1, lecture 2) and the highest $38.6\%$ (week 2, lecture 2). In turn, $78.2\%$ of the students were recorded as participating in some of the group study sessions, with an average attendance of $28.9\%$ students per week; the lowest attendance was $14.6\%$ (week 10) and the highest $49.5\%$ (week 7).

\subsection{The study}

The study was implemented in two phases.
The initial study entailed completing the course and using the available learning resources.
During the study, the trace log data, task points, exam points, and final grades of the participants were collected from the virtual learning environment.
After the course, an additional elective post-course questionnaire was administered to the original participants to gain additional self-report data.

We followed the university's guidelines for conducting studies involving human participants.
Prior to participation, students were provided with a research notification and
a GDPR-compliant privacy notice detailing how their data were to be used in this study.
Participating in both studies was fully elective, and participants could leave the study at any time.
Participation was based on informed consent, and whether a student participated in the study carried no incentives or penalties.
At our university, informed consent is sufficient except when the collected data are categorized as sensitive (race or ethnic origin; political opinions; religious conviction; philosophical beliefs; trade union membership; health information; sexual orientation or behavior; or genetic and biometric data when used to identify a person).
Explicit ethics approval is required only when data collection deviates from the informed-consent process, when research requires physical intervention, when the subjects are minors without parental consent, when the research exposes participants to strong stimuli, when it risks causing mental harm, or when it could threaten the safety of participants. As we did not collect sensitive data and none of the other criteria apply, we did not seek a third-party ethics review.

Altogether, $N=47$ students participated in the initial study.
None of the students formally withdrew from the study during the course.
However, only 29 students also answered the post-course questionnaire, potentially because participation was not mandatory and some participants dropped out of the course. We still included the post-course questionnaire data, as it provided valuable information for explaining student learning behaviors. As such, we used trace log data from all 47 students to
investigate the learning behaviors (\ref{rq1} and \ref{rq2}) and the questionnaire data from the 29 students to compare self-reports and behavior data (\ref{rq3}).

\subsubsection{Data collection} \label{sec:data_collection}

The main data of the study are the trace logs of the participating students.
The initial logs were collected as raw action logs which contained timestamps of every action a student performed in the virtual learning environment during the 11 weeks of the course.
The raw logs were processed into final trace logs by removing actions irrelevant to the course and by recoding the actions into context-relevant event codes.
The specific approach for processing the raw logs and the description of all event codes are provided in Appendix~\ref{appendix:event_codes}.
The event codes relate to the use of the learning resources available to the students during CS1, such as the coursebook, lectures, supplementary examples, weekly tasks, group sessions, one-on-one sessions, model answers and answer review sessions (see Section \ref{subsec:course_context}).
For the weekly tasks, the separate event codes were given for interacting with the different task types described in Section \ref{subsec:course_context} (introductory, basic, core, bonus, guru), as different tasks play a different role in the course.
Additionally, the events for interacting with the current week's tasks and the previous week's tasks were differentiated, as students could potentially review and revise the previous week's tasks based on the tasks review sessions and the model answers.
Overall, 82 distinct event codes were used to describe student behavior during CS1.

In addition to the trace log data, task completion and performance data was also collected for the duration of the whole course.
The data included timestamps of all task submission attempts sent to the virtual learning environment and the correctness of each submission.
General performance data, such as weekly task points before the weekly deadline, task points after the deadline (as students could edit their answers based on the model answers), final exam points and final course grade were also collected.

Finally, self-reports of students' SRL were collected via a questionnaire after the course but before the final exam.
The questionnaire included seven open-ended questions to which students could answer in brief sentences.
Such an approach was selected to reduce the length of the questionnaire while allowing students to more broadly reflect on their SRL during CS1.
The questions were modelled after MSLQ \citep{Pintrich1991} which is a commonly used self-report tool for assessing SRL skills \citep{Duncan2005}.
The full questionnaire is included in Appendix~\ref{appendix:questionnaire}.

\section{Data analysis pipeline}
\label{sec:data_analysis}

\begin{figure}[ht]
  \centering
  \includegraphics[width=\textwidth]{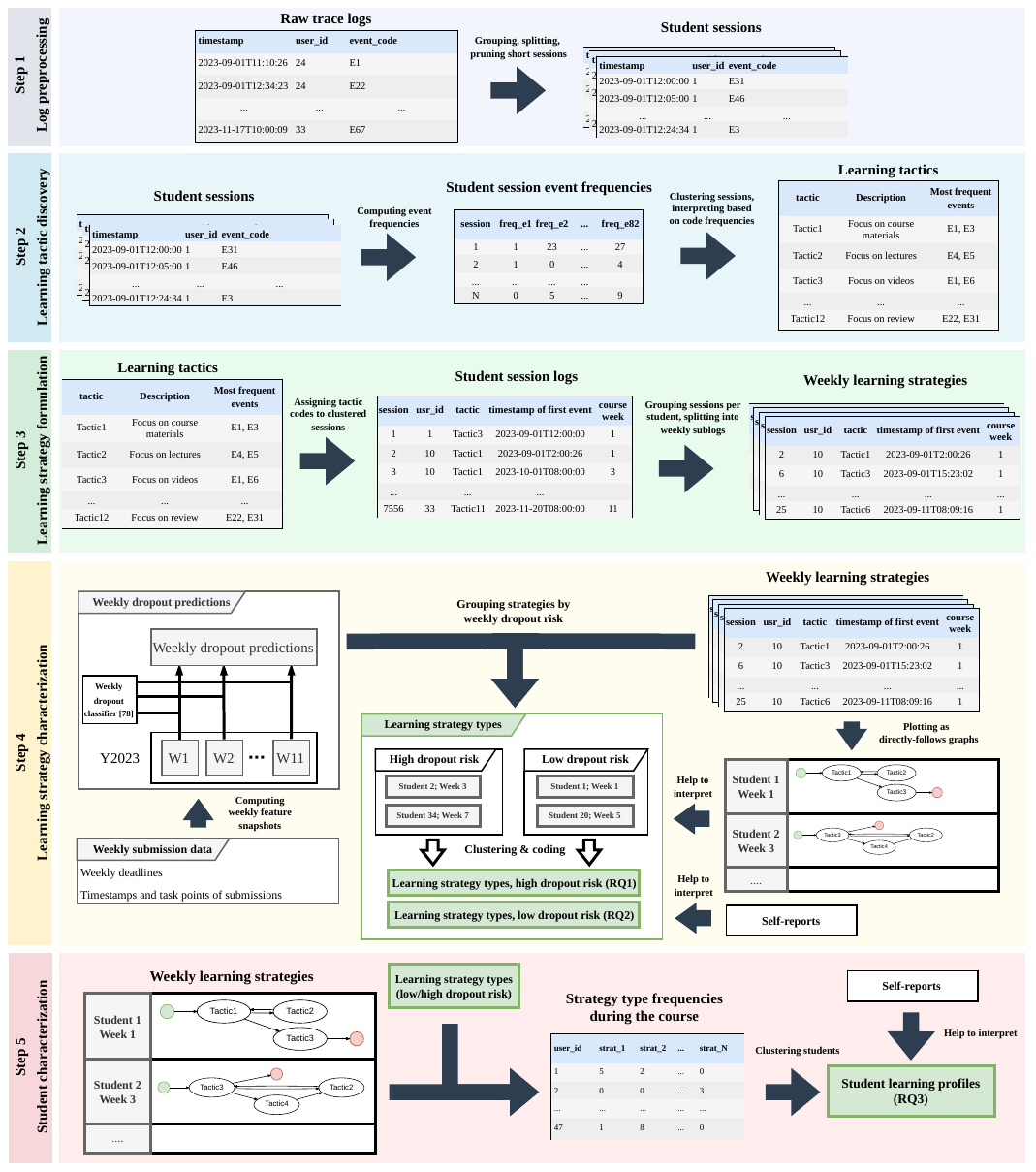}
  \caption{
    The pipeline for analyzing trace logs in the present paper.
    The pipeline consists of five steps, where the outputs of the previous steps are used as inputs for the subsequent steps.
    Analysis follows \citet{Winne2019LearningStrategiesSelfRegulateda}'s characterization of SRL, whereby student's learning strategy are characterized as transitions of learning tactics.
    \vspace{3em}
  } \label{fig:cs1_srl_analysis}
\end{figure}

To answer the research questions, we employ exploratory methods, such as cluster analysis and conventional content analysis, to identify patterns of student behavior and compare them with self-report data.
Our aim was not to establish statistically significant generalizations, but to examine the phenomenon through learning analytics supported by qualitative self-reports.

The overall data analysis pipeline is depicted in Figure \ref{fig:cs1_srl_analysis}.
The pipeline is based mainly on the prior analysis procedures by \citet{Matcha2019DetectionLearningStrategies} with modifications to focus more on learning strategies and their association with dropout risk.
The pipeline consists of five steps, where the outputs of the previous steps are used as inputs for the subsequent steps.
Steps 1--3 reduce the size and dimensionality of the logs by recoding raw events as learning tactics, while steps 4--5 form the main analysis used to answer the research questions.
We outline the steps in detail next.
The data used and the full analysis code are available online\footnote{URL available upon request from the first author.}.

\subsection{Log preprocessing}\label{subsec:log_preprocessing}

Before the main analysis, the input trace log is split based on the time distance between events.
This starting step is adapted from the prior analysis procedures of \citet{Jovanovic2017} and \citet{Matcha2019DetectionLearningStrategies} to preprocess logs into student sessions that represent students' consecutive active interactions with course materials, such as time spent on a single task.
First, we grouped all events in the logs by student and sorted them by timestamp. Next, we split the logs for each student: if the time difference between two events was longer than the cut-off threshold, the first event was set as the end of the previous session and the second as the start of a new session. This yields a set of student session logs, which is the step's output.

A critical parameter for this step is the cut-off threshold, as it affects the granularity and generalizability of the subsequent analysis \citep{Kovanovic2015DoesTimetaskEstimation}. However, there is no set rule for choosing a threshold: \citet{Jovanovic2017} opted for 30-minute sessions citing a previous study, while \citet{Matcha2019DetectionLearningStrategies} selected 30 minutes based on a review by \citet{Kovanovic2015DoesTimetaskEstimation} noting that it was one of the most common choices. In turn, \citet{Zhidkikh2023MeasuringSelfRegulated} used a 45-minute cut-off based on the study context and the nature of the tasks.
Here, we selected the cut-off based on context: given the number of weekly tasks available to the students, we estimated that a student was allocated 25 minutes for each task on average. As such, we chose 25 minutes as the cut-off threshold in this study.

Before proceeding to the next step, we removed all student sessions that consisted of a single event. This additional filtering step was suggested by \citet{Jovanovic2017} to remove potential outliers caused, for instance, by a student leaving the learning environment open and returning to it later or opening the course environment by mistake. \citet{Jovanovic2017} also suggested removing sessions whose length falls outside the 95th percentile to focus on more general patterns. However, we opted not to remove long sessions, as our goal was to characterize the full range of students' learning behaviors, including those who engaged more deeply with the course materials.

\subsection{Learning tactic discovery}

Once the student sessions are obtained, they are used as input to identify common learning tactics used throughout the course. The core idea of this step is based on the work of \citet{Matcha2019DetectionLearningStrategies}, whereby learning tactics are identified by applying cluster analysis to student sessions and interpreting the resulting clusters. In addition to identifying learning tactics, this reduces the dimensionality of the data by grouping similar sessions together. The output of this step is a set of clusters of student sessions and their characterizations as learning tactics.

In the original approach by \citet{Matcha2019DetectionLearningStrategies}, each student session is first converted into a First-Order Markov Model (FOMM) and represented as a transition matrix containing the transition frequencies between consecutive events. These FOMM representations are then clustered using the Expectation Maximization (EM) algorithm, which is useful as it does not require an explicit similarity metric between two student sessions \citep{Aggarwal2014}.
A potential concern with using FOMMs relates to the number of events: as the number of events grows, FOMM representations grow quadratically with the number of distinct event types. This can lead to sparsity issues, as each individual session likely contains only a small subset of possible event transitions. Reliable clustering may therefore require a larger number of student sessions.
For example, while \citet{Zhidkikh2023MeasuringSelfRegulated} included 68 sessions with 14 event types,
\citet{Matcha2019DetectionLearningStrategies} included nearly 5,000 sessions with only 22 event types. Many other studies similarly limit their analysis to roughly a dozen event types \citep[e.g.,][]{Villalobos2022SupportingSelfregulatedLearning,Fincham2019StudyTacticsLearning,Fan2021LearningAnalyticsReveal}, a number that rarely matches the range of actions a student can actually perform.
In this study, we opted for 82 event codes to obtain more detailed logs (see Appendix \ref{appendix:event_codes}).
This presents a possible concern for FOMM-based clustering, as the number of student sessions could be much lower than the number of potential event transitions.
Further, a potential limitation of EM is its computational complexity and scalability to large datasets \citep[see e.g.,][]{Karimi2019GlobalConvergenceFast}.

Given these concerns, we considered an alternative approach for identifying learning tactics. Instead of clustering FOMM representations, we represented each student session as an event frequency vector $(\text{fE}_1, \text{fE}_2, \ldots, \text{fE}_{82})$, where $\text{fE}_i$ indicates the relative frequency of an event type in the session. While this discards information about event transitions, this representation substantially reduces the dimensionality and sparsity of the student sessions. It also allowed us to employ a $k$-means-based algorithm for clustering the student sessions.
Because each session vector represents a discrete distribution of event frequencies, we used the $k$-medoids algorithm \citep{Schubert2022FastKmedoidsClustering}, which allows applying the $\ell_1$ distance metric that is well-suited to such distributions.
Once the clusters were obtained, they were reviewed and interpreted by the first author to describe the learning tactics.

\subsection{Learning strategy formulation}

Using the identified learning tactics as input, the original student sessions are labeled based on the cluster to which they were assigned. We also extracted the timestamp of the first event of each student session and determined the course week in which the session occurred, yielding a trace log of learning tactics that is essentially a higher-level view of the original trace log.
Since we defined learning strategies as meaningful sequences of learning tactics (see Section \ref{subsec:srl_theory}), we repeated the splitting process of Step 1 on the new learning tactics trace log.
However, instead of using a cut-off threshold, we split the trace logs into weekly sublogs. This was done because the course was structured into weekly modules and students were likely to adjust their strategies weekly in response to the weekly context and prior performance \citep[e.g.,][]{Pintrich2000}.
This additional splitting yielded per-student weekly learning tactic logs that capture how each student adjusted their tactics throughout the week.
The output of this step is thus a set of weekly learning tactic logs, i.e., learning strategies (see Section \ref{subsec:srl_theory}).

\subsection{Learning strategy characterization}\label{subsec:weekly_strategy_types_method}

This step aims to characterize the weekly learning strategies and group them into high- and low-risk strategies based on their possible association with dropout.
The output of the step is a set of descriptions of the learning strategy types students employed in CS1 and their association with dropout risk, addressing \ref{rq1} and \ref{rq2}.

Assessing dropout risk for weekly strategies can be challenging. For one, true dropout can be known only after the course ends. Further, all students adjust their learning strategies in response to the environment and prior experiences \citep[e.g.,][]{Pintrich2000}, which means that even a student who eventually drops out will likely have employed some of the same strategies as students who persist. Similar weekly strategies may therefore be associated with both dropout and persistence.
Finally, while manual evaluation of learning strategies is possible \citep[see e.g.,][]{Liao2019BehaviorsHigherLowera}, it is not scalable to a large number of students or strategies.

To address the issue of labeling strategies as high- or low-risk, we used a pre-trained dropout risk prediction model based on task completion data. Specifically, we used a pre-trained logistic regression models for classifying dropout developed by \citet{Zhidkikh2024ReproducingPredictiveLearning}. These models were trained on a large dataset in a similar CS1 context and were demonstrated to be transferable to other similar course contexts. Importantly, \citep{Zhidkikh2024ReproducingPredictiveLearning} include models for predicting dropout using only the task performance data available during a specific week of the course. In other words, these models can predict a student's dropout risk during a given week.
The models take as input the weekly task performance data described in Section \ref{sec:data_collection} and output a dropout prediction value $p_\text{drop}$ that ranges from 0 (likely to persist) to 1 (likely to drop out).

In this step, we computed the weekly dropout predictions for each student with the models and used them as a measure of dropout risk associated with the week's learning strategy.
We then labeled a student's weekly learning strategy as high risk if the model's prediction for that week was greater than 0.5 and as low risk otherwise.
In other words, we classify a strategy as low or high risk by taking the most probable outcome predicted by the model---this is a common approach to deciding the class based on a probability \citep[][p. 21]{Hastie2017ElementsStatisticalLearning}, and the method originally used to evaluate these models in \citep{Zhidkikh2024ReproducingPredictiveLearning}.
Overall, using a prediction model as a measure of dropout risk allowed us to group the weekly learning strategies into high and low dropout risk strategies while accounting for the challenges outlined above.

Once the weekly learning strategies were grouped into high- and low-risk, we clustered the strategies within each group. Here, we proceeded with the same approach \citet{Matcha2019DetectionLearningStrategies} use to cluster learning tactics. Each weekly strategy---described as a log of learning tactics---was converted into a FOMM and then clustered using the EM algorithm. We opted for this approach here because reducing the original trace logs into learning tactics significantly reduced the number of potential transitions between tactics, making the EM algorithm suitable for the task.
This yielded two groups of clusters: clusters of low dropout risk strategies and clusters of high dropout risk strategies.

\begin{figure}[t]
  \centering
  \hfill
  \begin{subfigure}[t]{0.35\textwidth}
    \centering
    \includegraphics[width=0.4\textwidth]{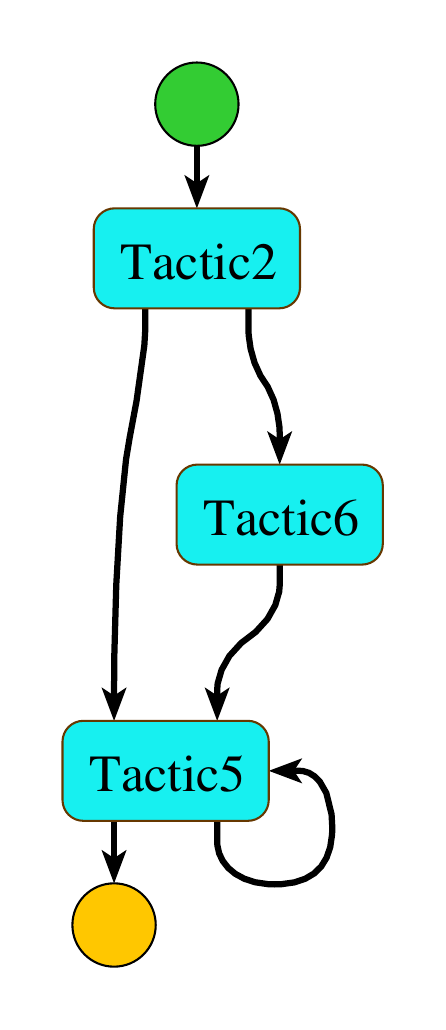}
    \caption{Heuristic Net computed and optimized from all strategies in a single learning strategy cluster. The net displays the most relevant transitions between learning tactics from the start of the week (green node) to the end of the week (red node). The net is an aggregation of all learning strategies in a learning strategy cluster.} \label{fig:heuristic_net}
  \end{subfigure}
  \hfill
  \begin{subfigure}[t]{0.6\textwidth}
    \centering
    \includegraphics[width=\textwidth]{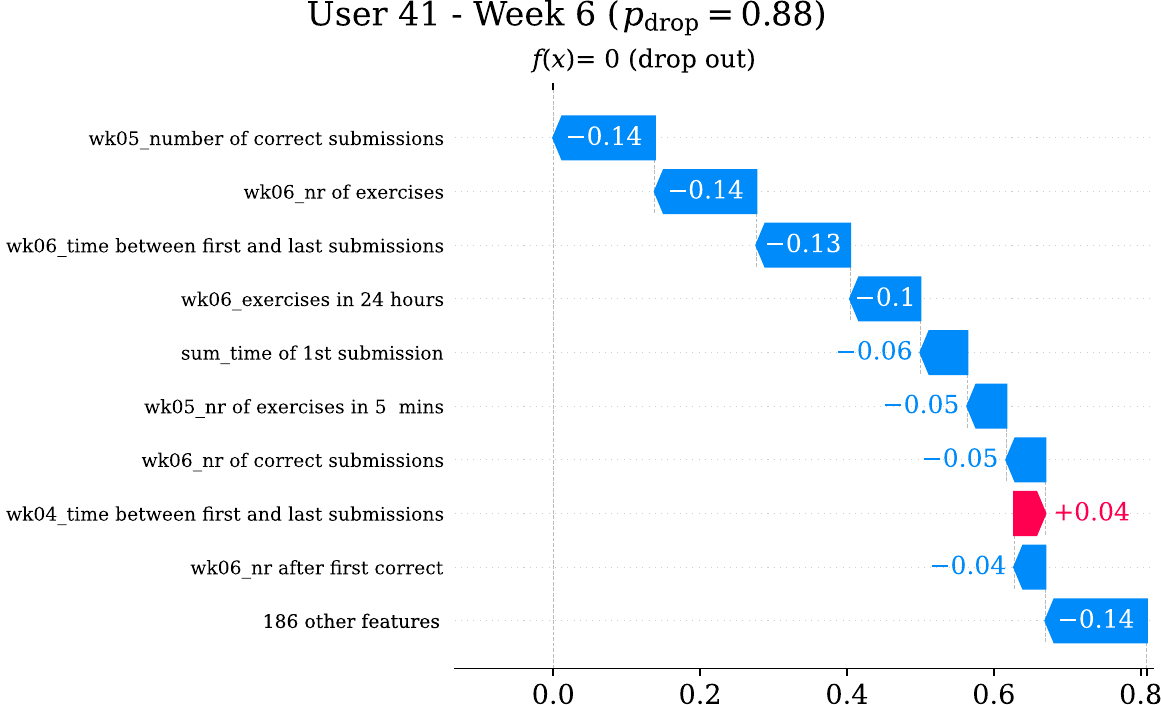}
    \caption{SHAP waterfall plot for the dropout prediction associated with one of the learning strategies (identified via student identifier and week number) in a learning strategy cluster.
      The waterfall plot displays dropout probability used to classify a specific learning strategy and how each of the feature contributed to the prediction. Negative values indicate features that explain towards dropout and positive values indicate features that explain towards persistence.
    } \label{fig:waterfall_plot}
  \end{subfigure}
  \hfill
  \caption{
    Example of graphs used to identify and characterize learning strategy types from clusters of learning strategies.
  } \label{fig:example_strategy_graphs}
\end{figure}

To interpret and characterize the clusters as learning strategy types, we produced aiding visualizations. For each cluster, these visualizations displayed the median dropout probability of the associated weekly strategies, a frequency diagram of the tactics present in the cluster, and a visual representation of the most relevant learning tactic transitions across all learning strategies in the cluster. For the latter, we generated a heuristic net \citep{weijters2006process}, a process mining algorithm used to visualize patterns in event logs. A core benefit of heuristic nets is the ability to filter out less frequent transitions, simplifying the visualization of event logs \citep{weijters2006process}. For each learning strategy cluster, we generated a heuristic net from the union of all learning tactic transitions present in the cluster. An example of a heuristic net is shown in Figure \ref{fig:heuristic_net}.
In addition to heuristic nets, we also generated classic directly-follows graphs for each weekly strategy to represent the exact transitions between learning tactics \citep{Berti2023PM4PyProcessMining}.
Together, heuristic nets and directly-follows graphs served as a graphical aid in characterizing the learning strategy clusters.

To further aid interpretation, we also examined how the dropout prediction model classified the weekly strategies in each cluster.
For this, we used SHAP analysis \citep{Lundberg2017UnifiedApproachInterpreting}, a general method for explaining the predictions of machine learning models. For each prediction, SHAP analysis assigns an importance value to each feature, indicating how much and in which direction a feature contributed to the dropout prediction. Here, we used SHAP analysis to compute how weekly metrics related to task performance contributed to the weekly dropout prediction.
We viewed the feature importances using a waterfall plot (see Figure \ref{fig:waterfall_plot} for an example) for each learning strategy in the cluster.
Waterfall plots provided a quick overview of how a dropout prediction for each student-week was formulated by the prediction model, particularly in terms of which features contributed most. For example, in Figure \ref{fig:waterfall_plot} the strongest explanators of the dropout prediction were the number of correct answers in the previous week and the number of submissions during the current course week. These explanations allowed us to more effectively characterize students' learning strategies without relying solely on the learning tactics.

The interpretation was performed by the first two authors through two hours of joint work, during which the clusters were reviewed and the interpretations agreed upon using the visualizations described above. In unclear cases, the authors also referred to students' self-reports to better understand the context of a specific student's strategy. The resulting strategy types were then used to describe the learning behaviors associated with low dropout risk (\ref{rq1}) and with high dropout risk (\ref{rq2}).

\subsection{Student characterization}

The final step of the analysis aims to characterize students' general activity throughout the course in order to compare this activity with self-reports (\ref{rq3}) and to provide further information for characterizing the learning behaviors in \ref{rq1} and \ref{rq2}.
This step follows the analysis used by \citet{Jovanovic2017} and \citet{Zhidkikh2023MeasuringSelfRegulated}.
First, the weekly learning strategies identified in Step 3 and the learning strategy types from Step 4 are taken as input.
Each student is then assigned a learning profile represented as a frequency vector $(\text{fStrat}_1, \text{fStrat}_2, \ldots, \text{fStrat}_N)$, where $\text{fStrat}_i$ represents the frequency of a student's weekly strategies corresponding to a specific learning strategy type and $N$ is the number of strategy types identified earlier.
These profiles capture students' overall activity and their weekly strategy adjustments during the course.
The use of frequencies simplified the profiles and enhanced the robustness of the analysis.
Following the analysis described in \citep{Jovanovic2017,Zhidkikh2023MeasuringSelfRegulated}, we used agglomerative clustering with the $\ell_1$ metric to cluster students based on their learning profiles. Agglomerative clustering allows drawing a dendrogram that visualizes the hierarchical structure of the clusters, supporting the selection of a suitable cluster count and the detection of potential outliers.
These clusters were then interpreted and characterized as student learning profile types that capture different patterns of learning strategy use throughout the course.

To aid in interpreting the student learning profile types, the open-ended questionnaire answers were preprocessed at this stage.
The last two authors of this study individually coded the questionnaire answers into themes. The coding was not guided but emerged from the answers \citep[cf. \textit{conventional content analysis} in][]{hsieh2005three}. The two authors also recorded whether each student expressed a positive, neutral or negative opinion of their own learning behavior during the course. After individual coding, the two authors convened in a joint two-hour session to construct shared interpretations across the entire dataset, resolving discrepancies rather than taking votes. The resulting codes were thus jointly decided between the last two authors, not merely reconciled.

In addition to the self-report codes, we used performance-related metrics for each cluster, such as the dropout probability across all student sessions in each cluster and the final course grades of the students in the cluster.
These values, along with the self-reported opinions about learning behavior, were used to compare the clusters and characterize the student learning profile types.
To support this mostly qualitative comparison, we employed a statistical test to gauge the differences between the three metrics across the clusters.
Here, we sought a test that can accommodate a small number of data points and is robust enough to provide a numerical measure of difference between clusters as an aid.
As such, we opted for the Brunner-Munzel test \citep{Neubert2007StudentizedPermutationTest}, a robust non-parametric alternative to the Mann-Whitney U test that is suitable for small numbers of observations without requiring assumptions about their distribution \citep{Wilcox2017IntroductionRobustEstimation,Karch2021PsychologistsShouldUse}.
Like the Mann-Whitney U test, the Brunner-Munzel test assesses stochastic equality---that is, whether the probability of an observation from one distribution being greater than an observation from another distribution is not random \citep{Karch2021PsychologistsShouldUse}.
Importantly, the Brunner-Munzel test statistic, the stochastic superiority $\hat{p}''$, can be interpreted as the probability that random variable $X$ is lower than random variable $Y$ when ties (i.e., observations of equal value) are split evenly, making it a useful effect size measure \citep{Karch2021PsychologistsShouldUse}.
We therefore use the Brunner-Munzel test and its effect size $\hat{p}''$ to provide a numerical comparison between clusters that aids interpretation.

Overall, using the self-report codes, the self-reports, predicted dropout and course grades as a guide, the first two authors iteratively interpreted the resulting clusters in up to three hours of joint work.
These characterizations of student learning profile types were then used to compare self-reports with the behaviors obtained from the trace logs (\ref{rq3}).

\section{Results}

We present the results of each step of the analysis in the order of the pipeline depicted in Figure \ref{fig:cs1_srl_analysis} and described in Section \ref{sec:data_analysis}.

\subsection{Log preprocessing}

The raw trace log consisted of 86,738 events in total. After splitting the events using the 25-minute cut-off rule, the log was divided into 7,226 individual student sessions. After removing sessions with just one event, 5,445 student sessions were included in the subsequent analysis.

\subsection{Learning tactic discovery}\label{subsec:learning_tactics}

\begin{figure}[t]
  \centering
  \includegraphics[width=0.9\linewidth]{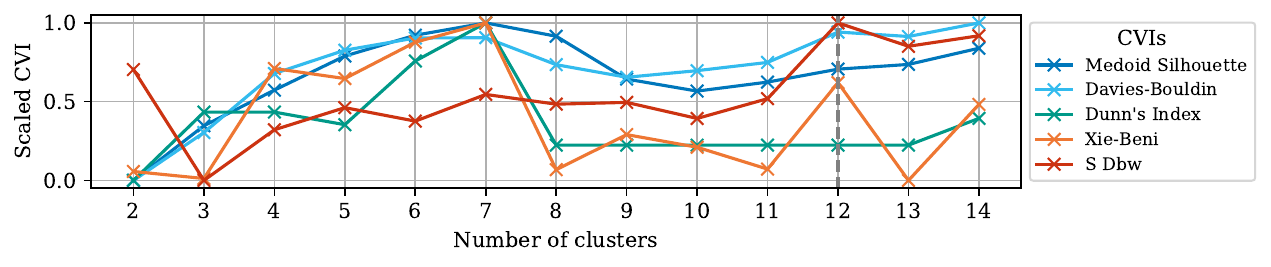}
  \caption{
    Cluster validation for clustering of student sessions.
    Cluster validation indices are rescaled to fit the figure and adjusted so that higher values are better. The selected number of clusters ($K_{\text{sess}}=12$) is indicated with a dashed line.
  } \label{fig:session_type_cvis}
\end{figure}

To select a reasonable number of clusters for grouping raw events into learning tactics, we computed multiple cluster validation indices (CVIs).
The clustering performance of the $k$-medoids clustering on the student sessions is depicted in Figure \ref{fig:session_type_cvis}.
Based on the figure, three CVIs suggest 7 clusters (Medoid Silhouette, Xie-Beni and Dunn's Index). However, a slight elbow appears at 12 clusters for several CVIs (S Dbw, Davies-Bouldin, Xie-Beni), suggesting a potentially interesting alternative clustering.

Based on the weekly schedule of the course (see Figure \ref{fig:cs1_weekly_schedule} and Section \ref{subsec:course_context}), we expected to select at least 7 session types related to reading course materials, attending lectures, completing different types of tasks (introductory, core, basic, bonus/guru), and reviewing previous weeks' tasks.
We were also open to selecting a more fine-grained clustering to uncover more specific patterns in students' learning activities. Being familiar with the course context, we therefore reviewed the session clusters for both 7 and 12 clusters.
We ultimately selected $K_{\text{sess}} = 12$ session clusters, as this clustering captured more detailed activities. Importantly, unlike the 7-cluster solution, the 12-cluster solution captured one-on-one TA sessions as a separate tactic.

From the session clusters, we interpreted 12 tactics that the participants used during CS1. We referred to the frequency of each event code in each cluster to form our interpretation. The event codes and frequencies used for interpretation are provided in detail in Appendix \ref{appendix:learning_tactic_clusters}.
Overall, we identified the following learning tactics:

\begin{itemize}
  \item \emph{Focusing on course materials and examples} (\texttt{F\_CourseMat\_Examples}, $N = 509$): This tactic is characterized by a focus on the course materials, such as the course book. Students read the course book and interacted with the programming examples provided in it.
    These are mostly runnable code examples that students were free to adjust, but they also include interactive simulators for visualizing abstract concepts, such as memory references.
  \item \emph{Focusing on the lecture and lecture materials} (\texttt{F\_Lec\_Engaged}, $N=766$): In this tactic, a student engaged actively with the lecture materials for a specific course week. The tactic includes events related to attending the lecture, although the relative frequency of these events is lower than that of interacting with the lecture materials. The lecture materials included additional interactive examples and questions intended for use during the lecture; compared to simply watching lecture videos, this tactic is characterized by higher interaction with the lecture examples and with questions intended for the live lecture.
  \item \emph{Focusing on lecture videos} (\texttt{F\_Lec\_Video}, $N=469$): Compared with the previous tactic, this tactic shows a higher relative frequency of watching lecture videos and a lower relative frequency of interacting with the lecture materials. We therefore characterize it as a student watching lecture videos, either after the lecture to revisit it or in place of the original live lecture. This tactic also captures events related to viewing the previous week's lecture materials, further emphasizing a focus on the lecture contents after the live lecture.
  \item \emph{Focusing on the current week's introductory tasks} (\texttt{F\_CurTasks\_Intro}, $N=350$): This tactic relates to focusing on introductory tasks for the current week. Since the introductory tasks are not mandatory for passing and are meant as a scaffolding tool, the number of sessions in this cluster is lower than for tactics focusing on basic or core tasks.
  \item \emph{Focusing on the current week's core or basic tasks, attempts and correct submissions}: The focus on the course's core and basic tasks is split into four separate tactics. The first two relate to core tasks: the \emph{attempting} tactic (\texttt{F\_CurTasks\_Core\_Attempt}, $N=356$) is characterized by a high number of incorrect submissions of core tasks, suggesting a tactic in which a student either tests the tasks, is in the process of solving them, or solves them only after several initial attempts; the \emph{correct submissions} tactic (\texttt{F\_CurTasks\_Core\_Correct}, $N=410$) in turn contains a high number of correct submissions of core tasks, suggesting a tactic in which a student either quickly solves the core tasks after an initial incorrect attempt or solves them after multiple attempts. Similarly, two tactics exist for basic tasks: one for \emph{attempting} (\texttt{F\_CurTasks\_Basic\_Attempt}, $N=369$) and one for \emph{correct submissions} (\texttt{F\_CurTasks\_Basic\_Correct}, $N=529$).
  \item \emph{Focusing on the current week's extra or supplementary tasks} (\texttt{F\_CurTasks\_Extra}, $N=640$): This tactic captures students' focus on materials beyond the course book and mandatory exercises. These materials are primarily aimed at supporting students with the core tasks, for example through simplified guided exercises. The tactic also includes sessions focused on advanced bonus or guru tasks, but the relative frequency of these events is small.
  \item \emph{Attending a one-on-one session with a TA} (\texttt{TA\_Sess}, $N=57$): This tactic is characterized by attending one-on-one sessions with TAs as described in Section \ref{subsec:course_context}. While the primary events of this tactic are starting and ending the session, the tactic also includes a small number of other events, suggesting that the TA and the student occasionally looked at course materials beyond the course mini-project.
  \item \emph{Reviewing and fixing previous week's basic answers} (\texttt{F\_PrevTasks\_Basic}, $N=361$): In this tactic, a student reviews and fixes answers from the previous week. It primarily includes submitting answers to the previous week's basic questions, watching the task review session videos and viewing model answers. Overall, the tactic suggests a basic review of the previous week's tasks, inspecting the student's own answers and trying out model answers.
  \item \emph{Deeply reviewing and fixing the previous week's tasks} (\texttt{F\_PrevTasks\_Deep}, $N=629$): The final identified tactic relates to reviewing and likely fixing answers to the previous week's tasks. Given that students were required to review and fix their wrong answers in this course (see Section \ref{subsec:course_context}), this tactic includes a variety of events related to submitting answers to various old tasks.
\end{itemize}

We note that while each tactic is associated with distinct event codes, in practice student sessions may contain a mix of similar events.
For example, both the \emph{correct submissions} and \emph{attempting} tactics for basic tasks include events where students answered wrongly (\texttt{answer-wrong:tasks-cur-basic}) and correctly (\texttt{answer:tasks-cur-basic}), but the proportion of correct answers in the sessions is much higher for the former tactic (32\% correct vs.\ 7\% wrong) than for the latter (10\% correct vs.\ 37\% wrong).
As such, the interpretation focused only on the most prominent event codes in each cluster, with less frequent events treated as noise.

\subsection{Learning strategy formulation}

Using the above clusters, we recoded each student session into its associated learning tactic and extracted the timestamp of the first event in each session. We then grouped the sessions by student, and for each student's sessions we further grouped them by course week using the time of the first event. This yielded 458 weekly learning strategies in total.

\subsection{Learning strategy characterization}\label{subsec:weekly_strategy_types}

\begin{figure}[t]
  \centering
  \includegraphics[width=0.9\linewidth]{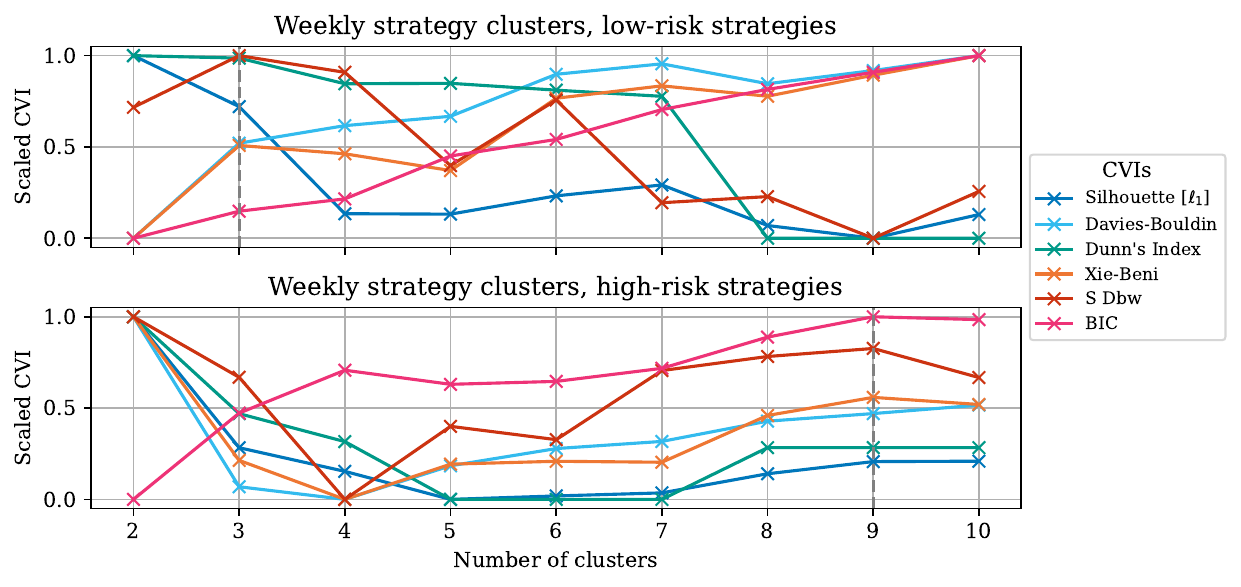}
  \caption{
    Cluster validation for clustering of the weekly learning strategies.
    Cluster validation indices are rescaled to fit the figure and adjusted so that higher values are better. The selected number of clusters ($K_{\text{drop}}=3$ and $K_{\text{pass}}=9$) are indicated with dashed lines.
  } \label{fig:weekly_strategy_clusters_cvis}
\end{figure}

\begin{figure}[t]
  \centering
  \includegraphics[width=\textwidth, height=\textheight, keepaspectratio]{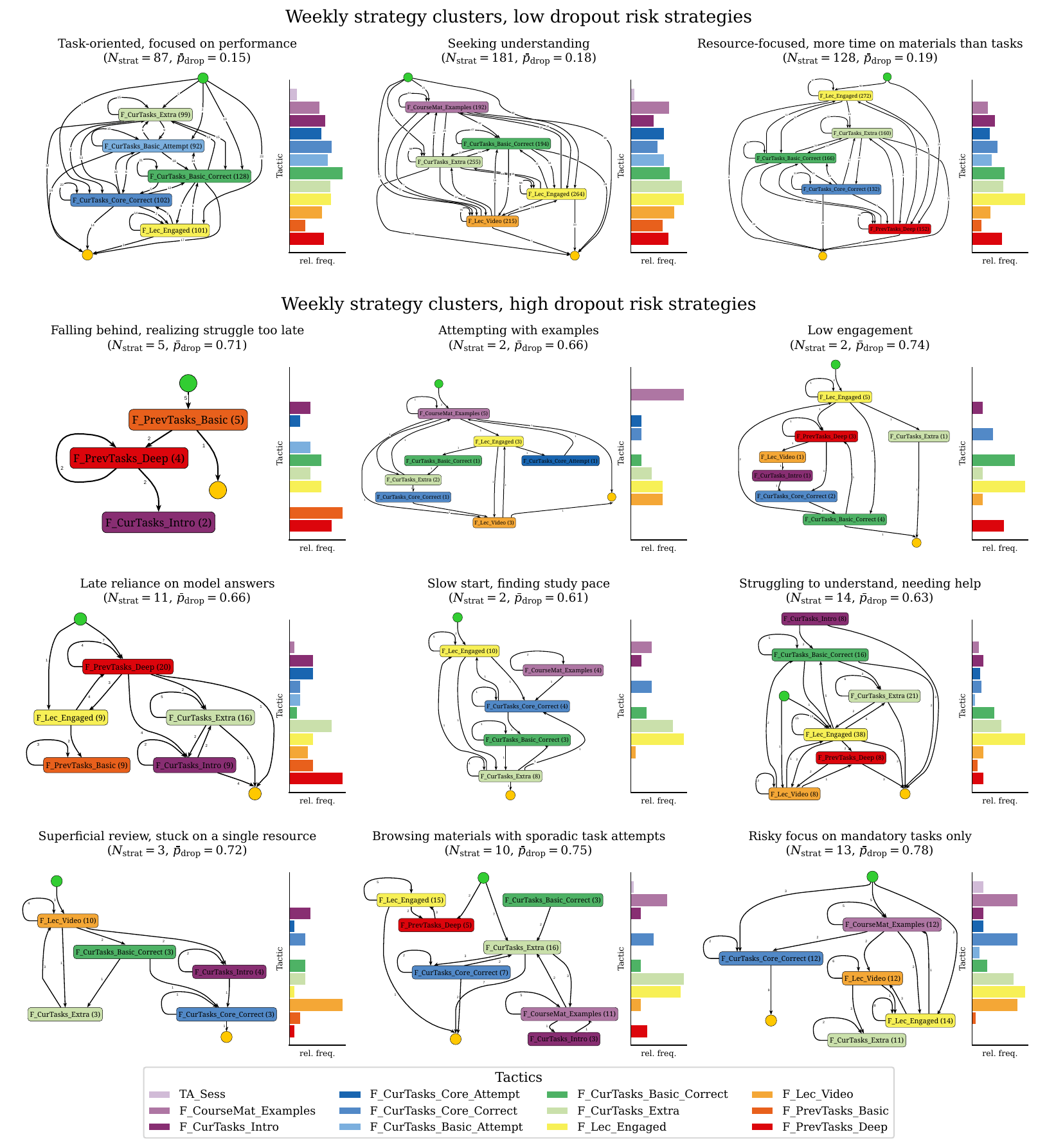}
  \caption{
    Summary of the identified weekly strategy clusters.
    For each cluster, the interpreted strategy type, the number of strategies in the cluster $N_\text{strat}$, and the average predicted dropout probability $\bar{p}_\text{drop}$ are given.
    Each cluster includes a heuristic net that depicts the most important transitions between tactics within the cluster.
    The relative frequencies of all learning tactics in the cluster are shown as bar plots; the colors representing the learning tactics are listed in the figure legend and described in Section~\ref{subsec:learning_tactics}.
    \vspace{4em}
  } \label{fig:weekly_strategy_clusters}
\end{figure}

Using the weekly dropout predictor described in Section \ref{subsec:weekly_strategy_types_method} for each student, we grouped the weekly learning strategy logs into low- and high-dropout risk strategies.
In total, 396 strategies were categorized as low dropout risk ($p_{\text{drop}} \leq 0.5$) and 62 as high dropout risk ($p_{\text{drop}} > 0.5$).

As with the identification of learning tactics, we used cluster analysis with multiple CVIs to explore the types of weekly study behavior.
Initial cluster validation for the EM clustering algorithm using BIC suggested using a diagonal covariance matrix.
The clustering performance of the EM algorithm on low- and high-dropout risk strategies is presented in Figure \ref{fig:weekly_strategy_clusters_cvis}.
For low-risk strategies, three CVIs suggest 10 clusters. However, these indices appear to exhibit a linear monotonic increase after $K=8$ while the other three CVI values decrease, suggesting possible over-segmentation rather than true structural discovery \citep{Liu2010UnderstandingInternalClustering}. Because we had no prior expectations about the number of behavior types we expected to see, we aimed to select the highest cluster number supported by multiple CVIs in order to obtain as fine-grained a clustering as possible while avoiding over-segmentation. Overall, we selected $K_\text{pass} = 3$ clusters for low dropout risk strategies, as two CVIs attain the highest value there (Dunn's Index, S Dbw) and it is a notable elbow point for two other CVIs (Xie-Beni, Davies-Bouldin).
Similarly, we selected $K_\text{drop} = 9$ clusters for high dropout risk strategies, as multiple CVIs attain a high value and it is an elbow point for some of the CVIs.

Figure \ref{fig:weekly_strategy_clusters} summarizes the clusters of low- and high-dropout risk strategies. For each cluster, we include the distribution of learning tactics.
We also generated heuristic nets for each cluster to aid interpretation. The nets were heuristically simplified to include only transitions between the most frequent events.
When interpreting the clusters, we also referred to dropout predictions and performance data (see Section \ref{subsec:weekly_strategy_types_method} for a description of the available data).

Overall, we interpreted the following learning strategy types from clusters of low dropout risk strategies:

\begin{enumerate}
  \item \emph{Task-oriented, focused on performance}: Student invests significant time in completing mandatory and basic tasks, ensuring the main weekly goal is finished.
  \item \emph{Seeking understanding}: Student prioritizes gaining understanding through extensive use of lecture materials and supplementary resources before and during task completion.
  \item \emph{Resource-focused, more time on materials than tasks}: Student focuses first on course materials rather than task completion, consistently attending lectures and reviewing model answers.
\end{enumerate}
In turn, we interpreted the learning strategy types below from clusters of high dropout risk strategies:
\begin{enumerate}[resume]
  \item \emph{Falling behind, realizing struggle too late}: Student views previous weeks' model answers without engaging in current tasks.
    The median grade of students inside the cluster is 0, indicating that this strategy type is used by students near dropout.
  \item \emph{Attempting with examples}: Student relies on examples to attempt tasks, possibly experimenting with different tactics to find a suitable approach.
  \item \emph{Low engagement}: Student relies solely on lectures as a learning resource, completing few tasks with points below the weekly goal.
    No effort used to seek help from other course materials.
  \item \emph{Late reliance on model answers}: Student depends on viewing model answers from previous weeks retroactively to catch up with the course pace, with minimal engagement with other materials.
  \item \emph{Slow start, finding study pace}: Student initially completes few random tasks despite using a variety of materials.
    Occurred only in the first week, representing an initial struggle to find the study rhythm. 
  \item \emph{Struggling to understand, needing help}: Student spends considerable time searching through course materials in an effort to understand (cf. \emph{Seeking understanding}) but fails to grasp the concepts, completing fewer tasks.
  \item \emph{Superficial review, stuck on a single resource}: Student engages primarily in watching videos without utilizing other support materials or progressing to task completion.
  \item \emph{Browsing materials with sporadic task attempts}: Student focuses on supplementary materials and tasks as the main activity, making random attempts at tasks, possibly to grasp weekly content but generally completing few or no actual tasks.
  \item \emph{Risky focus on mandatory tasks only}: Student concentrates solely on mandatory tasks with minimal involvement in other course activities, either only attending one-on-one supervision sessions or engaging very little otherwise, representing a risky approach to course completion.
\end{enumerate}

\subsection{Student characterization}\label{subsec:student_learning_profiles}

\begin{figure}[t]
  \centering
  \begin{subfigure}[b]{0.9\textwidth}
    \centering
    \includegraphics[width=\textwidth]{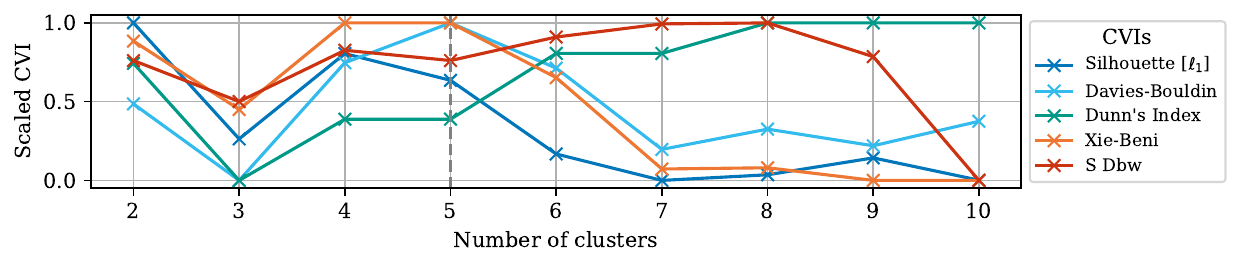}
    \caption{Cluster validation indices for student profile clusters. Cluster validation indices are rescaled to fit the figure and adjusted so that higher values are better.}
    \label{fig:sutdent_profile_clusters_cvis}
  \end{subfigure}
  \hfill
  \begin{subfigure}[b]{0.7\textwidth}
    \centering
    \includegraphics[width=\textwidth]{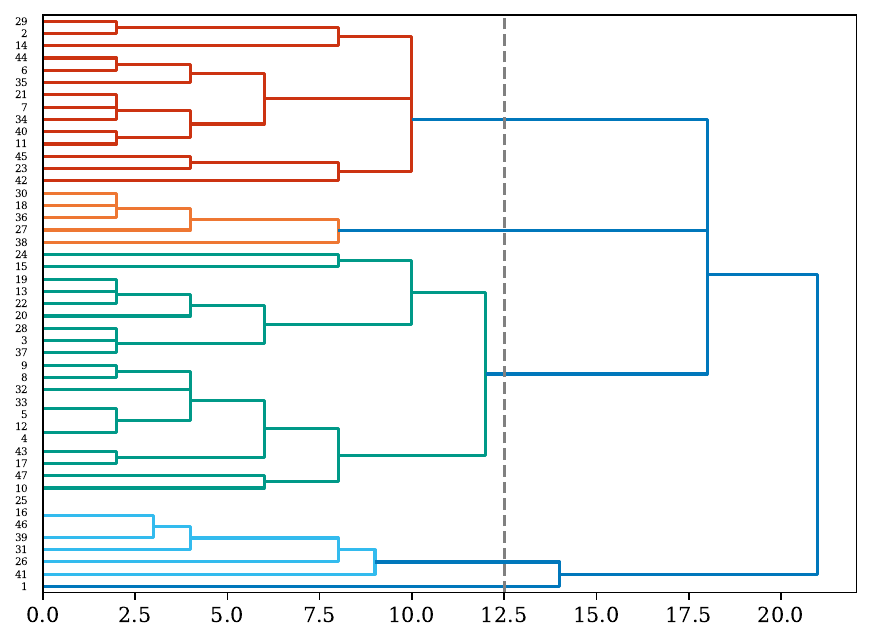}
    \caption{Dendrogram of the clustered student profiles.
      The vertical axis includes identification numbers of each student.
    }\label{fig:student_strategy_profiles_dendrogram}
  \end{subfigure}
  \caption{
    Cluster validation for student profile clusters using CVIs and a visual approach using a dendrogram. The selected number of clusters ($K_{\text{student}}=5$) is indicated with a dashed line.
  } \label{fig:student_profile_cluster_validation}
\end{figure}

\begin{figure}[h!]
  \centering
  \includegraphics[width=0.9\textwidth]{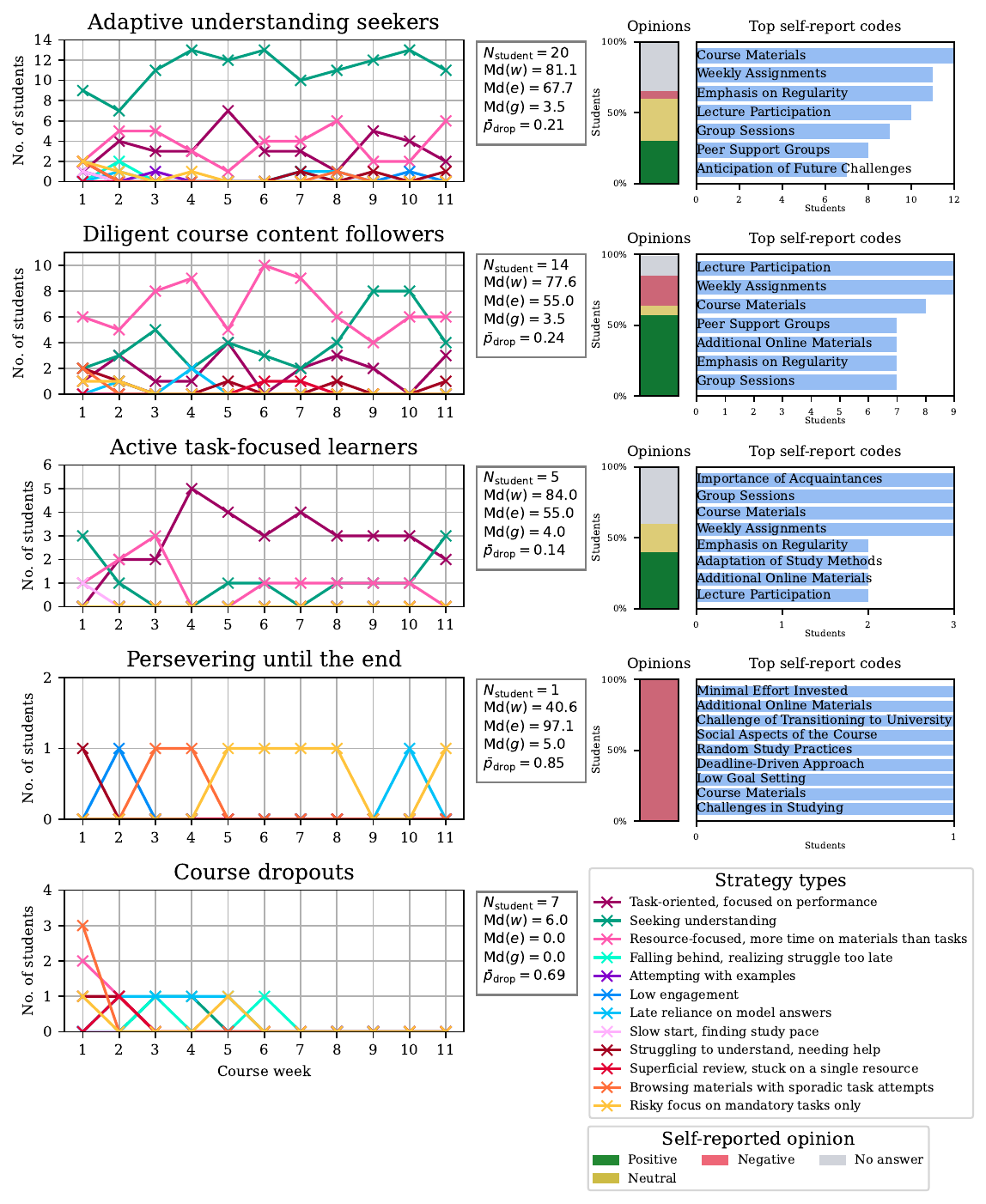}
  \caption{
    Student profile clusters displayed as the number of students that used each weekly strategy type throughout all course weeks.
    For each cluster, the interpreted profile type name, number of students in the cluster $N_\text{student}$ median percentage of the completed weekly tasks $\text{Md}(w)$, median percentage of exam points $\text{Md}(w)$, median course grade on scale 0--5 $\text{Md}(g)$ and mean dropout prediction over all weekly strategies $\bar{p}_\text{drop}$, is listed.
    For clusters with students who filled out the self-report, themes mentioned by at least half of the answerers are listed (see Appendix~\ref{appendix:themes} for a list of all themes and descriptions). Additionally, a distribution of self-reported opinions of own effort are shown. Cluster interpreted as \emph{Course dropouts} does not include any students with self-reports.
    \vspace{1.5em}
  } \label{fig:student_profile_clusters}
\end{figure}

Initial validation suggested using the complete linkage criterion in agglomerative clustering.
To select the appropriate number of student profile clusters, we used both CVIs (Figure \ref{fig:sutdent_profile_clusters_cvis}) and a dendrogram (Figure \ref{fig:student_strategy_profiles_dendrogram}).
On closer inspection of the CVIs, $K=4$ and $K=5$ emerged as potential cluster numbers, as some CVIs either reach the maximum value or show a clear increase.
Based on an additional visual review of the dendrogram, one student's profile appears clearly separate from the others. We therefore selected $K_\text{student} = 5$ as the cluster number, with four clusters representing learning profile types and one cluster containing the outlier student.

An overview of the interpreted learning profile types, along with a summary of the available self-report data, is shown in Figure \ref{fig:student_profile_clusters}. A full listing and description of all themes that emerged in the study's analysis is provided in Appendix~\ref{appendix:themes}.
Three of the learning profiles represent various levels of activity during CS1, while one learning profile includes all students who dropped out mid-course. The profile of the single outlier student represents a low-activity student who managed to pass the course by compensating for low task performance with good exam results.
All clusters include some participants without submitted self-reports, and the cluster of course dropouts contains no self-reports at all. The profile of the single outlier contains one self-report that was used extensively to characterize the student.
Nevertheless, visually, there appear to be some differences between clusters in the self-reported opinions of students' own performance.
Next, we describe the learning profile types interpreted from the clusters based on the visualizations, descriptive statistics, confirmatory tests, and excerpts from the self-reports.
When reporting the excerpts, we use student identifiers in relation to all original participants in the study ($N=47$). The same identifiers appear in Figure~\ref{fig:student_strategy_profiles_dendrogram}.

\subsubsection{Adaptive understanding seekers}

The first interpreted profile type ($N_\text{student} = 20$) contains students who primarily sought understanding and adapted their study strategies as needed. In general, these students stuck to the understanding-seeking strategy throughout the course, but they also adjusted their strategies on a weekly basis. Based on the available reports (13 students), most of the adjustment was made in response to personal circumstances, such as switching from on-site lectures to online videos owing to a lack of time or changes in personal life. Some students, however, adjusted their strategies based on reflections on their learning, as exemplified in the quotes below:
\begin{quote}
  [Student 10]: \emph{[My] routines for setting the week's rhythm improved during the first course week. Initially, I did double the work a bit too much, e.g., re-watching lecture recordings multiple times because I did not read the course materials first.}
\end{quote}
\begin{quote}
  [Student 33]: \emph{As the course progressed, I read the materials less and less. I started reading theory texts only when completing the weekly tasks. I felt I learned the best by applying the written information right away.}
\end{quote}
\begin{quote}
  [Student 4]: \emph{I noticed that theory-focused learning is more natural and familiar to me [...] I made a lot of notes to the tasks and course book. [...] Compared to my previous experiences I changed my study methods in a way by using more time on exercises, as this is clearly the best way to learn...}
\end{quote}
Nevertheless, most of the self-reports in this cluster (11 out of 13) emphasized the importance of developing regular study habits, indicating that the adaptation was mostly a response to the weekly topic or changed circumstances.
Based on the learning strategies, some students occasionally slipped into high dropout risk strategies but eventually managed to recover by adjusting their strategy. However, there is no mention of this in the self-reports.

All of the students with self-reports in this cluster reported using multiple learning resources, with the course book (12 reported), lectures (10 reported) and weekly group learning sessions (9 reported) being the most commonly used. Some students also used materials external to the course, as illustrated in the quote below:
\begin{quote}
  [Student 3]: \emph{I survived [the course] by mainly using the course materials and attending weekly group sessions. Also, the peer support group motivated and ``tied'' me to completing the weekly tasks. I peeked at some info via a search engine; I hope was able to differentiate between valid and invalid guides.}
\end{quote}
As such, this cluster represents students who really wanted to ensure understanding of the topic while progressing through the course.
It must be noted that not all students in this cluster succeeded with high grades; those who self-reported bad experiences with the course attributed it to external factors such as course workload and personal life:
\begin{quote}
  [Student 19]: \emph{The course has a high workload and for me required much more effort than 10--12 hours weekly. I enrolled in the course out of interest, which is why other studies took priority.}
\end{quote}
\begin{quote}
  [Student 12]: \emph{When doing the course while employed, you can't use all your time for the course. On the other hand, studying is a good counterbalance to work.}
\end{quote}
In addition, two students (Student 8 and Student 28) did not complete all course requirements in time, which prevented them from getting the final grade despite being otherwise active during the course. Student 28 filled a self-report, where they explicated the struggles as follows:
\begin{quote}
  [Student 28]: \emph{This fall was very busy, and during some weeks I did not make as much effort as needed to deeply learn the topic.}
\end{quote}

While the self-reported opinion of own progress appears visually positive, the Brunner-Munzel test suggests that the opinions in this profile type's cluster are stochastically equal to the opinions of other students who answered the self-report ($\hat{p}^*(25.886) = 0.283, p = .779$). In other words, the opinions of students in this profile type and of the other students tended to be the same. Similarly, the final course grades of this profile type's students and other study participants are stochastically equal ($\hat{p}^*(44.684) = -0.953, p = .346$).
On the other hand, the dropout predictions for the weekly strategies used by students of this profile type and those of the other study participants are not stochastically equal ($\hat{p}^*(434.690) = -3.351, p < .001, \hat{p}'' = 0.589$); the effect size indicates that students of this profile type tended to use strategies with lower dropout probability than other participants.

\subsubsection{Diligent course content followers}

The second profile type ($N_\text{student} = 14$) represents students who conscientiously used the course materials and appeared to follow the suggested weekly schedule described in Section \ref{subsec:course_context}.
Based on the frequency of the learning strategies (Figure \ref{fig:student_profile_clusters}), their primary strategy was material-focused for most of the course. However, toward the end of the course, more students shifted to the understanding-seeking strategy. This may indicate a need to strengthen comprehension before the final exam, as the suggested course schedule may not have been enough to gain a sufficient understanding of the topics.
All students with self-reports in this profile type (12 students) reported following the course schedule, with most attending lectures on-site and some watching lecture recordings. Only one student reported skipping lectures at the start of the course:
\begin{quote}
  [Student 44]: \emph{At the beginning, I attended lectures, but right during the first weeks of the course I realized that I absolutely cannot concentrate there and that it was a waste of time. After that, I went directly to the lecture examples and to the weekly exercises.}
\end{quote}

All students with self-reports in this profile type also reported relying on some form of help-seeking resource, with weekly group learning sessions (7 reported), peer support groups (7 reported) and friends or relatives (6 reported) being the most commonly mentioned.
Looking more closely at the reports, most students emphasized the importance of seeking help:
\begin{quote}
  [Student 2]: \emph{I read course materials, I made a learning map, I sought help via Google and asked for help from the people I know.}
\end{quote}
\begin{quote}
  [Student 5]: \emph{...my partner who knows how to code helped me a lot.}
\end{quote}
\begin{quote}
  [Student 23]: \emph{I was in a peer support group, which helped me to schedule the weekly tasks. I also asked for help from friends a lot.}
\end{quote}
Some students also reported relying heavily on the teaching assistants, suggesting some dependence on them to complete tasks and even to self-regulate:
\begin{quote}
  [Student 40]: \emph{I used over 20 hours per week on the course but didn't understand anything without group sessions and teaching assistants.
  [...] I don't even dare to attend CS2 this spring since I'm so mentally exhausted.}
\end{quote}
\begin{quote}
  [Student 43]: \emph{I felt I was left by myself with my problems the first time studying CS1, so now I tried to make use of all the available help channels.}
\end{quote}
Overall, while following the ideal schedule worked for some students, others reported the course to be challenging and struggled to adapt their strategies, resulting in varied outcomes.

In this profile, the self-reported opinions are visually more negative than in other profiles. However, similar to adaptive understanding seekers, the opinions of students of this profile type appear to be stochastically equal to opinions of other answerers ($\hat{p}^*(17.963) = -0.481, p = .637$). Further, the final course grade is also stochastically equal to that of other students ($\hat{p}^*(39.376) = -1.297, p = .202$). In turn, predicted dropout probability of weekly strategies that the students of this profile type used is not stochastically equal to those of other study participants ($\hat{p}^*(453.337) = 2.143, p = .033, \hat{p}'' = 0.443$). The effect size suggests the students in this profile tended to use strategies with higher dropout probability than other students.

\subsubsection{Active task-focused learners}

The active task-focused learner profile type ($N_\text{student} = 5$) represents students who focused primarily on completing the weekly tasks.
Unlike the two aforementioned profiles, these students consistently employed effective, low dropout risk strategies throughout the course, with only one student showing initial adjustment struggles during the first week.
That student did not specify the reason behind the slower start and mentioned only a generally slower learning pace:
\begin{quote}
  [Student 27]: \emph{[...] even though I have slipped through my schedule a bit, I was able to get the course assignment and weekly exercises done well.}
\end{quote}
Overall, the active focus on course tasks is reflected in the low dropout predictions of the weekly strategies and the high median percentage of completed weekly tasks.
The task-centered approach is also reflected in the self-reports (3 students), exemplified by the following quote:
\begin{quote}
  [Student 36]: \emph{At the start of the week I read the coursebook, studied the weekly tasks and completed them since I knew how to do them.}
\end{quote}
However, high task activity does not guarantee a high grade. One student mentions a language barrier as an external obstacle to their performance:
\begin{quote}
  [Student 18]: \emph{This was my first year studying CS. Because of this and the language barrier, I had a lot of problems. Because of this, self-study worked better for me. I did not find help from group supervision sessions.}
\end{quote}

By visual inspection, the active task-focused learners' self-reports appear more positive than those of the other students. However, the overall self-reported opinions of students in this profile type appear stochastically equal to those of the other self-report answerers (permuted Brunner-Munzel, $p = .713$). While the median grade appears higher, there is no statistically significant support for the difference (permuted Brunner-Munzel, $p = .350$).
The dropout predictions of the strategies used by students in this profile type are not stochastically equal to those of the other study participants ($\hat{p}^*(73.978) = -4.379, p < .001, \hat{p}'' = 0.662$). Based on the effect size, students in this profile tended to use more low dropout risk strategies than the other study participants, agreeing with the visualization of the learning strategy types used.

\subsubsection{Persevering until the end}\label{subsubsec:perservering}

A notable outlier among all students is Student 1, who completed the course with the highest grade despite only minimally interacting with the weekly tasks. Even though they relied only on high dropout risk strategies, the student achieved the top grade by obtaining near-best points on the final exam.
While the final grade can be seen as an indicator of good performance, the student's self-evaluation of their own learning methods is much more critical, setting a low bar for their own performance:
\begin{quote}
  [Student 1]: \emph{[The study style] was not really working. The worst for me was time management, which required extra effort. For now, my goal is to just pass the course for which my approach should be good enough.}
\end{quote}
When further reflecting on regulating own effort, the student brought up a larger issue with transitioning to university:
\begin{quote}
  [Student 1]: \emph{...I had difficulties getting accustomed to university studies. At the same time I had to also get used to living by myself.}
\end{quote}
Visually, the student's strategy was focused on a quick, risky attempt to complete the mandatory tasks, and in course week 9 there are no observed student sessions at all. This risky approach to studying is confirmed in the self-report:
\begin{quote}
  [Student 1]: \emph{I usually used one day for programming stuff. With the course assignment I used a bit more time in places.}
\end{quote}
When answering the question about scheduling learning, the student notes a reactive approach to deadlines:
\begin{quote}
  [Student 1]: \emph{Actually, I went by deadlines and personal life, depending on how I felt or had time. }
\end{quote}
Further, the student studied alone, citing avoidance of social events as the main reason:
\begin{quote}
  [Student 1]: \emph{I mostly studied remotely, to which I am not accustomed to, but large lecture halls don't really seem attractive to me.}
\end{quote}
Nevertheless, the main reason for passing the course could be attributed to the student's ability to find the necessary information:
\begin{quote}
  [Student 1]: \emph{I mainly used course materials and the internet. I was doing pretty well in terms of finding the right information. [...] The course materials were good.}
\end{quote}

\subsubsection{Course dropouts}

The final profile type ($N_\text{student} = 7$) includes all participants
who dropped out during the course. While no students in this profile completed the self-report questionnaire, Brunner-Munzel tests confirm that both the grades ($\hat{p}^*(39) = 27.221, p < .001, \hat{p}'' = 0.975$) and the dropout predictions ($\hat{p}^*(20.362) = 14.408, p < .001, \hat{p}'' = 0.085$) of the students in this profile are not stochastically equal to those of the other students. Based on the effect sizes, the students in this profile tend to have lower grades and higher predicted dropout than the other students.

Upon a manual review of all the students, we identified three paths to dropout:
\begin{itemize}
  \item four students dropped out during the first course weeks, using only high dropout risk strategies;
  \item two held on to the course until the difficulty increased considerably in weeks 3 and 5;
  \item one student started with low dropout risk strategies and managed to continue through the initial complex topics but quickly shifted to high dropout risk strategies and dropped out right after a particularly difficult topic was introduced.
\end{itemize}
In other words, even when a student in this profile managed to use low dropout risk strategies, all students reverted to high-risk strategies 1--2 weeks prior to dropout. All students in this profile eventually dropped out by the midpoint of the course, when the programming tasks became more complex.

\section{Discussion}

This study aimed to examine the behavioral patterns of CS1 students throughout the course by analyzing trace logs and student dropout predictions, interpreting the results through the lens of SRL.
The findings demonstrate that combining event-based logs, dropout predictions, and basic self-reports reveals clear differences between high and low dropout risk CS1 student behaviors, providing detailed insights into their successes and struggles throughout the course.
Although detecting and understanding learning patterns associated with different dropout risks can help address course attrition, research into automatic pattern detection remains scarce in CSEd \citep{Omer2023LearningAnalyticsProgramming}.
This study therefore contributes to the research area by demonstrating a method for semi-automated learning pattern analysis that complements and extends qualitative inquiries into learning patterns of programming students, such as those by \citet{Liao2019BehaviorsHigherLowera}.
Furthermore, by relating trace log data to self-reports to obtain a more comprehensive narrative of student behavior during CS1, this study contributes to the discussions regarding practical applications of event-based measures \citep[e.g.,][]{Moreno-Marcos2020TemporalAnalysisDropout,Fan2023FullerPictureTriangulation,Han2023LevelConsistencyStudents,Choi2023LogsSelfReportsMisalignment}.

\subsection{Addressing the research questions}

Overall, we identified and interpreted 12 distinct weekly learning strategy types of various dropout risk (see Section \ref{subsec:weekly_strategy_types}) and categorized student course behaviors into five learning profile types (see Section \ref{subsec:student_learning_profiles}). The results bear resemblance to those of \citet{Liao2019BehaviorsHigherLowera} and \citet{Moreno-Marcos2020TemporalAnalysisDropout}, confirming the validity of the study. Further, the weekly learning strategy types relate to other studies in which SRL learning strategies were mapped in CS1 context \citep{falkner2014identifying,Silva2024WhatLearningStrategies,Garcia2018SystematicLiteratureReview}.

Regarding the learning behaviors associated with low dropout risk (\ref{rq1}), we identified three clusters among strategies with low dropout probability and interpreted them as the following weekly learning strategy types: \emph{task-oriented, prioritizing main task performance}; \emph{seeking understanding}; and \emph{resource-focused, prioritizing course materials over tasks}. These strategies align with \citet{Liao2019BehaviorsHigherLowera}, who reported that high-performing students tend to use more varied course resources. Furthermore, information seeking and experimentation with simpler tasks before solving more complex problems are common SRL strategies in computer science education, as exemplified by \citet{falkner2014identifying} and \citet{Silva2024WhatLearningStrategies}. Although prioritizing task completion is not generally reported as an SRL strategy itself, it can be considered a form of goal-setting and planning, as students allocate more time and effort to tasks \citep[e.g.,][]{Garcia2018SystematicLiteratureReview}. Additionally, from clustering student behaviors over the entire course, we identified three clusters that we interpreted as learning profile types associated with course success: \emph{adaptive understanding seekers}, \emph{diligent course content followers}, and \emph{active task-focused learners}. These profile types relate to each other in that most of the students of these types applied low dropout risk strategies throughout the course, and often actively varying strategies depending on the weekly topic. This aligns with \citet{Liao2019BehaviorsHigherLowera}, who noted that high-performing students tend to vary their study approaches depending on the task.

Regarding the learning behaviors associated with high dropout risk (\ref{rq2}), we identified nine distinct clusters among strategies with high dropout probability from which we interpreted high dropout risk strategy types. Unlike the low dropout risk strategy types, these strategy types show greater variation in behavior and dropout risk. For instance, one of the interpreted strategy types, \emph{slow start, finding study pace}, does not by itself indicate dropout, as study participants who employed this strategy improved their learning style as the course progressed. This strategy type aligns with the common observation that students may initially struggle to self-regulate in a new context but learn it over time \citep[e.g.,][]{Loksa2016RoleSelfregulationProgramminga,Winne2019LearningStrategiesSelfRegulateda}. Most of the interpreted strategy types, such as \emph{attempting with examples}; \emph{low engagement}; \emph{late reliance on model answers}; and \emph{struggling to understand, needing help}, represent in our view basic temporary unsuccessful strategies that even successful students may experience and recover from, as observed in the identified student profile clusters (cf. Figure \ref{fig:weekly_strategy_clusters}). Such notion has been noted as well by \citet{Liao2019BehaviorsHigherLowera}, who noted that lower-performing students can overcome problems just as high-performing students may temporarily struggle. However, the final interpreted strategy types --- \emph{falling behind, realizing struggle too late}; \emph{superficial review, stuck on a single resource}; \emph{browsing materials with sporadic task attempts}; and \emph{risky focus on mandatory tasks only} --- seem to describe behaviors of students on the verge of dropping out. In the case of the course dropouts present in this study, these strategy types were among the last ones those students attempted before dropping out. In our view, these strategy types can be related to issues with time management and difficulty assessment, both identified by \citet{falkner2014identifying} as the most frequent unsuccessful SRL strategies in CS. On the other hand, these strategy types can be simply seen as acts of giving up prior to dropout. In other words, those students who persevere through problems recovered from high-risk strategies, while dropouts remained locked in high-risk behavior until they were overwhelmed by the workload --- a common narrative for dropping CS1 \citep{Petersen2016RevisitingWhyStudents,Kinnunen2006WhyStudentsDrop}.

Regarding the interplay between the self-reports and trace log data (\ref{rq3}), we found that SRL as we observed and interpreted via trace logs and students' self-reported evaluations of their own SRL are generally consistent and complementary. Although we initially mapped learning strategies and student profiles using only trace logs and dropout prediction, self-reports enhanced the interpretation of the types. Specifically, both trace log-based data and self-reports were equally valuable but provided different perspectives.
While \citet{Liao2019BehaviorsHigherLowera} and our self-reports offer a snapshot of students' cognition at the end of the course, trace logs allow following the full learning path of a student and capture their overall learning profile as transitions between learning strategies (cf. Figure \ref{fig:weekly_strategy_clusters}).
Although the somewhat open-ended self-report tool we used may not be appropriate for predicting dropout on its own, it aids describing student behavior.
In this study, the self-reports were instrumental in being able to differentiate the three main learning profile types: \emph{adaptive understanding seekers}, \emph{diligent course content followers}, and \emph{active task-focused learners}. Furthermore, the profile of a student who passed the course with full marks despite being identified as high-risk (see Section \ref{subsubsec:perservering}) would not have a reasonable explanation without the student's self-report.
Thus, our results confirm an argument made in an earlier study \citep{Zhidkikh2023MeasuringSelfRegulated}: using predictive learning analytics and trace logs alongside self-reports allows for a more comprehensive understanding of how and why a student uses the available learning resources.
Aligning with numerous prior studies \citep{Tempelaar2020SubjectiveDataObjective,Zhidkikh2024ReproducingPredictiveLearning,Fan2023FullerPictureTriangulation,Lim2021ImpactLearningAnalytics}, we posit that the inclusion of self-reports can improve the interpretation of purely trace log-based analyses by contextualizing them and bringing in cognitive aspects that cannot be easily captured as atomic events.

\subsection{Practical implications}

We believe the results demonstrate students' study habits and progress in a way that can be useful for educating TAs. This knowledge can, of course, be shared among teachers, while TAs are recruited to various roles in which they interact with students, including autonomous positions and providing guidance on the coursework \citep{MirEtAl19}. In using the results as educative material, specific attention could be given to illustration of dropouts, as these cases show the point in the course over which students with unproductive strategies hardly progress and encourage early interventions. On the other hand, the outlier student profile (\emph{Persevering until the end}) demonstrates that, regardless of strategies that appear risky, a student may achieve a good result and is hence likely to have good potential for finding and grasping the material (see illustration at the end of Section \ref{subsubsec:perservering}). From this perspective, a view that the course structure ``upholds'' beginner students (through allowing a focus on mandatory exercises only or late rework with them) may pay off for some students. In effect, the student in the outlier profile acknowledged difficulties in transitioning to university, illustrating that student difficulties need to be screened and addressed in a broader scope than strategies with learning to program.

An intervention between a TA and a student could begin gently, with the TA asking the student to reflect on their study habits in relation to the learning profile types (clusters in Figure~\ref{fig:student_profile_clusters}). Beginning from the student's reflection conforms to good dialogue-based counseling practice \citep{Pea98}, with TAs avoiding the situation of needing to be able to give ``best'' advice to students. The goal would be to prompt the student to consider whether their current strategies seem risky or successful. When necessary, the profiles identified as successful can be offered as suggestions for improving learning strategies. Such a use of the results would also serve to validate them, revealing to what extent the current profiles are applicable in practice or if more varied profiles exist among student cohorts.

We would also pay attention to students within the learning profile types generally considered to reflect successful strategies. We observed that they may yet achieve rather different grades. We anticipate that this condition derives from students' varied backgrounds, which also calls for sensitive dialogue in supervision sessions in the scope broader than navigation through course exercises.

Learning analytics in programming courses have been noted to lack granularity, as efforts are often based on limited data, such as assessment results \citep{Omer2023LearningAnalyticsProgramming}. The present study revealed paths of students progress through clustered profiles that became available by employing multiple types of data, including trace log data and weekly prediction data, along with the students' end-of-course reflections. Above, we described one scenario of using these results as an \emph{informative} source for supervision \emph{during} subsequent courses. A future goal that naturally intrigues us as teachers is the possibility of seeing timely, automated characterizations of students' current progress. This leads to the well-acknowledged challenges of how teachers can effectively use learning analytics data and how effective automated feedback can be generated \citep{Shibani2020EducatorPerspectivesLearning}. We plan to continue by replicating the present study with larger samples and varied contexts, and then explore methods to automatically map new datasets to previously developed clusters. \citet{Shibani2020EducatorPerspectivesLearning} noted the need to inform users of learning analytics outcomes about potentially misleading information. If the suggested future direction turns out fruitful, the automated results could be used safely in dialogue-based supervision sessions, where the nature of such analytics can be critically reflected on.

\subsection{Trustworthiness and limitations}

We recognize several considerations regarding trustworthiness and limitations in our study design. In this subsection, we have argued for our chosen research setting, its limitations, and how we have strived to mitigate potential threats to validity.

\textit{Number of participants}: The number of student participants in this study is a potential limitation considering the total number of students in the course.
The sample size poses potential threats to validity with regard to sampling bias, as participation was voluntary and we cannot ensure a truly random sample of CS1 students.
However, despite the limited number of participants (\textit{N} = 47), the number of \textit{observations} relevant to the study is larger: 5,445 student sessions grouped into 458 weekly learning strategies, collected over 11 weeks of the course.
The number of participants is apparent primarily in the characterization of students' study profiles, where statistical analysis was used as an additional aid rather than as the primary method of characterization.
Because our data did not meet the independence assumption due to the study design, we used the Brunner-Munzel test for statistical comparison. This test has been shown to be robust with smaller sample sizes than other comparable non-parametric tests. SRL trace log analysis has also been applied successfully with a relatively small number of participants in earlier work \citep{Zhidkikh2023MeasuringSelfRegulated}.

\textit{Changes in study profile}: We categorized \textit{weekly} learning strategies in our analyses given the weekly schedule of the course (see Section \ref{subsec:course_context}), yet this level of abstraction might not capture all nuances of changes in a student's study profile. That is, changes in learning strategy are likely to change more often than on a weekly basis. This choice was by design to abstract the results to be more easily interpreted, yet the approach might overgeneralize the phenomenon of study profiles.

\textit{Selection of appropriate clustering}: The clustering performed in Section~\ref{subsec:learning_tactics} and Section~\ref{subsec:weekly_strategy_types} could be done with a higher or lower number of clusters, yielding clustering at different levels of granularity.
To support our choice of clustering, we computed multiple cluster validation indices as a guide. However, each index tends to be sensitive to different aspects of the clustering \citep{Liu2010UnderstandingInternalClustering,Schubert2023StopUsingElbowa}.
Cluster analysis is ultimately an exploratory method that can yield multiple valid and interesting solutions, and relying solely on statistical indices may not be enough to produce a meaningful result \citep{Schubert2023StopUsingElbowa,Salvador2004DeterminingNumberClusters}.
Our choice of clustering was therefore informed by the indices, by our domain knowledge of the studied CS1 course, and by the pragmatic need to capture a \textit{sufficient} level of abstraction for the results to be relatively easy to interpret. To ensure transparency, we reported our observations and our process for choosing the cluster numbers in the sections referenced above.

\textit{Selection of study session length}: In Section \ref{subsec:log_preprocessing}, we set a 25-minute length for a study session in the trace log analysis, which affected the number and the granularity of the study sessions. We chose this value based on the course context and our estimate of how much time a student would typically need to complete one exercise. Other heuristics could have been more accurate, but there are no standardized rules for determining study sessions from trace logs \citep{Kovanovic2015DoesTimetaskEstimation}.
Some other studies \citep[e.g.,][]{Jovanovic2017,Matcha2019DetectionLearningStrategies} have used a 30-minute cut-off without an explicit rationale or grounding in the course context.
Reliably determining when a study session ends would have required a more controlled experiment (e.g., a laboratory setting or capturing participant screens). We hypothesized that such a requirement would have yielded considerably less participants.

\textit{Self-report answer interpretation}: In addition to the statistical tests, we analyzed the self-reports of 29 participants using content analysis. While the results yielded by content analysis are subject to researcher interpretation and subjective bias, we strived to mitigate this by having two authors analyze the data independently, discussing possible discrepancies, and forming a consensus, as detailed in Section~\ref{sec:data_analysis}. Despite this approach, another group of researchers might have formed at least slightly different set of themes.

\textit{Nature of self-reports}: A potential concern is that self-reports, by nature, are prone to recall bias. In this study, self-reports were administered before the course exam, while students' memory of their study habits was still fresh.
Further, self-reports were used as only one of several data sources, with trace logs providing objective information on student performance. We did not observe any discrepancies between the two data sources that would suggest recall bias, although we acknowledge that some self-reports may omit information that could affect the interpretation of the student profiles.
Another potential source of bias is social desirability, i.e., the tendency of participants to respond in a favorable way.
In this study, this is mitigated by the voluntary nature of participation, which provided no direct benefit to the students.

\textit{Single-cohort study}: The participants were recruited from a single cohort in a single university course. This approach poses potential challenges to the generalizability of the results concerning, e.g., topics besides CS1 or computing in general, cross-cultural differences, different teaching approaches, or some intra-cohort intricacies not considered in this study.
For example, the studied CS1 course allowed students to compensate for low task performance with good exam results, yielding the \emph{Persevering until the end} profile that might not emerge under other course designs.
However, self-regulation is generally dependent on the course context \citep{Ben-Eliyahu2015}, and the syllabus, teaching methods, and assessment practices can all affect the specific approaches students take to learning.
From a methodological standpoint, we posit that grand generalizations of student behavior are difficult to make without assuming a closed system and absolute control over all external factors, which are considered problematic \citep[][p. 67]{Bhaskar1978}. Even with a different cohort and sample size, the behavior patterns identified in such a study would not negate those identified in the present cohort. We therefore emphasize the exploratory nature of our results and the value of the analysis pipeline to the course context in which the method is applied.
Finally, as with many other research settings, these potential differences may be examined in replication studies.

\section{Conclusion}

In this study, we investigated the behavioral patterns of students in a CS1 course and their interactions with the course's learning resources. By applying trace log analysis and student dropout prediction methods developed in two prior studies \citep{Zhidkikh2023MeasuringSelfRegulated,Zhidkikh2024ReproducingPredictiveLearning} to a cohort of CS1 students, we identified and interpreted 12 weekly learning strategy types based on students' interactions with course resources and four core learning profile types that describe how these interactions change over the course duration. Among these learning strategy types, three were interpreted from low dropout risk behaviors and nine from high risk behaviors, indicating that dropout risk can be observed in various forms and may require different types of interventions. We compared the interpreted strategy types with other studies on self-regulated learning, finding close connections to related works such as \citet{Liao2019BehaviorsHigherLowera}. Additionally, our observations of student behavior prior to dropout was consistent with prior studies that map how students drop CS1 \citep[e.g.,][]{Kinnunen2006WhyStudentsDrop,Petersen2016RevisitingWhyStudents}. Altogether, previous studies provide initial support and validation to our interpretations.

Further, we found points of contact between student learning profiles and self-reports, validating our interpretations of trace log data.
Overall, the combination of trace logs and dropout predictions enhanced the identification and interpretability of successful and at-risk student behaviors. Given that dropout prediction can be performed multiple times throughout the course, this study demonstrates a semi-automated method for obtaining detailed narratives of study behavior and associated dropout risk during the course. Notably, this research approach is easily applicable to large CS1 course implementations where trace logs and basic task performance data are available. Through this approach, educators can better understand student study behaviors to devise more timely and targeted interventions.

Nevertheless, this study has limitations that warrant further research. Most importantly, studies in different CS1 contexts and with larger cohorts are needed.
Given that the CS1 course from which we recruited participants had 321 enrolled students, there may be additional learning paths that our identified strategy types and learning profile types do not fully capture. Different CS1 contexts may also reveal further nuances in student behavior, particularly concerning dropout.
Additionally, we see potential for a reproduction study \citep[cf.][, Figure 2]{Ihantola2015EducationalDataMining} by examining paths to dropout using longitudinal analysis tools. For example, applying Latent Transition Analysis \citep{Lanza2003LatentClassLatent} to study sessions could uncover more intricate learning paths leading to dropout.

\bibliographystyle{ACM-Reference-Format}
\bibliography{main}


\begin{thebibliography}{98}


\ifx \showCODEN    \undefined \def \showCODEN     #1{\unskip}     \fi
\ifx \showISBNx    \undefined \def \showISBNx     #1{\unskip}     \fi
\ifx \showISBNxiii \undefined \def \showISBNxiii  #1{\unskip}     \fi
\ifx \showISSN     \undefined \def \showISSN      #1{\unskip}     \fi
\ifx \showLCCN     \undefined \def \showLCCN      #1{\unskip}     \fi
\ifx \shownote     \undefined \def \shownote      #1{#1}          \fi
\ifx \showarticletitle \undefined \def \showarticletitle #1{#1}   \fi
\ifx \showURL      \undefined \def \showURL       {\relax}        \fi
\providecommand\bibfield[2]{#2}
\providecommand\bibinfo[2]{#2}
\providecommand\natexlab[1]{#1}
\providecommand\showeprint[2][]{arXiv:#2}

\bibitem[Aggarwal and Reddy(2014)]%
        {Aggarwal2014}
\bibfield{author}{\bibinfo{person}{Charu~C Aggarwal} {and} \bibinfo{person}{Chandan~K Reddy}.} \bibinfo{year}{2014}\natexlab{}.
\newblock \bibinfo{booktitle}{\emph{Data Clustering : Algorithms and Applications}}.
\newblock \bibinfo{publisher}{CRC Press LLC}.
\newblock
\showISBNx{978-1-315-37351-5}
\href{https://doi.org/10.1201/9781315373515}{doi:\nolinkurl{10.1201/9781315373515}}


\bibitem[Arakawa et~al\mbox{.}(2021)]%
        {arakawa2021situ}
\bibfield{author}{\bibinfo{person}{Kai Arakawa}, \bibinfo{person}{Qiang Hao}, \bibinfo{person}{Tyler Greer}, \bibinfo{person}{Lu Ding}, \bibinfo{person}{Christopher~D. Hundhausen}, {and} \bibinfo{person}{Abigayle Peterson}.} \bibinfo{year}{2021}\natexlab{}.
\newblock \showarticletitle{In {{Situ Identification}} of {{Student Self-Regulated Learning Struggles}} in {{Programming Assignments}}}. In \bibinfo{booktitle}{\emph{Proceedings of the 52nd {{ACM Technical Symposium}} on {{Computer Science Education}}}} (Virtual Event USA, 2021-03-03). \bibinfo{publisher}{ACM}, \bibinfo{pages}{467--473}.
\newblock
\showISBNx{978-1-4503-8062-1}
\href{https://doi.org/10.1145/3408877.3432357}{doi:\nolinkurl{10.1145/3408877.3432357}}


\bibitem[Barr and Kallia(2022)]%
        {Barr2022WhyStudentsDrop}
\bibfield{author}{\bibinfo{person}{Matthew Barr} {and} \bibinfo{person}{Maria Kallia}.} \bibinfo{year}{2022}\natexlab{}.
\newblock \showarticletitle{Why {{Students Drop Computing Science}}: {{Using Models}} of {{Motivation}} to {{Understand Student Attrition}} and {{Retention}}}.
\newblock   \bibinfo{volume}{1} (\bibinfo{year}{2022}).
\newblock
\showISBNx{9781450396165}
\href{https://doi.org/10.1145/3564721}{doi:\nolinkurl{10.1145/3564721}}


\bibitem[Ben-Eliyahu and Bernacki(2015)]%
        {Ben-Eliyahu2015}
\bibfield{author}{\bibinfo{person}{Adar Ben-Eliyahu} {and} \bibinfo{person}{Matthew~L. Bernacki}.} \bibinfo{year}{2015}\natexlab{}.
\newblock \showarticletitle{Addressing Complexities in Self-Regulated Learning: A Focus on Contextual Factors, Contingencies, and Dynamic Relations}.
\newblock  \bibinfo{volume}{10}, \bibinfo{number}{1} (\bibinfo{year}{2015}), \bibinfo{pages}{1--13}.
\newblock
\href{https://doi.org/10.1007/s11409-015-9134-6}{doi:\nolinkurl{10.1007/s11409-015-9134-6}}


\bibitem[Bergin et~al\mbox{.}(2005)]%
        {bergin2005examining}
\bibfield{author}{\bibinfo{person}{Susan Bergin}, \bibinfo{person}{Ronan Reilly}, {and} \bibinfo{person}{Desmond Traynor}.} \bibinfo{year}{2005}\natexlab{}.
\newblock \showarticletitle{Examining the Role of Self-Regulated Learning on Introductory Programming Performance}. In \bibinfo{booktitle}{\emph{Proceedings of the 2005 International Workshop on {{Computing}} Education Research - {{ICER}} '05}} (Seattle, WA, USA, 2005). \bibinfo{publisher}{ACM Press}, \bibinfo{pages}{81--86}.
\newblock
\showISBNx{978-1-59593-043-9}
\href{https://doi.org/10.1145/1089786.1089794}{doi:\nolinkurl{10.1145/1089786.1089794}}


\bibitem[Berti et~al\mbox{.}(2023)]%
        {Berti2023PM4PyProcessMining}
\bibfield{author}{\bibinfo{person}{Alessandro Berti}, \bibinfo{person}{Sebastiaan van Zelst}, {and} \bibinfo{person}{Daniel Schuster}.} \bibinfo{year}{2023}\natexlab{}.
\newblock \showarticletitle{{{PM4Py}}: {{A}} Process Mining Library for {{Python}}}.
\newblock   \bibinfo{volume}{17} (\bibinfo{year}{2023}), \bibinfo{pages}{100556}.
\newblock
\showISSN{2665-9638}
\href{https://doi.org/10.1016/j.simpa.2023.100556}{doi:\nolinkurl{10.1016/j.simpa.2023.100556}}


\bibitem[Bhaskar(1978)]%
        {Bhaskar1978}
\bibfield{author}{\bibinfo{person}{Roy Bhaskar}.} \bibinfo{year}{1978}\natexlab{}.
\newblock \bibinfo{booktitle}{\emph{A Realist Theory of Science} (\bibinfo{edition}{2nd} ed.)}.
\newblock \bibinfo{publisher}{Harvester Press}.
\newblock


\bibitem[Bielaczyc et~al\mbox{.}(1995)]%
        {bielaczyc1995training}
\bibfield{author}{\bibinfo{person}{Katerine Bielaczyc}, \bibinfo{person}{Peter~L. Pirolli}, {and} \bibinfo{person}{Ann~L. Brown}.} \bibinfo{year}{1995}\natexlab{}.
\newblock \showarticletitle{Training in {{Self-Explanation}} and {{Self-Regulation Strategies}}: {{Investigating}} the {{Effects}} of {{Knowledge Acquisition Activities}} on {{Problem Solving}}}.
\newblock  \bibinfo{volume}{13}, \bibinfo{number}{2} (\bibinfo{year}{1995}), \bibinfo{pages}{221--252}.
\newblock
\showISSN{0737-0008}
\href{https://doi.org/10.1207/s1532690xci1302_3}{doi:\nolinkurl{10.1207/s1532690xci1302_3}}


\bibitem[Borrella et~al\mbox{.}(2022)]%
        {Borrella2022TakingActionReduce}
\bibfield{author}{\bibinfo{person}{Inma Borrella}, \bibinfo{person}{Sergio Caballero-Caballero}, {and} \bibinfo{person}{Eva Ponce-Cueto}.} \bibinfo{year}{2022}\natexlab{}.
\newblock \showarticletitle{Taking Action to Reduce Dropout in {{MOOCs}}: {{Tested}} Interventions}.
\newblock   \bibinfo{volume}{179} (\bibinfo{year}{2022}), \bibinfo{pages}{104412}.
\newblock
\showISSN{0360-1315}
\href{https://doi.org/10.1016/j.compedu.2021.104412}{doi:\nolinkurl{10.1016/j.compedu.2021.104412}}


\bibitem[Butler and Winne(1995)]%
        {butler1995feedback}
\bibfield{author}{\bibinfo{person}{Deborah~L. Butler} {and} \bibinfo{person}{Philip~H. Winne}.} \bibinfo{year}{1995}\natexlab{}.
\newblock \showarticletitle{Feedback and {{Self-Regulated Learning}}: {{A Theoretical Synthesis}}}.
\newblock  \bibinfo{volume}{65}, \bibinfo{number}{3} (\bibinfo{year}{1995}), \bibinfo{pages}{245--281}.
\newblock
\showISSN{0034-6543}
\href{https://doi.org/10.3102/00346543065003245}{doi:\nolinkurl{10.3102/00346543065003245}}


\bibitem[Choi et~al\mbox{.}(2023)]%
        {Choi2023LogsSelfReportsMisalignment}
\bibfield{author}{\bibinfo{person}{Heeryung Choi}, \bibinfo{person}{Philip~H. Winne}, \bibinfo{person}{Christopher Brooks}, \bibinfo{person}{Warren Li}, {and} \bibinfo{person}{Kerby Shedden}.} \bibinfo{year}{2023}\natexlab{}.
\newblock \showarticletitle{Logs or {{Self-Reports}}? {{Misalignment Between Behavioral Trace Data}} and {{Surveys When Modeling Learner Achievement Goal Orientation}}}. In \bibinfo{booktitle}{\emph{{{LAK23}}: 13th {{International Learning Analytics}} and {{Knowledge Conference}}}} (New York, NY, USA, 2023-03-13) \emph{(\bibinfo{series}{{{LAK2023}}})}. \bibinfo{publisher}{Association for Computing Machinery}, \bibinfo{pages}{11--21}.
\newblock
\showISBNx{978-1-4503-9865-7}
\href{https://doi.org/10.1145/3576050.3576052}{doi:\nolinkurl{10.1145/3576050.3576052}}


\bibitem[Cigdem(2015)]%
        {cigdem2015does}
\bibfield{author}{\bibinfo{person}{Harun Cigdem}.} \bibinfo{year}{2015}\natexlab{}.
\newblock \showarticletitle{How Does Self-Regulation Affect Computer-Programming Achievement in a Blended Context?}
\newblock  \bibinfo{volume}{6}, \bibinfo{number}{1} (\bibinfo{year}{2015}), \bibinfo{pages}{19--37}.
\newblock


\bibitem[Duncan and McKeachie(2005)]%
        {Duncan2005}
\bibfield{author}{\bibinfo{person}{Teresa~García Duncan} {and} \bibinfo{person}{Wilbert~J. McKeachie}.} \bibinfo{year}{2005}\natexlab{}.
\newblock \showarticletitle{The {{Making}} of the {{Motivated Strategies}} for {{Learning Questionnaire}}}.
\newblock  \bibinfo{volume}{40}, \bibinfo{number}{2} (\bibinfo{year}{2005}), \bibinfo{pages}{117--128}.
\newblock
\showISSN{00461520}
\href{https://doi.org/10.1207/s15326985ep4002_6}{doi:\nolinkurl{10.1207/s15326985ep4002_6}}


\bibitem[Emagnaw(2019)]%
        {emagnaw2019self}
\bibfield{author}{\bibinfo{person}{Alemayehu~Belay Emagnaw}.} \bibinfo{year}{2019}\natexlab{}.
\newblock \showarticletitle{Self-Regulated Learning Strategies and School Performance in Higher and Lower Students in Secondary and Preparatory School.}
\newblock  \bibinfo{volume}{14}, \bibinfo{number}{4} (\bibinfo{year}{2019}), \bibinfo{pages}{37--48}.
\newblock


\bibitem[Ergen and Kanadlı(2017)]%
        {ergen2017effect}
\bibfield{author}{\bibinfo{person}{Binnur Ergen} {and} \bibinfo{person}{Sedat Kanadlı}.} \bibinfo{year}{2017}\natexlab{}.
\newblock \showarticletitle{The Effect of Self-Regulated Learning Strategies on Academic Achievement: {{A}} Meta-Analysis Study}.
\newblock  \bibinfo{volume}{17}, \bibinfo{number}{69} (\bibinfo{year}{2017}), \bibinfo{pages}{55--74}.
\newblock


\bibitem[Falkner et~al\mbox{.}(2014)]%
        {falkner2014identifying}
\bibfield{author}{\bibinfo{person}{Katrina Falkner}, \bibinfo{person}{Rebecca Vivian}, {and} \bibinfo{person}{Nickolas~J.G. Falkner}.} \bibinfo{year}{2014}\natexlab{}.
\newblock \showarticletitle{Identifying Computer Science Self-Regulated Learning Strategies}. In \bibinfo{booktitle}{\emph{Proceedings of the 2014 Conference on {{Innovation}} \& Technology in Computer Science Education - {{ITiCSE}} '14}} (Uppsala, Sweden, 2014). \bibinfo{publisher}{ACM Press}, \bibinfo{pages}{291--296}.
\newblock
\showISBNx{978-1-4503-2833-3}
\href{https://doi.org/10.1145/2591708.2591715}{doi:\nolinkurl{10.1145/2591708.2591715}}


\bibitem[Fan et~al\mbox{.}(2021a)]%
        {Fan2021LearningAnalyticsReveal}
\bibfield{author}{\bibinfo{person}{Yizhou Fan}, \bibinfo{person}{Wannisa Matcha}, \bibinfo{person}{Nora'ayu~Ahmad Uzir}, \bibinfo{person}{Qiong Wang}, {and} \bibinfo{person}{Dragan Gašević}.} \bibinfo{year}{2021}\natexlab{a}.
\newblock \showarticletitle{Learning Analytics to Reveal Links Between Learning Design and Self-Regulated Learning}.
\newblock  \bibinfo{volume}{31}, \bibinfo{number}{4} (\bibinfo{year}{2021}), \bibinfo{pages}{980--1021}.
\newblock
\showISSN{1560-4306}
\href{https://doi.org/10.1007/s40593-021-00249-z}{doi:\nolinkurl{10.1007/s40593-021-00249-z}}


\bibitem[Fan et~al\mbox{.}(2023b)]%
        {Fan2023FullerPictureTriangulation}
\bibfield{author}{\bibinfo{person}{Yizhou Fan}, \bibinfo{person}{Mladen Rakovic}, \bibinfo{person}{Joep van~der Graaf}, \bibinfo{person}{Lyn Lim}, \bibinfo{person}{Shaveen Singh}, \bibinfo{person}{Johanna Moore}, \bibinfo{person}{Inge Molenaar}, \bibinfo{person}{Maria Bannert}, {and} \bibinfo{person}{Dragan Gašević}.} \bibinfo{year}{2023}\natexlab{b}.
\newblock \showarticletitle{Towards a Fuller Picture: {{Triangulation}} and Integration of the Measurement of Self-Regulated Learning Based on Trace and Think Aloud Data}.
\newblock  \bibinfo{volume}{39}, \bibinfo{number}{4} (\bibinfo{year}{2023}), \bibinfo{pages}{1303--1324}.
\newblock
\showISSN{1365-2729}
\href{https://doi.org/10.1111/jcal.12801}{doi:\nolinkurl{10.1111/jcal.12801}}


\bibitem[Fan et~al\mbox{.}(2023c)]%
        {Fan2023DissectingLearningTactics}
\bibfield{author}{\bibinfo{person}{Yizhou Fan}, \bibinfo{person}{Yuanru Tan}, \bibinfo{person}{Mladen Raković}, \bibinfo{person}{Yeyu Wang}, \bibinfo{person}{Zhiqiang Cai}, \bibinfo{person}{David~Williamson Shaffer}, {and} \bibinfo{person}{Dragan Gašević}.} \bibinfo{year}{2023}\natexlab{c}.
\newblock \showarticletitle{Dissecting Learning Tactics in MOOC Using Ordered Network Analysis}.
\newblock  \bibinfo{volume}{39}, \bibinfo{number}{1} (\bibinfo{year}{2023}), \bibinfo{pages}{154--166}.
\newblock
\showISSN{1365-2729}
\href{https://doi.org/10.1111/jcal.12735}{doi:\nolinkurl{10.1111/jcal.12735}}


\bibitem[Fincham et~al\mbox{.}(2019)]%
        {Fincham2019StudyTacticsLearning}
\bibfield{author}{\bibinfo{person}{Ed Fincham}, \bibinfo{person}{Dragan Gašević}, \bibinfo{person}{Jelena Jovanović}, {and} \bibinfo{person}{Abelardo Pardo}.} \bibinfo{year}{2019}\natexlab{}.
\newblock \showarticletitle{From Study Tactics to Learning Strategies: An Analytical Method for Extracting Interpretable Representations}.
\newblock  \bibinfo{volume}{12}, \bibinfo{number}{1} (\bibinfo{year}{2019}), \bibinfo{pages}{59--72}.
\newblock
\showISSN{1939-1382}
\href{https://doi.org/10.1109/TLT.2018.2823317}{doi:\nolinkurl{10.1109/TLT.2018.2823317}}


\bibitem[Flanigan et~al\mbox{.}(2023)]%
        {flanigan2023relationship}
\bibfield{author}{\bibinfo{person}{Abraham~E. Flanigan}, \bibinfo{person}{Markeya~S. Peteranetz}, \bibinfo{person}{Duane~F. Shell}, {and} \bibinfo{person}{Leen-Kiat Soh}.} \bibinfo{year}{2023}\natexlab{}.
\newblock \showarticletitle{Relationship {{Between Implicit Intelligence Beliefs}} and {{Maladaptive Self-Regulation}} of {{Learning}}}.
\newblock  \bibinfo{volume}{23}, \bibinfo{number}{3} (\bibinfo{year}{2023}), \bibinfo{pages}{1--23}.
\newblock
\showISSN{1946-6226, 1946-6226}
\href{https://doi.org/10.1145/3595187}{doi:\nolinkurl{10.1145/3595187}}


\bibitem[Garcia et~al\mbox{.}(2018)]%
        {Garcia2018SystematicLiteratureReview}
\bibfield{author}{\bibinfo{person}{Rita Garcia}, \bibinfo{person}{Katrina Falkner}, {and} \bibinfo{person}{Rebecca Vivian}.} \bibinfo{year}{2018}\natexlab{}.
\newblock \showarticletitle{Systematic Literature Review: {{Self-Regulated Learning}} Strategies Using e-Learning Tools for {{Computer Science}}}.
\newblock   \bibinfo{volume}{123} (\bibinfo{year}{2018}), \bibinfo{pages}{150--163}.
\newblock
\showISSN{0360-1315}
\href{https://doi.org/10.1016/J.COMPEDU.2018.05.006}{doi:\nolinkurl{10.1016/J.COMPEDU.2018.05.006}}


\bibitem[Greene and Azevedo(2007)]%
        {greene2007theoretical}
\bibfield{author}{\bibinfo{person}{Jeffrey~Alan Greene} {and} \bibinfo{person}{Roger Azevedo}.} \bibinfo{year}{2007}\natexlab{}.
\newblock \showarticletitle{A {{Theoretical Review}} of {{Winne}} and {{Hadwin}}’s {{Model}} of {{Self-Regulated Learning}}: {{New Perspectives}} and {{Directions}}}.
\newblock  \bibinfo{volume}{77}, \bibinfo{number}{3} (\bibinfo{year}{2007}), \bibinfo{pages}{334--372}.
\newblock
\showISSN{0034-6543}
\href{https://doi.org/10.3102/003465430303953}{doi:\nolinkurl{10.3102/003465430303953}}


\bibitem[Han(2023)]%
        {Han2023LevelConsistencyStudents}
\bibfield{author}{\bibinfo{person}{Feifei Han}.} \bibinfo{year}{2023}\natexlab{}.
\newblock \showarticletitle{Level of Consistency between Students' Self-Reported and Observed Study Approaches in Flipped Classroom Courses: {{How}} Does It Influence Students' Academic Learning Outcomes?}
\newblock  \bibinfo{volume}{18}, \bibinfo{number}{6} (\bibinfo{year}{2023}), \bibinfo{pages}{e0286549}.
\newblock
\showISSN{1932-6203}
\href{https://doi.org/10.1371/journal.pone.0286549}{doi:\nolinkurl{10.1371/journal.pone.0286549}}


\bibitem[Hastie et~al\mbox{.}(2009)]%
        {Hastie2017ElementsStatisticalLearning}
\bibfield{author}{\bibinfo{person}{Trevor Hastie}, \bibinfo{person}{Robert Tibshirani}, {and} \bibinfo{person}{Jerome~H. Friedman}.} \bibinfo{year}{2009}\natexlab{}.
\newblock \bibinfo{booktitle}{\emph{The Elements of Statistical Learning: Data Mining, Inference, and Prediction} (\bibinfo{edition}{2} ed.)}.
\newblock \bibinfo{publisher}{Springer}.
\newblock
\showISBNx{978-0-387-84857-0}


\bibitem[Hellas et~al\mbox{.}(2018)]%
        {Hellas2018PredictingAcademicPerformance}
\bibfield{author}{\bibinfo{person}{Arto Hellas}, \bibinfo{person}{Petri Ihantola}, \bibinfo{person}{Andrew Petersen}, \bibinfo{person}{Vangel~V. Ajanovski}, \bibinfo{person}{Mirela Gutica}, \bibinfo{person}{Timo Hynninen}, \bibinfo{person}{Antti Knutas}, \bibinfo{person}{Juho Leinonen}, \bibinfo{person}{Chris Messom}, {and} \bibinfo{person}{Soohyun~Nam Liao}.} \bibinfo{year}{2018}\natexlab{}.
\newblock \showarticletitle{Predicting Academic Performance: A Systematic Literature Review}. In \bibinfo{booktitle}{\emph{Proceedings {{Companion}} of the 23rd {{Annual ACM Conference}} on {{Innovation}} and {{Technology}} in {{Computer Science Education}}}} (New York, NY, USA, 2018-07-02) \emph{(\bibinfo{series}{{{ITiCSE}} 2018 {{Companion}}})}. \bibinfo{publisher}{Association for Computing Machinery}, \bibinfo{pages}{175--199}.
\newblock
\showISBNx{978-1-4503-6223-8}
\href{https://doi.org/10.1145/3293881.3295783}{doi:\nolinkurl{10.1145/3293881.3295783}}


\bibitem[Hsieh and Shannon(2005)]%
        {hsieh2005three}
\bibfield{author}{\bibinfo{person}{Hsiu-Fang Hsieh} {and} \bibinfo{person}{Sarah~E. Shannon}.} \bibinfo{year}{2005}\natexlab{}.
\newblock \showarticletitle{Three {{Approaches}} to {{Qualitative Content Analysis}}}.
\newblock  \bibinfo{volume}{15}, \bibinfo{number}{9} (\bibinfo{year}{2005}), \bibinfo{pages}{1277--1288}.
\newblock
\showISSN{1049-7323}
\href{https://doi.org/10.1177/1049732305276687}{doi:\nolinkurl{10.1177/1049732305276687}}


\bibitem[Hämäläinen and Isomöttönen(2019)]%
        {Hamalainen2019}
\bibfield{author}{\bibinfo{person}{Ville Hämäläinen} {and} \bibinfo{person}{Ville Isomöttönen}.} \bibinfo{year}{2019}\natexlab{}.
\newblock \showarticletitle{What {{Did CS Students Recognize}} as {{Study Difficulties}}?}. In \bibinfo{booktitle}{\emph{2019 {{IEEE Frontiers}} in {{Education Conference}} ({{FIE}})}} (2019-10), Vol.~\bibinfo{volume}{2019-Octob}. \bibinfo{publisher}{IEEE}, \bibinfo{pages}{1--9}.
\newblock
\showISBNx{978-1-72811-746-1}
\href{https://doi.org/10.1109/FIE43999.2019.9028714}{doi:\nolinkurl{10.1109/FIE43999.2019.9028714}}


\bibitem[Ifenthaler et~al\mbox{.}(2022)]%
        {Ifenthaler2022-hb}
\bibfield{author}{\bibinfo{person}{Dirk Ifenthaler}, \bibinfo{person}{Clara Schumacher}, {and} \bibinfo{person}{Jakub Kuzilek}.} \bibinfo{year}{2022}\natexlab{}.
\newblock \showarticletitle{Investigating Students' Use of Self-assessments in Higher Education Using Learning Analytics}.
\newblock  \bibinfo{volume}{39}, \bibinfo{number}{1} (\bibinfo{year}{2022}), \bibinfo{pages}{255--268}.
\newblock
\showISSN{0266-4909, 1365-2729}
\href{https://doi.org/10.1111/jcal.12744}{doi:\nolinkurl{10.1111/jcal.12744}}


\bibitem[Ifenthaler and Yau(2020)]%
        {Ifenthaler2020-di}
\bibfield{author}{\bibinfo{person}{Dirk Ifenthaler} {and} \bibinfo{person}{Jane Yin-Kim Yau}.} \bibinfo{year}{2020}\natexlab{}.
\newblock \showarticletitle{Reflections on Different Learning Analytics Indicators for Supporting Study Success}.
\newblock  \bibinfo{volume}{2}, \bibinfo{number}{2} (\bibinfo{year}{2020}), \bibinfo{pages}{4--23}.
\newblock
\showISSN{2706-7564, 2706-7564}
\href{https://doi.org/10.3991/ijai.v2i2.15639}{doi:\nolinkurl{10.3991/ijai.v2i2.15639}}


\bibitem[Ihantola et~al\mbox{.}(2015)]%
        {Ihantola2015EducationalDataMining}
\bibfield{author}{\bibinfo{person}{Petri Ihantola}, \bibinfo{person}{Arto Vihavainen}, \bibinfo{person}{Alireza Ahadi}, \bibinfo{person}{Matthew Butler}, \bibinfo{person}{Jürgen Börstler}, \bibinfo{person}{Stephen~H. Edwards}, \bibinfo{person}{Essi Isohanni}, \bibinfo{person}{Ari Korhonen}, \bibinfo{person}{Andrew Petersen}, \bibinfo{person}{Kelly Rivers}, \bibinfo{person}{Miguel~Ángel Rubio}, \bibinfo{person}{Judy Sheard}, \bibinfo{person}{Bronius Skupas}, \bibinfo{person}{Jaime Spacco}, \bibinfo{person}{Claudia Szabo}, {and} \bibinfo{person}{Daniel Toll}.} \bibinfo{year}{2015}\natexlab{}.
\newblock \showarticletitle{Educational {{Data Mining}} and {{Learning Analytics}} in {{Programming}}: {{Literature Review}} and {{Case Studies}}}. In \bibinfo{booktitle}{\emph{Proceedings of the 2015 {{ITiCSE}} on {{Working Group Reports}}}} (New York, NY, USA, 2015-07-04) \emph{(\bibinfo{series}{{{ITICSE-WGR}} '15})}. \bibinfo{publisher}{Association for Computing Machinery}, \bibinfo{pages}{41--63}.
\newblock
\showISBNx{978-1-4503-4146-2}
\href{https://doi.org/10.1145/2858796.2858798}{doi:\nolinkurl{10.1145/2858796.2858798}}


\bibitem[Jovanović et~al\mbox{.}(2017)]%
        {Jovanovic2017}
\bibfield{author}{\bibinfo{person}{Jelena Jovanović}, \bibinfo{person}{Dragan Gašević}, \bibinfo{person}{Shane Dawson}, \bibinfo{person}{Abelardo Pardo}, {and} \bibinfo{person}{Negin Mirriahi}.} \bibinfo{year}{2017}\natexlab{}.
\newblock \showarticletitle{Learning Analytics to Unveil Learning Strategies in a Flipped Classroom}.
\newblock   \bibinfo{volume}{33} (\bibinfo{year}{2017}), \bibinfo{pages}{74--85}.
\newblock
\showISSN{10967516}
\href{https://doi.org/10.1016/j.iheduc.2017.02.001}{doi:\nolinkurl{10.1016/j.iheduc.2017.02.001}}


\bibitem[Järvinen et~al\mbox{.}(2024)]%
        {Jarvinen2024AcademicExperiencesInformation}
\bibfield{author}{\bibinfo{person}{Miitta Järvinen}, \bibinfo{person}{Katriina Sipiläinen}, \bibinfo{person}{Janne Roslöf}, \bibinfo{person}{Sami Lehesvuori}, \bibinfo{person}{Lauri Kettunen}, {and} \bibinfo{person}{Raija Hämäläinen}.} \bibinfo{year}{2024}\natexlab{}.
\newblock \showarticletitle{Academic Experiences of Information Technology Students: Uncovering First-Year Challenges}.
\newblock  (\bibinfo{year}{2024}), \bibinfo{pages}{1--26}.
\newblock
\showISSN{0304-3797, 1469-5898}
\href{https://doi.org/10.1080/03043797.2024.2377304}{doi:\nolinkurl{10.1080/03043797.2024.2377304}}


\bibitem[Kangas et~al\mbox{.}(2017)]%
        {Kangas2017HowFacilitateFreshmen}
\bibfield{author}{\bibinfo{person}{Jari Kangas}, \bibinfo{person}{Elisa Rantanen}, {and} \bibinfo{person}{Lauri Kettunen}.} \bibinfo{year}{2017}\natexlab{}.
\newblock \showarticletitle{How to Facilitate Freshmen Learning and Support Their Transition to a University Study Environment}.
\newblock  \bibinfo{volume}{42}, \bibinfo{number}{6} (\bibinfo{year}{2017}), \bibinfo{pages}{668--683}.
\newblock
\showISSN{0304-3797}
\href{https://doi.org/10.1080/03043797.2016.1214818}{doi:\nolinkurl{10.1080/03043797.2016.1214818}}


\bibitem[Karch(2021)]%
        {Karch2021PsychologistsShouldUse}
\bibfield{author}{\bibinfo{person}{Julian~D. Karch}.} \bibinfo{year}{2021}\natexlab{}.
\newblock \showarticletitle{Psychologists {{Should Use Brunner-Munzel}}'s {{Instead}} of {{Mann-Whitney}}'s {{U Test}} as the {{Default Nonparametric Procedure}}}.
\newblock  \bibinfo{volume}{4}, \bibinfo{number}{2} (\bibinfo{year}{2021}), \bibinfo{pages}{2515245921999602}.
\newblock
\showISSN{2515-2459}
\href{https://doi.org/10.1177/2515245921999602}{doi:\nolinkurl{10.1177/2515245921999602}}


\bibitem[Karimi et~al\mbox{.}(2019)]%
        {Karimi2019GlobalConvergenceFast}
\bibfield{author}{\bibinfo{person}{Belhal Karimi}, \bibinfo{person}{Hoi-To Wai}, \bibinfo{person}{Eric Moulines}, {and} \bibinfo{person}{Marc Lavielle}.} \bibinfo{year}{2019}\natexlab{}.
\newblock \bibinfo{booktitle}{\emph{On the Global Convergence of (Fast) Incremental Expectation Maximization Methods}}.
\newblock
\showeprint[arXiv]{1910.12521}~[stat]
\href{https://doi.org/10.48550/arXiv.1910.12521}{doi:\nolinkurl{10.48550/arXiv.1910.12521}}


\bibitem[Kim et~al\mbox{.}(2020)]%
        {Kim2020ExploringStudentTeacher}
\bibfield{author}{\bibinfo{person}{Dongho Kim}, \bibinfo{person}{Yongseok Lee}, \bibinfo{person}{Walter~L. Leite}, {and} \bibinfo{person}{Anne~Corinne Huggins-Manley}.} \bibinfo{year}{2020}\natexlab{}.
\newblock \showarticletitle{Exploring Student and Teacher Usage Patterns Associated with Student Attrition in an Open Educational Resource-Supported Online Learning Platform}.
\newblock   \bibinfo{volume}{156} (\bibinfo{year}{2020}), \bibinfo{pages}{103961}.
\newblock
\showISSN{03601315}
\href{https://doi.org/10.1016/j.compedu.2020.103961}{doi:\nolinkurl{10.1016/j.compedu.2020.103961}}


\bibitem[Kinnunen and Malmi(2006)]%
        {Kinnunen2006WhyStudentsDrop}
\bibfield{author}{\bibinfo{person}{Päivi Kinnunen} {and} \bibinfo{person}{Lauri Malmi}.} \bibinfo{year}{2006}\natexlab{}.
\newblock \showarticletitle{Why Students Drop out {{CS1}} Course?}. In \bibinfo{booktitle}{\emph{Proceedings of the 2006 International Workshop on {{Computing}} Education Research - {{ICER}} '06}} (New York, New York, USA, 2006). \bibinfo{publisher}{ACM Press}, \bibinfo{pages}{97}.
\newblock
\showISBNx{1-59593-494-4}
\href{https://doi.org/10.1145/1151588.1151604}{doi:\nolinkurl{10.1145/1151588.1151604}}


\bibitem[Ko and Leu(2016)]%
        {ko2016applying}
\bibfield{author}{\bibinfo{person}{Chia-Yin Ko} {and} \bibinfo{person}{Fang-Yie Leu}.} \bibinfo{year}{2016}\natexlab{}.
\newblock \showarticletitle{Applying {{Data Mining}} to {{Explore Students}}' {{Self-Regulation}} in {{Learning Contexts}}}. In \bibinfo{booktitle}{\emph{2016 {{IEEE}} 30th {{International Conference}} on {{Advanced Information Networking}} and {{Applications}} ({{AINA}})}} (2016-03). \bibinfo{pages}{74--78}.
\newblock
\showISSN{1550-445X}
\href{https://doi.org/10.1109/AINA.2016.123}{doi:\nolinkurl{10.1109/AINA.2016.123}}


\bibitem[Kovanovic et~al\mbox{.}(2015)]%
        {Kovanovic2015DoesTimetaskEstimation}
\bibfield{author}{\bibinfo{person}{Vitomir Kovanovic}, \bibinfo{person}{Dragan Gašević}, \bibinfo{person}{Shane Dawson}, \bibinfo{person}{Srećko Joksimovic}, {and} \bibinfo{person}{Ryan Baker}.} \bibinfo{year}{2015}\natexlab{}.
\newblock \showarticletitle{Does {{Time-on-task Estimation Matter}}? {{Implications}} on {{Validity}} of {{Learning Analytics Findings}}}.
\newblock  \bibinfo{volume}{2}, \bibinfo{number}{3} (\bibinfo{year}{2015}), \bibinfo{pages}{81--110}.
\newblock
Issue 3.
\showISSN{1929-7750}
\href{https://doi.org/10.18608/jla.2015.23.6}{doi:\nolinkurl{10.18608/jla.2015.23.6}}


\bibitem[Lanza et~al\mbox{.}(2003)]%
        {Lanza2003LatentClassLatent}
\bibfield{author}{\bibinfo{person}{Stephanie~T. Lanza}, \bibinfo{person}{Brian~P. Flaherty}, {and} \bibinfo{person}{Linda~M. Collins}.} \bibinfo{year}{2003}\natexlab{}.
\newblock \showarticletitle{Latent {{Class}} and {{Latent Transition Analysis}}}.
\newblock In \bibinfo{booktitle}{\emph{Handbook of {{Psychology}}}}. \bibinfo{publisher}{John Wiley \& Sons, Ltd}, \bibinfo{pages}{663--685}.
\newblock
\showISBNx{978-0-471-26438-5}
\href{https://doi.org/10.1002/0471264385.wei0226}{doi:\nolinkurl{10.1002/0471264385.wei0226}}


\bibitem[Leese(2010)]%
        {Leese2010BridgingGapSupporting}
\bibfield{author}{\bibinfo{person}{Maggie Leese}.} \bibinfo{year}{2010}\natexlab{}.
\newblock \showarticletitle{Bridging the Gap: Supporting Student Transitions into Higher Education}.
\newblock  \bibinfo{volume}{34}, \bibinfo{number}{2} (\bibinfo{year}{2010}), \bibinfo{pages}{239--251}.
\newblock
\showISSN{0309-877X}
\href{https://doi.org/10.1080/03098771003695494}{doi:\nolinkurl{10.1080/03098771003695494}}


\bibitem[Liao et~al\mbox{.}(2019)]%
        {Liao2019BehaviorsHigherLowera}
\bibfield{author}{\bibinfo{person}{Soohyun~Nam Liao}, \bibinfo{person}{Sander Valstar}, \bibinfo{person}{Kevin Thai}, \bibinfo{person}{Christine Alvarado}, \bibinfo{person}{Daniel Zingaro}, \bibinfo{person}{William~G. Griswold}, {and} \bibinfo{person}{Leo Porter}.} \bibinfo{year}{2019}\natexlab{}.
\newblock \showarticletitle{Behaviors of {{Higher}} and {{Lower Performing Students}} in {{CS1}}}. In \bibinfo{booktitle}{\emph{Proceedings of the 2019 {{ACM Conference}} on {{Innovation}} and {{Technology}} in {{Computer Science Education}}}} (Aberdeen Scotland Uk, 2019-07-02). \bibinfo{publisher}{ACM}, \bibinfo{pages}{196--202}.
\newblock
\showISBNx{978-1-4503-6895-7}
\href{https://doi.org/10.1145/3304221.3319740}{doi:\nolinkurl{10.1145/3304221.3319740}}


\bibitem[Lim et~al\mbox{.}(2021)]%
        {Lim2021ImpactLearningAnalytics}
\bibfield{author}{\bibinfo{person}{Lisa-Angelique Lim}, \bibinfo{person}{Dragan Gasevic}, \bibinfo{person}{Wannisa Matcha}, \bibinfo{person}{Nora'Ayu Ahmad~Uzir}, {and} \bibinfo{person}{Shane Dawson}.} \bibinfo{year}{2021}\natexlab{}.
\newblock \showarticletitle{Impact of Learning Analytics Feedback on Self-Regulated Learning: {{Triangulating}} Behavioural Logs with Students' Recall}. In \bibinfo{booktitle}{\emph{{{LAK21}}: 11th {{International Learning Analytics}} and {{Knowledge Conference}}}} (New York, NY, USA, 2021-04-12) \emph{(\bibinfo{series}{{{LAK21}}})}. \bibinfo{publisher}{Association for Computing Machinery}, \bibinfo{pages}{364--374}.
\newblock
\showISBNx{978-1-4503-8935-8}
\href{https://doi.org/10.1145/3448139.3448174}{doi:\nolinkurl{10.1145/3448139.3448174}}


\bibitem[Liu et~al\mbox{.}(2010)]%
        {Liu2010UnderstandingInternalClustering}
\bibfield{author}{\bibinfo{person}{Yanchi Liu}, \bibinfo{person}{Zhongmou Li}, \bibinfo{person}{Hui Xiong}, \bibinfo{person}{Xuedong Gao}, {and} \bibinfo{person}{Junjie Wu}.} \bibinfo{year}{2010}\natexlab{}.
\newblock \showarticletitle{Understanding of Internal Clustering Validation Measures}. In \bibinfo{booktitle}{\emph{2010 IEEE International Conference on Data Mining}} (2010). \bibinfo{pages}{911--916}.
\newblock
\showISSN{2374-8486}
\href{https://doi.org/10.1109/ICDM.2010.35}{doi:\nolinkurl{10.1109/ICDM.2010.35}}


\bibitem[Loksa and Ko(2016)]%
        {Loksa2016RoleSelfregulationProgramminga}
\bibfield{author}{\bibinfo{person}{Dastyni Loksa} {and} \bibinfo{person}{Andrew~J. Ko}.} \bibinfo{year}{2016}\natexlab{}.
\newblock \showarticletitle{The Role of Self-Regulation in Programming Problem Solving Process and Success}.
\newblock  (\bibinfo{year}{2016}), \bibinfo{pages}{83--91}.
\newblock
\showISBNx{9781450344494}
\href{https://doi.org/10.1145/2960310.2960334}{doi:\nolinkurl{10.1145/2960310.2960334}}


\bibitem[Loksa et~al\mbox{.}(2022b)]%
        {Loksa2022MetacognitionSelfRegulationProgrammingb}
\bibfield{author}{\bibinfo{person}{Dastyni Loksa}, \bibinfo{person}{Lauren Margulieux}, \bibinfo{person}{Brett~A. Becker}, \bibinfo{person}{Michelle Craig}, \bibinfo{person}{Paul Denny}, \bibinfo{person}{Raymond Pettit}, {and} \bibinfo{person}{James Prather}.} \bibinfo{year}{2022}\natexlab{b}.
\newblock \showarticletitle{Metacognition and {{Self-Regulation}} in {{Programming Education}}: {{Theories}} and {{Exemplars}} of {{Use}}}.
\newblock  \bibinfo{volume}{22}, \bibinfo{number}{4} (\bibinfo{year}{2022}), \bibinfo{pages}{39:1--39:31}.
\newblock
\href{https://doi.org/10.1145/3487050}{doi:\nolinkurl{10.1145/3487050}}


\bibitem[Loksa et~al\mbox{.}(2020a)]%
        {Loksa2020InvestigatingNovicesSitu}
\bibfield{author}{\bibinfo{person}{Dastyni Loksa}, \bibinfo{person}{Benjamin Xie}, \bibinfo{person}{Harrison Kwik}, {and} \bibinfo{person}{Amy~J. Ko}.} \bibinfo{year}{2020}\natexlab{a}.
\newblock \showarticletitle{Investigating {{Novices}}' {{In Situ Reflections}} on {{Their Programming Process}}}. In \bibinfo{booktitle}{\emph{Proceedings of the 51st {{ACM Technical Symposium}} on {{Computer Science Education}}}} (New York, NY, USA, 2020-02-26) \emph{(\bibinfo{series}{{{SIGCSE}} '20})}. \bibinfo{publisher}{Association for Computing Machinery}, \bibinfo{pages}{149--155}.
\newblock
\showISBNx{978-1-4503-6793-6}
\href{https://doi.org/10.1145/3328778.3366846}{doi:\nolinkurl{10.1145/3328778.3366846}}


\bibitem[Lorås et~al\mbox{.}(2022)]%
        {Loras2022StudyBehaviorComputinga}
\bibfield{author}{\bibinfo{person}{Madeleine Lorås}, \bibinfo{person}{Guttorm Sindre}, \bibinfo{person}{Hallvard Trætteberg}, {and} \bibinfo{person}{Trond Aalberg}.} \bibinfo{year}{2022}\natexlab{}.
\newblock \showarticletitle{Study {{Behavior}} in {{Computing Education}}---{{A Systematic Literature Review}}}.
\newblock  \bibinfo{volume}{22}, \bibinfo{number}{1} (\bibinfo{year}{2022}), \bibinfo{pages}{1--40}.
\newblock
\showISSN{1946-6226, 1946-6226}
\href{https://doi.org/10.1145/3469129}{doi:\nolinkurl{10.1145/3469129}}


\bibitem[Lundberg and Lee(2017)]%
        {Lundberg2017UnifiedApproachInterpreting}
\bibfield{author}{\bibinfo{person}{Scott~M Lundberg} {and} \bibinfo{person}{Su-In Lee}.} \bibinfo{year}{2017}\natexlab{}.
\newblock \showarticletitle{A {{Unified Approach}} to {{Interpreting Model Predictions}}}. In \bibinfo{booktitle}{\emph{Advances in {{Neural Information Processing Systems}}}} (2017), Vol.~\bibinfo{volume}{30}. \bibinfo{publisher}{Curran Associates, Inc.}
\newblock
\urldef\tempurl%
\url{https://papers.nips.cc/paper_files/paper/2017/hash/8a20a8621978632d76c43dfd28b67767-Abstract.html}
\showURL{%
\tempurl}


\bibitem[Maldonado-Mahauad et~al\mbox{.}(2018)]%
        {maldonado2018predicting}
\bibfield{author}{\bibinfo{person}{Jorge Maldonado-Mahauad}, \bibinfo{person}{Mar Pérez-Sanagustín}, \bibinfo{person}{Pedro~Manuel Moreno-Marcos}, \bibinfo{person}{Carlos Alario-Hoyos}, \bibinfo{person}{Pedro~J. Muñoz-Merino}, {and} \bibinfo{person}{Carlos Delgado-Kloos}.} \bibinfo{year}{2018}\natexlab{}.
\newblock \showarticletitle{Predicting {{Learners}}’ {{Success}} in a {{Self-paced MOOC Through Sequence Patterns}} of {{Self-regulated Learning}}}. In \bibinfo{booktitle}{\emph{Lifelong {{Technology-Enhanced Learning}}}} (Cham, 2018), \bibfield{editor}{\bibinfo{person}{Viktoria Pammer-Schindler}, \bibinfo{person}{Mar Pérez-Sanagustín}, \bibinfo{person}{Hendrik Drachsler}, \bibinfo{person}{Raymond Elferink}, {and} \bibinfo{person}{Maren Scheffel}} (Eds.). \bibinfo{publisher}{Springer International Publishing}, \bibinfo{pages}{355--369}.
\newblock
\showISBNx{978-3-319-98572-5}
\href{https://doi.org/10.1007/978-3-319-98572-5_27}{doi:\nolinkurl{10.1007/978-3-319-98572-5_27}}


\bibitem[Matcha et~al\mbox{.}(2019a)]%
        {Matcha2019DetectionLearningStrategies}
\bibfield{author}{\bibinfo{person}{Wannisa Matcha}, \bibinfo{person}{Dragan Gašević}, \bibinfo{person}{Nora'ayu Ahmad~Uzir}, \bibinfo{person}{Jelena Jovanović}, \bibinfo{person}{Abelardo Pardo}, \bibinfo{person}{Jorge Maldonado-Mahauad}, {and} \bibinfo{person}{Mar Pérez-Sanagustín}.} \bibinfo{year}{2019}\natexlab{a}.
\newblock \showarticletitle{Detection of {{Learning Strategies}}: {{A Comparison}} of {{Process}}, {{Sequence}} and {{Network Analytic Approaches}}}. In \bibinfo{booktitle}{\emph{Transforming {{Learning}} with {{Meaningful Technologies}}}} (Cham, 2019) \emph{(\bibinfo{series}{Lecture {{Notes}} in {{Computer Science}}})}, \bibfield{editor}{\bibinfo{person}{Maren Scheffel}, \bibinfo{person}{Julien Broisin}, \bibinfo{person}{Viktoria Pammer-Schindler}, \bibinfo{person}{Andri Ioannou}, {and} \bibinfo{person}{Jan Schneider}} (Eds.). \bibinfo{publisher}{Springer International Publishing}, \bibinfo{pages}{525--540}.
\newblock
\showISBNx{978-3-030-29736-7}
\href{https://doi.org/10.1007/978-3-030-29736-7_39}{doi:\nolinkurl{10.1007/978-3-030-29736-7_39}}


\bibitem[Matcha et~al\mbox{.}(2019b)]%
        {Matcha2019AnalyticsLearningStrategies}
\bibfield{author}{\bibinfo{person}{Wannisa Matcha}, \bibinfo{person}{Dragan Gašević}, \bibinfo{person}{Nora'Ayu~Ahmad Uzir}, \bibinfo{person}{Jelena Jovanović}, {and} \bibinfo{person}{Abelardo Pardo}.} \bibinfo{year}{2019}\natexlab{b}.
\newblock \showarticletitle{Analytics of Learning Strategies: Associations with Academic Performance and Feedback}. In \bibinfo{booktitle}{\emph{Proceedings of the 9th International Conference on Learning Analytics \& Knowledge}} (New York, NY, USA, 2019) \emph{(\bibinfo{series}{LAK19})}. \bibinfo{publisher}{Association for Computing Machinery}, \bibinfo{pages}{461--470}.
\newblock
\showISBNx{978-1-4503-6256-6}
\href{https://doi.org/10.1145/3303772.3303787}{doi:\nolinkurl{10.1145/3303772.3303787}}


\bibitem[Mirza et~al\mbox{.}(2019)]%
        {MirEtAl19}
\bibfield{author}{\bibinfo{person}{Diba Mirza}, \bibinfo{person}{Phillip~T. Conrad}, \bibinfo{person}{Christian Lloyd}, \bibinfo{person}{Ziad Matni}, {and} \bibinfo{person}{Arthur Gatin}.} \bibinfo{year}{2019}\natexlab{}.
\newblock \showarticletitle{Undergraduate {{Teaching Assistants}} in {{Computer Science}}: {{A Systematic Literature Review}}}. In \bibinfo{booktitle}{\emph{Proceedings of the 2019 {{ACM Conference}} on {{International Computing Education Research}}}} (Toronto ON Canada, 2019-07-30). \bibinfo{publisher}{ACM}, \bibinfo{pages}{31--40}.
\newblock
\href{https://doi.org/10.1145/3291279.3339422}{doi:\nolinkurl{10.1145/3291279.3339422}}


\bibitem[Moreno-Marcos et~al\mbox{.}(2020)]%
        {Moreno-Marcos2020TemporalAnalysisDropout}
\bibfield{author}{\bibinfo{person}{Pedro~Manuel Moreno-Marcos}, \bibinfo{person}{Pedro~J. Muñoz-Merino}, \bibinfo{person}{Jorge Maldonado-Mahauad}, \bibinfo{person}{Mar Pérez-Sanagustín}, \bibinfo{person}{Carlos Alario-Hoyos}, {and} \bibinfo{person}{Carlos Delgado~Kloos}.} \bibinfo{year}{2020}\natexlab{}.
\newblock \showarticletitle{Temporal Analysis for Dropout Prediction Using Self-Regulated Learning Strategies in Self-Paced {{MOOCs}}}.
\newblock   \bibinfo{volume}{145} (\bibinfo{year}{2020}), \bibinfo{pages}{103728}.
\newblock
\showISSN{0360-1315}
\href{https://doi.org/10.1016/j.compedu.2019.103728}{doi:\nolinkurl{10.1016/j.compedu.2019.103728}}


\bibitem[Naseem et~al\mbox{.}(2019)]%
        {Naseem2019UsingEnsembleDecision}
\bibfield{author}{\bibinfo{person}{Mohammed Naseem}, \bibinfo{person}{Kaylash Chaudhary}, \bibinfo{person}{Bibhya Sharma}, {and} \bibinfo{person}{Aman~Goel Lal}.} \bibinfo{year}{2019}\natexlab{}.
\newblock \showarticletitle{Using {{Ensemble Decision Tree Model}} to {{Predict Student Dropout}} in {{Computing Science}}}. In \bibinfo{booktitle}{\emph{2019 {{IEEE Asia-Pacific Conference}} on {{Computer Science}} and {{Data Engineering}} ({{CSDE}})}} (2019-12). \bibinfo{pages}{1--8}.
\newblock
\href{https://doi.org/10.1109/CSDE48274.2019.9162389}{doi:\nolinkurl{10.1109/CSDE48274.2019.9162389}}


\bibitem[Neubert and Brunner(2007)]%
        {Neubert2007StudentizedPermutationTest}
\bibfield{author}{\bibinfo{person}{Karin Neubert} {and} \bibinfo{person}{Edgar Brunner}.} \bibinfo{year}{2007}\natexlab{}.
\newblock \showarticletitle{A Studentized Permutation Test for the Non-Parametric {{Behrens}}--{{Fisher}} Problem}.
\newblock  \bibinfo{volume}{51}, \bibinfo{number}{10} (\bibinfo{year}{2007}), \bibinfo{pages}{5192--5204}.
\newblock
\showISSN{0167-9473}
\href{https://doi.org/10.1016/j.csda.2006.05.024}{doi:\nolinkurl{10.1016/j.csda.2006.05.024}}


\bibitem[Omer et~al\mbox{.}(2023)]%
        {Omer2023LearningAnalyticsProgramming}
\bibfield{author}{\bibinfo{person}{Uzma Omer}, \bibinfo{person}{Rabia Tehseen}, \bibinfo{person}{Muhammad~Shoaib Farooq}, {and} \bibinfo{person}{Adnan Abid}.} \bibinfo{year}{2023}\natexlab{}.
\newblock \showarticletitle{Learning Analytics in Programming Courses: {{Review}} and Implications}.
\newblock  \bibinfo{volume}{28}, \bibinfo{number}{9} (\bibinfo{year}{2023}), \bibinfo{pages}{11221--11268}.
\newblock
\showISSN{1573-7608}
\href{https://doi.org/10.1007/s10639-023-11611-0}{doi:\nolinkurl{10.1007/s10639-023-11611-0}}


\bibitem[Panadero(2017)]%
        {panadero2017review}
\bibfield{author}{\bibinfo{person}{Ernesto Panadero}.} \bibinfo{year}{2017}\natexlab{}.
\newblock \showarticletitle{A {{Review}} of {{Self-regulated Learning}}: {{Six Models}} and {{Four Directions}} for {{Research}}}.
\newblock   \bibinfo{volume}{8} (\bibinfo{year}{2017}).
\newblock
Issue April.
\showISSN{1664-1078}
\href{https://doi.org/10.3389/fpsyg.2017.00422}{doi:\nolinkurl{10.3389/fpsyg.2017.00422}}


\bibitem[Peavy(1998)]%
        {Pea98}
\bibfield{author}{\bibinfo{person}{R.~Vance Peavy}.} \bibinfo{year}{1998}\natexlab{}.
\newblock \bibinfo{booktitle}{\emph{Sociodynamic Counselling: A Constructivist Perspective} (\bibinfo{edition}{nachdr.} ed.)}.
\newblock \bibinfo{publisher}{Trafford}.
\newblock
\showISBNx{978-1-55212-094-1}


\bibitem[Petersen et~al\mbox{.}(2016)]%
        {Petersen2016RevisitingWhyStudents}
\bibfield{author}{\bibinfo{person}{Andrew Petersen}, \bibinfo{person}{Michelle Craig}, \bibinfo{person}{Jennifer Campbell}, {and} \bibinfo{person}{Anya Tafliovich}.} \bibinfo{year}{2016}\natexlab{}.
\newblock \showarticletitle{Revisiting Why Students Drop {{CS1}}}. In \bibinfo{booktitle}{\emph{Proceedings of the 16th {{Koli Calling International Conference}} on {{Computing Education Research}}}} (New York, NY, USA, 2016). \bibinfo{publisher}{ACM}, \bibinfo{pages}{71--80}.
\newblock
\showISBNx{978-1-4503-4770-9}
\href{https://doi.org/10.1145/2999541.2999552}{doi:\nolinkurl{10.1145/2999541.2999552}}


\bibitem[Pintrich(2000)]%
        {Pintrich2000}
\bibfield{author}{\bibinfo{person}{Paul~R. Pintrich}.} \bibinfo{year}{2000}\natexlab{}.
\newblock \showarticletitle{The {{Role}} of {{Goal Orientation}} in {{Self-Regulated Learning}}}.
\newblock In \bibinfo{booktitle}{\emph{Handbook of {{Self-Regulation}}}}, \bibfield{editor}{\bibinfo{person}{Monique Boekaerts}, \bibinfo{person}{Paul~R. Pintrich}, {and} \bibinfo{person}{Moshe Zeidner}} (Eds.). \bibinfo{publisher}{Elsevier}, \bibinfo{pages}{451--502}.
\newblock
\showISBNx{0-12-369519-8}
\href{https://doi.org/10.1016/b978-012109890-2/50043-3}{doi:\nolinkurl{10.1016/b978-012109890-2/50043-3}}


\bibitem[Pintrich et~al\mbox{.}(1991)]%
        {Pintrich1991}
\bibfield{author}{\bibinfo{person}{Paul~R. Pintrich}, \bibinfo{person}{Davig A.~F. Smith}, \bibinfo{person}{Teresa Garcia}, {and} \bibinfo{person}{Wilbert~J. McKeachie}.} \bibinfo{year}{1991}\natexlab{}.
\newblock \showarticletitle{A Manual for the Use of the {{Motivated Strategies}} for {{Learning Questionnaire}} ({{MSLQ}}).}
\newblock  (\bibinfo{year}{1991}).
\newblock
\urldef\tempurl%
\url{https://eric.ed.gov/?id=ED338122}
\showURL{%
\tempurl}


\bibitem[Piotrkowicz et~al\mbox{.}(2021)]%
        {Piotrkowicz2021DatadrivenExplorationEngagement}
\bibfield{author}{\bibinfo{person}{Alicja Piotrkowicz}, \bibinfo{person}{Kaiwen Wang}, \bibinfo{person}{Jennifer Hallam}, {and} \bibinfo{person}{Vania Dimitrova}.} \bibinfo{year}{2021}\natexlab{}.
\newblock \showarticletitle{Data-Driven Exploration of Engagement with Workplace-Based Assessment in the Clinical Skills Domain}.
\newblock  \bibinfo{volume}{31}, \bibinfo{number}{4} (\bibinfo{year}{2021}), \bibinfo{pages}{1022--1052}.
\newblock
\showISSN{1560-4306}
\href{https://doi.org/10.1007/s40593-021-00264-0}{doi:\nolinkurl{10.1007/s40593-021-00264-0}}


\bibitem[Pon-Barry et~al\mbox{.}(2017)]%
        {Pon-Barry2017ExpandingCapacityPromotinga}
\bibfield{author}{\bibinfo{person}{Heather Pon-Barry}, \bibinfo{person}{Becky Wai-Ling Packard}, {and} \bibinfo{person}{Audrey St.~John}.} \bibinfo{year}{2017}\natexlab{}.
\newblock \showarticletitle{Expanding Capacity and Promoting Inclusion in Introductory Computer Science: A Focus on near-Peer Mentor Preparation and Code Review}.
\newblock  \bibinfo{volume}{27}, \bibinfo{number}{1} (\bibinfo{year}{2017}), \bibinfo{pages}{54--77}.
\newblock
\showISSN{0899-3408}
\href{https://doi.org/10.1080/08993408.2017.1333270}{doi:\nolinkurl{10.1080/08993408.2017.1333270}}


\bibitem[Prather et~al\mbox{.}(2020)]%
        {Prather2020WhatWeThinka}
\bibfield{author}{\bibinfo{person}{James Prather}, \bibinfo{person}{Brett~A. Becker}, \bibinfo{person}{Michelle Craig}, \bibinfo{person}{Paul Denny}, \bibinfo{person}{Dastyni Loksa}, {and} \bibinfo{person}{Lauren Margulieux}.} \bibinfo{year}{2020}\natexlab{}.
\newblock \showarticletitle{What {{Do We Think We Think We Are Doing}}?: {{Metacognition}} and {{Self-Regulation}} in {{Programming}}}. In \bibinfo{booktitle}{\emph{Proceedings of the 2020 {{ACM Conference}} on {{International Computing Education Research}}}} (Virtual Event New Zealand, 2020-08-10). \bibinfo{publisher}{ACM}, \bibinfo{pages}{2--13}.
\newblock
\showISBNx{978-1-4503-7092-9}
\href{https://doi.org/10.1145/3372782.3406263}{doi:\nolinkurl{10.1145/3372782.3406263}}


\bibitem[Prenkaj et~al\mbox{.}(2020)]%
        {Prenkaj2020-af}
\bibfield{author}{\bibinfo{person}{Bardh Prenkaj}, \bibinfo{person}{Paola Velardi}, \bibinfo{person}{Giovanni Stilo}, \bibinfo{person}{Damiano Distante}, {and} \bibinfo{person}{Stefano Faralli}.} \bibinfo{year}{2020}\natexlab{}.
\newblock \showarticletitle{A Survey of Machine Learning Approaches for Student Dropout Prediction in Online Courses}.
\newblock  \bibinfo{volume}{53}, \bibinfo{number}{3} (\bibinfo{year}{2020}), \bibinfo{pages}{1--34}.
\newblock
\showISSN{0360-0300}
\href{https://doi.org/10.1145/3388792}{doi:\nolinkurl{10.1145/3388792}}


\bibitem[Quille and Bergin(2019)]%
        {Quille2019-ta}
\bibfield{author}{\bibinfo{person}{Keith Quille} {and} \bibinfo{person}{Susan Bergin}.} \bibinfo{year}{2019}\natexlab{}.
\newblock \showarticletitle{{{CS1}}: {{How}} Will They Do? {{How}} Can We Help? {{A}} Decade of Research and Practice}.
\newblock  \bibinfo{volume}{29}, \bibinfo{number}{2-3} (\bibinfo{year}{2019}), \bibinfo{pages}{254--282}.
\newblock
\showISSN{0899-3408}
\href{https://doi.org/10.1080/08993408.2019.1612679}{doi:\nolinkurl{10.1080/08993408.2019.1612679}}


\bibitem[Rakovic et~al\mbox{.}(2023)]%
        {Rakovic2023NetworkAnalyticsUnveil}
\bibfield{author}{\bibinfo{person}{Mladen Rakovic}, \bibinfo{person}{Nora'ayu~Ahmad Uzir}, \bibinfo{person}{Wannisa Matcha}, \bibinfo{person}{Brendan Eagan}, \bibinfo{person}{Jelena Jovanović}, \bibinfo{person}{David~Williamson Shaffer}, \bibinfo{person}{Abelardo Pardo}, {and} \bibinfo{person}{Dragan Gašević}.} \bibinfo{year}{2023}\natexlab{}.
\newblock \showarticletitle{Network {{Analytics}} to {{Unveil Links}} of {{Learning Strategies}}, {{Time Management}}, and {{Academic Performance}} in a {{Flipped Classroom}}}.
\newblock  (\bibinfo{year}{2023}), \bibinfo{pages}{1--23}.
\newblock
\showISSN{1929-7750}
\href{https://doi.org/10.18608/jla.2023.7843}{doi:\nolinkurl{10.18608/jla.2023.7843}}


\bibitem[Roth et~al\mbox{.}(2016)]%
        {Roth2016}
\bibfield{author}{\bibinfo{person}{Anne Roth}, \bibinfo{person}{Sabine Ogrin}, {and} \bibinfo{person}{Bernhard Schmitz}.} \bibinfo{year}{2016}\natexlab{}.
\newblock \showarticletitle{Assessing Self-Regulated Learning in Higher Education: A Systematic Literature Review of Self-Report Instruments}.
\newblock  \bibinfo{volume}{28}, \bibinfo{number}{3} (\bibinfo{year}{2016}), \bibinfo{pages}{225--250}.
\newblock
\showISSN{18748600}
\href{https://doi.org/10.1007/s11092-015-9229-2}{doi:\nolinkurl{10.1007/s11092-015-9229-2}}


\bibitem[Salvador and Chan(2004)]%
        {Salvador2004DeterminingNumberClusters}
\bibfield{author}{\bibinfo{person}{S. Salvador} {and} \bibinfo{person}{P. Chan}.} \bibinfo{year}{2004}\natexlab{}.
\newblock \showarticletitle{Determining the Number of Clusters/Segments in Hierarchical Clustering/Segmentation Algorithms}. In \bibinfo{booktitle}{\emph{16th IEEE International Conference on Tools with Artificial Intelligence}} (2004). \bibinfo{pages}{576--584}.
\newblock
\showISSN{1082-3409}
\href{https://doi.org/10.1109/ICTAI.2004.50}{doi:\nolinkurl{10.1109/ICTAI.2004.50}}


\bibitem[Saqr et~al\mbox{.}(2020)]%
        {Saqr2020UsingPsychologicalNetworks}
\bibfield{author}{\bibinfo{person}{Mohammed Saqr}, \bibinfo{person}{Olga Viberg}, {and} \bibinfo{person}{Ward Peteers}.} \bibinfo{year}{2020}\natexlab{}.
\newblock \showarticletitle{Using Psychological Networks to Reveal the Interplay between Foreign Language Students' Self-Regulated Learning Tactics}.
\newblock  (\bibinfo{year}{2020}).
\newblock


\bibitem[Schraw et~al\mbox{.}(2006)]%
        {schraw2006promoting}
\bibfield{author}{\bibinfo{person}{Gregory Schraw}, \bibinfo{person}{Kent~J. Crippen}, {and} \bibinfo{person}{Kendall Hartley}.} \bibinfo{year}{2006}\natexlab{}.
\newblock \showarticletitle{Promoting {{Self-Regulation}} in {{Science Education}}: {{Metacognition}} as {{Part}} of a {{Broader Perspective}} on {{Learning}}}.
\newblock  \bibinfo{volume}{36}, \bibinfo{number}{1} (\bibinfo{year}{2006}), \bibinfo{pages}{111--139}.
\newblock
\showISSN{1573-1898}
\href{https://doi.org/10.1007/s11165-005-3917-8}{doi:\nolinkurl{10.1007/s11165-005-3917-8}}


\bibitem[Schubert(2023)]%
        {Schubert2023StopUsingElbowa}
\bibfield{author}{\bibinfo{person}{Erich Schubert}.} \bibinfo{year}{2023}\natexlab{}.
\newblock \showarticletitle{Stop Using the Elbow Criterion for K-Means and How to Choose the Number of Clusters Instead}.
\newblock  \bibinfo{volume}{25}, \bibinfo{number}{1} (\bibinfo{year}{2023}), \bibinfo{pages}{36--42}.
\newblock
\showISSN{1931-0145, 1931-0153}
\href{https://doi.org/10.1145/3606274.3606278}{doi:\nolinkurl{10.1145/3606274.3606278}}


\bibitem[Schubert and Lenssen(2022)]%
        {Schubert2022FastKmedoidsClustering}
\bibfield{author}{\bibinfo{person}{Erich Schubert} {and} \bibinfo{person}{Lars Lenssen}.} \bibinfo{year}{2022}\natexlab{}.
\newblock \showarticletitle{Fast K-Medoids {{Clustering}} in {{Rust}} and {{Python}}}.
\newblock  \bibinfo{volume}{7}, \bibinfo{number}{75} (\bibinfo{year}{2022}), \bibinfo{pages}{4183}.
\newblock
\showISSN{2475-9066}
\href{https://doi.org/10.21105/joss.04183}{doi:\nolinkurl{10.21105/joss.04183}}


\bibitem[Schunk(2005)]%
        {Schunk2005SelfRegulatedLearningEducational}
\bibfield{author}{\bibinfo{person}{Dale~H. Schunk}.} \bibinfo{year}{2005}\natexlab{}.
\newblock \showarticletitle{Self-{{Regulated Learning}}: {{The Educational Legacy}} of {{Paul R}}. {{Pintrich}}}.
\newblock  \bibinfo{volume}{40}, \bibinfo{number}{2} (\bibinfo{year}{2005}), \bibinfo{pages}{85--94}.
\newblock
\showISSN{0046-1520}
\href{https://doi.org/10.1207/s15326985ep4002_3}{doi:\nolinkurl{10.1207/s15326985ep4002_3}}


\bibitem[Sghir et~al\mbox{.}(2022)]%
        {Sghir2022-ei}
\bibfield{author}{\bibinfo{person}{Nabila Sghir}, \bibinfo{person}{Amina Adadi}, {and} \bibinfo{person}{Mohammed Lahmer}.} \bibinfo{year}{2022}\natexlab{}.
\newblock \showarticletitle{Recent Advances in {{Predictive Learning Analytics}}: {{A}} Decade Systematic Review (2012--2022)}.
\newblock   \bibinfo{volume}{28} (\bibinfo{year}{2022}), \bibinfo{pages}{8299--8333}.
\newblock
\showISSN{1573-7608}
\href{https://doi.org/10.1007/s10639-022-11536-0}{doi:\nolinkurl{10.1007/s10639-022-11536-0}}


\bibitem[Sharma et~al\mbox{.}(2015)]%
        {Sharma2015IdentifyingStylesPaths}
\bibfield{author}{\bibinfo{person}{Kshitij Sharma}, \bibinfo{person}{Patrick Jermann}, {and} \bibinfo{person}{Pierre Dillenbourg}.} \bibinfo{year}{2015}\natexlab{}.
\newblock \bibinfo{title}{Identifying {{Styles}} and {{Paths}} toward {{Success}} in {{MOOCs}}}.
\newblock
\urldef\tempurl%
\url{https://eric.ed.gov/?id=ED560766}
\showURL{%
\tempurl}


\bibitem[Shibani et~al\mbox{.}(2020)]%
        {Shibani2020EducatorPerspectivesLearning}
\bibfield{author}{\bibinfo{person}{Antonette Shibani}, \bibinfo{person}{Simon Knight}, {and} \bibinfo{person}{Simon Buckingham~Shum}.} \bibinfo{year}{2020}\natexlab{}.
\newblock \showarticletitle{Educator Perspectives on Learning Analytics in Classroom Practice}.
\newblock   \bibinfo{volume}{46} (\bibinfo{year}{2020}), \bibinfo{pages}{100730}.
\newblock
\showISSN{1096-7516}
\href{https://doi.org/10.1016/j.iheduc.2020.100730}{doi:\nolinkurl{10.1016/j.iheduc.2020.100730}}


\bibitem[Silva et~al\mbox{.}(2024)]%
        {Silva2024WhatLearningStrategies}
\bibfield{author}{\bibinfo{person}{Leonardo Silva}, \bibinfo{person}{António Mendes}, \bibinfo{person}{Anabela Gomes}, {and} \bibinfo{person}{Gabriel Fortes}.} \bibinfo{year}{2024}\natexlab{}.
\newblock \showarticletitle{What {{Learning Strategies}} Are {{Used}} by {{Programming Students}}? {{A Qualitative Study Grounded}} on the {{Self-regulation}} of {{Learning Theory}}}.
\newblock  \bibinfo{volume}{24}, \bibinfo{number}{1} (\bibinfo{year}{2024}), \bibinfo{pages}{9:1--9:26}.
\newblock
\href{https://doi.org/10.1145/3635720}{doi:\nolinkurl{10.1145/3635720}}


\bibitem[Tempelaar et~al\mbox{.}(2018a)]%
        {Tempelaar2018StudentProfilingDispositional}
\bibfield{author}{\bibinfo{person}{Dirk Tempelaar}, \bibinfo{person}{Bart Rienties}, \bibinfo{person}{Jenna Mittelmeier}, {and} \bibinfo{person}{Quan Nguyen}.} \bibinfo{year}{2018}\natexlab{a}.
\newblock \showarticletitle{Student Profiling in a Dispositional Learning Analytics Application Using Formative Assessment}.
\newblock   \bibinfo{volume}{78} (\bibinfo{year}{2018}), \bibinfo{pages}{408--420}.
\newblock
\showISSN{0747-5632}
\href{https://doi.org/10.1016/j.chb.2017.08.010}{doi:\nolinkurl{10.1016/j.chb.2017.08.010}}


\bibitem[Tempelaar et~al\mbox{.}(2020b)]%
        {Tempelaar2020SubjectiveDataObjective}
\bibfield{author}{\bibinfo{person}{Dirk Tempelaar}, \bibinfo{person}{Bart Rienties}, {and} \bibinfo{person}{Quan Nguyen}.} \bibinfo{year}{2020}\natexlab{b}.
\newblock \showarticletitle{Subjective Data, Objective Data and the Role of Bias in Predictive Modelling: {{Lessons}} from a Dispositional Learning Analytics Application}.
\newblock  \bibinfo{volume}{15}, \bibinfo{number}{6} (\bibinfo{year}{2020}), \bibinfo{pages}{e0233977}.
\newblock
\showISSN{1932-6203}
\href{https://doi.org/10.1371/journal.pone.0233977}{doi:\nolinkurl{10.1371/journal.pone.0233977}}


\bibitem[Turner et~al\mbox{.}(2017)]%
        {Turner2017EasingTransitionFirsta}
\bibfield{author}{\bibinfo{person}{Rebecca Turner}, \bibinfo{person}{David Morrison}, \bibinfo{person}{Debby Cotton}, \bibinfo{person}{Samantha Child}, \bibinfo{person}{Sebastian Stevens}, \bibinfo{person}{Patricia Nash}, {and} \bibinfo{person}{Pauline Kneale}.} \bibinfo{year}{2017}\natexlab{}.
\newblock \showarticletitle{Easing the Transition of First Year Undergraduates through an Immersive Induction Module}.
\newblock  \bibinfo{volume}{22}, \bibinfo{number}{7} (\bibinfo{year}{2017}), \bibinfo{pages}{805--821}.
\newblock
\showISSN{1356-2517}
\href{https://doi.org/10.1080/13562517.2017.1301906}{doi:\nolinkurl{10.1080/13562517.2017.1301906}}


\bibitem[Uzir et~al\mbox{.}(2020)]%
        {Uzir2020}
\bibfield{author}{\bibinfo{person}{Nora'ayu~Ahmad Uzir}, \bibinfo{person}{Dragan Gašević}, \bibinfo{person}{Wannisa Matcha}, \bibinfo{person}{Jelena Jovanović}, {and} \bibinfo{person}{Abelardo Pardo}.} \bibinfo{year}{2020}\natexlab{}.
\newblock \showarticletitle{Analytics of Time Management Strategies in a Flipped Classroom}.
\newblock  \bibinfo{volume}{36}, \bibinfo{number}{1} (\bibinfo{year}{2020}), \bibinfo{pages}{70--88}.
\newblock
\showISSN{1365-2729}
\href{https://doi.org/10.1111/JCAL.12392}{doi:\nolinkurl{10.1111/JCAL.12392}}


\bibitem[Van~Petegem et~al\mbox{.}(2022)]%
        {VanPetegem2022}
\bibfield{author}{\bibinfo{person}{Charlotte Van~Petegem}, \bibinfo{person}{Louise Deconinck}, \bibinfo{person}{Dieter Mourisse}, \bibinfo{person}{Rien Maertens}, \bibinfo{person}{Niko Strijbol}, \bibinfo{person}{Bart Dhoedt}, \bibinfo{person}{Bram De~Wever}, \bibinfo{person}{Peter Dawyndt}, {and} \bibinfo{person}{Bart Mesuere}.} \bibinfo{year}{2022}\natexlab{}.
\newblock \showarticletitle{Pass/{{Fail}} Prediction in Programming Courses}.
\newblock  \bibinfo{volume}{61}, \bibinfo{number}{1} (\bibinfo{year}{2022}), \bibinfo{pages}{68--95}.
\newblock
\showISSN{0735-6331}
\href{https://doi.org/10.1177/07356331221085595}{doi:\nolinkurl{10.1177/07356331221085595}}


\bibitem[Villalobos et~al\mbox{.}(2022)]%
        {Villalobos2022SupportingSelfregulatedLearning}
\bibfield{author}{\bibinfo{person}{Esteban Villalobos}, \bibinfo{person}{Mar Pérez-Sanagustin}, \bibinfo{person}{Cédric Sanza}, \bibinfo{person}{André Tricot}, {and} \bibinfo{person}{Julien Broisin}.} \bibinfo{year}{2022}\natexlab{}.
\newblock \showarticletitle{Supporting Self-Regulated Learning in~BL: Exploring Learners' Tactics and~Strategies}. In \bibinfo{booktitle}{\emph{Educating for a New Future: Making Sense of Technology-Enhanced Learning Adoption}} (Cham, 2022), \bibfield{editor}{\bibinfo{person}{Isabel Hilliger}, \bibinfo{person}{Pedro~J. Muñoz-Merino}, \bibinfo{person}{Tinne De~Laet}, \bibinfo{person}{Alejandro Ortega-Arranz}, {and} \bibinfo{person}{Tracie Farrell}} (Eds.). \bibinfo{publisher}{Springer International Publishing}, \bibinfo{pages}{407--420}.
\newblock
\showISBNx{978-3-031-16290-9}
\href{https://doi.org/10.1007/978-3-031-16290-9_30}{doi:\nolinkurl{10.1007/978-3-031-16290-9_30}}


\bibitem[Waheed et~al\mbox{.}(2023)]%
        {Waheed2023EarlyPredictionLearnersa}
\bibfield{author}{\bibinfo{person}{Hajra Waheed}, \bibinfo{person}{Saeed-Ul Hassan}, \bibinfo{person}{Raheel Nawaz}, \bibinfo{person}{Naif~R. Aljohani}, \bibinfo{person}{Guanliang Chen}, {and} \bibinfo{person}{Dragan Gasevic}.} \bibinfo{year}{2023}\natexlab{}.
\newblock \showarticletitle{Early Prediction of Learners at Risk in Self-Paced Education: {{A}} Neural Network Approach}.
\newblock   \bibinfo{volume}{213} (\bibinfo{year}{2023}), \bibinfo{pages}{118868}.
\newblock
\showISSN{0957-4174}
\href{https://doi.org/10.1016/j.eswa.2022.118868}{doi:\nolinkurl{10.1016/j.eswa.2022.118868}}


\bibitem[Weijters et~al\mbox{.}(2006)]%
        {weijters2006process}
\bibfield{author}{\bibinfo{person}{A.J.M.M. Weijters}, \bibinfo{person}{van der Aalst, W.M.P.}, {and} \bibinfo{person}{A.K. Alves De~Medeiros}.} \bibinfo{year}{2006}\natexlab{}.
\newblock \bibinfo{booktitle}{\emph{Process Mining with the {{HeuristicsMiner}} Algorithm}}.
\newblock \bibinfo{publisher}{Technische Universiteit Eindhoven}.
\newblock
\showISBNx{978-90-386-0813-6}


\bibitem[Wilcox(2017)]%
        {Wilcox2017IntroductionRobustEstimation}
\bibfield{author}{\bibinfo{person}{Rand Wilcox}.} \bibinfo{year}{2017}\natexlab{}.
\newblock \bibinfo{booktitle}{\emph{Introduction to Robust Estimation and Hypothesis Testing}}.
\newblock \bibinfo{publisher}{Academic Press}.
\newblock
\showISBNx{978-0-12-804733-0}
\href{https://doi.org/10.1016/B978-0-12-804733-0.00014-7}{doi:\nolinkurl{10.1016/B978-0-12-804733-0.00014-7}}


\bibitem[Winne(2001)]%
        {Winne2001}
\bibfield{author}{\bibinfo{person}{Philip~H Winne}.} \bibinfo{year}{2001}\natexlab{}.
\newblock \showarticletitle{Self-Regulated Learning Viewed from Models of Information Processing}.
\newblock In \bibinfo{booktitle}{\emph{Self-Regulated Learning and Academic Achievement: {{Theoretical}} Perspectives}}, \bibfield{editor}{\bibinfo{person}{Barry~J. Zimmerman} {and} \bibinfo{person}{Dale~H. Schunk}} (Eds.). \bibinfo{publisher}{Routledge}, \bibinfo{pages}{153--189}.
\newblock
\showISBNx{978-1-135-65913-4}


\bibitem[Winne and Hadwin(1998)]%
        {Winne1998}
\bibfield{author}{\bibinfo{person}{Philip~H Winne} {and} \bibinfo{person}{Allyson~F Hadwin}.} \bibinfo{year}{1998}\natexlab{}.
\newblock \showarticletitle{Studying as Self-Regulated Learning}.
\newblock In \bibinfo{booktitle}{\emph{Metacognition in Educational Theory and Practice}}, \bibfield{editor}{\bibinfo{person}{Douglas~J. Hacker}, \bibinfo{person}{John Dunlosky}, {and} \bibinfo{person}{Arthur~C. Graesser}} (Eds.). \bibinfo{publisher}{Erlbaum}, \bibinfo{pages}{277--304}.
\newblock
\showISBNx{0-8058-2481-2}


\bibitem[Winne and Marzouk(2019)]%
        {Winne2019LearningStrategiesSelfRegulateda}
\bibfield{author}{\bibinfo{person}{Philip~H. Winne} {and} \bibinfo{person}{Zahia Marzouk}.} \bibinfo{year}{2019}\natexlab{}.
\newblock \showarticletitle{Learning {{Strategies}} and {{Self-Regulated Learning}}}.
\newblock In \bibinfo{booktitle}{\emph{The {{Cambridge Handbook}} of {{Cognition}} and {{Education}}}}, \bibfield{editor}{\bibinfo{person}{John Dunlosky} {and} \bibinfo{person}{Katherine~A Rawson}} (Eds.). \bibinfo{publisher}{Cambridge University Press}, \bibinfo{pages}{696--715}.
\newblock
\href{https://doi.org/10.1017/9781108235631.028}{doi:\nolinkurl{10.1017/9781108235631.028}}


\bibitem[Yang et~al\mbox{.}(2025)]%
        {Yang2025SelfRegulatedLearningProcesses}
\bibfield{author}{\bibinfo{person}{Kaixun Yang}, \bibinfo{person}{Yizhou Fan}, \bibinfo{person}{Luzhen Tang}, \bibinfo{person}{Mladen Raković}, \bibinfo{person}{Xinyu Li}, \bibinfo{person}{Dragan Gašević}, {and} \bibinfo{person}{Guanliang Chen}.} \bibinfo{year}{2025}\natexlab{}.
\newblock \bibinfo{booktitle}{\emph{Beyond Self-Regulated Learning Processes: Unveiling Hidden Tactics in Generative AI-Assisted Writing}}.
\newblock
\showeprint[arXiv]{2508.10310}~[cs]
\href{https://doi.org/10.48550/arXiv.2508.10310}{doi:\nolinkurl{10.48550/arXiv.2508.10310}}


\bibitem[Zhidkikh et~al\mbox{.}(2024b)]%
        {Zhidkikh2024ReproducingPredictiveLearning}
\bibfield{author}{\bibinfo{person}{Denis Zhidkikh}, \bibinfo{person}{Ville Heilala}, \bibinfo{person}{Charlotte Van~Petegem}, \bibinfo{person}{Peter Dawyndt}, \bibinfo{person}{Miitta Järvinen}, \bibinfo{person}{Sami Viitanen}, \bibinfo{person}{Bram De~Wever}, \bibinfo{person}{Bart Mesuere}, \bibinfo{person}{Vesa Lappalainen}, \bibinfo{person}{Lauri Kettunen}, {and} \bibinfo{person}{Raija Hämäläinen}.} \bibinfo{year}{2024}\natexlab{b}.
\newblock \showarticletitle{Reproducing {{Predictive Learning Analytics}} in {{CS1}}: {{Toward Generalizable}} and {{Explainable Models}} for {{Enhancing Student Retention}}}.
\newblock  \bibinfo{volume}{11}, \bibinfo{number}{1} (\bibinfo{year}{2024}), \bibinfo{pages}{132--150}.
\newblock
\showISSN{1929-7750}
\href{https://doi.org/10.18608/jla.2024.7979}{doi:\nolinkurl{10.18608/jla.2024.7979}}


\bibitem[Zhidkikh et~al\mbox{.}(2023a)]%
        {Zhidkikh2023MeasuringSelfRegulated}
\bibfield{author}{\bibinfo{person}{Denis Zhidkikh}, \bibinfo{person}{Mirka Saarela}, {and} \bibinfo{person}{Tommi Kärkkäinen}.} \bibinfo{year}{2023}\natexlab{a}.
\newblock \showarticletitle{Measuring Self‐regulated Learning in a Junior High School Mathematics Classroom: {{Combining}} Aptitude and Event Measures in Digital Learning Materials}.
\newblock  (\bibinfo{year}{2023}), \bibinfo{pages}{jcal.12842}.
\newblock
\showISSN{0266-4909, 1365-2729}
\href{https://doi.org/10.1111/jcal.12842}{doi:\nolinkurl{10.1111/jcal.12842}}


\bibitem[Zimmerman(2008)]%
        {Zimmerman2008InvestigatingSelfRegulationMotivation}
\bibfield{author}{\bibinfo{person}{Barry~J. Zimmerman}.} \bibinfo{year}{2008}\natexlab{}.
\newblock \showarticletitle{Investigating {{Self-Regulation}} and {{Motivation}}: {{Historical Background}}, {{Methodological Developments}}, and {{Future Prospects}}}.
\newblock  \bibinfo{volume}{45}, \bibinfo{number}{1} (\bibinfo{year}{2008}), \bibinfo{pages}{166--183}.
\newblock
\showISSN{0002-8312}
\href{https://doi.org/10.3102/0002831207312909}{doi:\nolinkurl{10.3102/0002831207312909}}


\bibitem[Zimmerman and Martinez-Pons(1986)]%
        {Zimmerman1986}
\bibfield{author}{\bibinfo{person}{Barry~J. Zimmerman} {and} \bibinfo{person}{Manuel Martinez-Pons}.} \bibinfo{year}{1986}\natexlab{}.
\newblock \showarticletitle{Development of a {{Structured Interview}} for {{Assessing Student Use}} of {{Self-Regulated Learning Strategies}}}.
\newblock  \bibinfo{volume}{23}, \bibinfo{number}{4} (\bibinfo{year}{1986}), \bibinfo{pages}{614--628}.
\newblock
\showISSN{0002-8312}
\href{https://doi.org/10.3102/00028312023004614}{doi:\nolinkurl{10.3102/00028312023004614}}


\bibitem[Zimmerman and Moylan(2009)]%
        {Zimmerman2009}
\bibfield{author}{\bibinfo{person}{Barry~J. Zimmerman} {and} \bibinfo{person}{Adam~R Moylan}.} \bibinfo{year}{2009}\natexlab{}.
\newblock \showarticletitle{Self-Regulation: {{Where}} Metacognition and Motivation Intersect}.
\newblock In \bibinfo{booktitle}{\emph{Handbook of {{Metacognition}} in {{Education}}}}. \bibinfo{pages}{299--315}.
\newblock
\showISBNx{978-0-203-87642-8}


\end{thebibliography}

\pagebreak

\appendix

\section{Trace event codes and the detailed coding approach}\label{appendix:event_codes}

\begin{table}[H]
    \centering
    \resizebox{0.895\textwidth}{!}{%
        \begin{tabular}
            {@{}lp{13cm}@{}}
            \toprule
            Event code                                                          & Description                                                                                                                   \\ \midrule
            \texttt{answer-wrong:lecture-cur-wk-example}                        & The student answered wrong to an example question in the current week's lecture materials.                                    \\ \addlinespace[0.5em]
            \texttt{answer-wrong:materials-book-example}                        & The student answered wrong to an example question in the book materials.                                                      \\ \addlinespace[0.5em]
            \texttt{answer-wrong:tasks-{[}prev|cur{]}-self-assessment}          & The student answered wrong to a self-assessment task in the previous/current week.                                            \\ \addlinespace[0.5em]
            \texttt{answer-wrong:tasks-{[}prev|cur{]}-intro}                    & The student answered wrong to an introductory task in the previous/current week.                                              \\ \addlinespace[0.5em]
            \texttt{answer-wrong:tasks-{[}prev|cur{]}-basic}                    & The student answered wrong to a basic task in the previous/current week.                                                      \\ \addlinespace[0.5em]
            \texttt{answer-wrong:tasks-{[}prev|cur{]}-core}                     & The student answered wrong to a core task in the previous/current week.                                                       \\ \addlinespace[0.5em]
            \texttt{answer-wrong:tasks-{[}prev|cur{]}-bonus}                    & The student answered wrong to a bonus task in the previous/current week.                                                      \\ \addlinespace[0.5em]
            \texttt{answer-wrong:tasks-{[}prev|cur{]}-guru}                     & The student answered wrong to a guru task in the previous/current week.                                                       \\ \addlinespace[0.5em]
            \texttt{answer-wrong:tasks-{[}prev|cur|extra|supp{]}-supplementary} & The student answered wrong to a supplementary task in the previous/current week or in extra/supplementary materials.          \\ \addlinespace[0.5em]
            \texttt{answer:lecture-{[}prev|cur|next{]}-wk-example}              & The student answered to an example question in the previous/current/next week's lecture materials.                            \\ \addlinespace[0.5em]
            \texttt{answer:materials-book-example}                              & The student answered to an example question in the book materials.                                                            \\ \addlinespace[0.5em]
            \texttt{answer:tasks-{[}prev|cur|next{]}-self-assessment}           & The student answered to a self-assessment task in the previous/current/next week.                                             \\ \addlinespace[0.5em]
            \texttt{answer:tasks-{[}prev|cur|next{]}-intro}                     & The student answered correctly to an introductory task in the previous/current/next week.                                     \\ \addlinespace[0.5em]
            \texttt{answer:tasks-{[}prev|cur|next{]}-basic}                     & The student answered correctly to a basic task in the previous/current/next week.                                             \\ \addlinespace[0.5em]
            \texttt{answer:tasks-{[}prev|cur|next{]}-core}                      & The student answered correctly to a core task in the previous/current/next week.                                              \\ \addlinespace[0.5em]
            \texttt{answer:tasks-{[}prev|cur|next{]}-bonus}                     & The student answered correctly to a bonus task in the previous/current/next week.                                             \\ \addlinespace[0.5em]
            \texttt{answer:tasks-{[}prev|cur|next{]}-guru}                      & The student answered correctly to a guru task in the previous/current/next week.                                              \\ \addlinespace[0.5em]
            \texttt{answer:tasks-{[}prev|cur|next|extra|supp{]}-supplementary}  & The student answered correctly to a supplementary task in the previous/current/next week or in extra/supplementary materials. \\ \addlinespace[0.5em]
            \texttt{check-model-answer:tasks-{[}prev|cur{]}-basic}              & The student checked the model answer for a basic task in the previous/current week.                                           \\ \addlinespace[0.5em]
            \texttt{check-model-answer:tasks-{[}prev|cur{]}-core}               & The student checked the model answer for a core task in the previous/current week.                                            \\ \addlinespace[0.5em]
            \texttt{check-model-answer:tasks-prev-bonus}                        & The student checked the model answer for a bonus task in the previous week.                                                   \\ \addlinespace[0.5em]
            \texttt{check-model-answer:tasks-prev-guru}                         & The student checked the model answer for a guru task in the previous week.                                                    \\ \addlinespace[0.5em]
            \texttt{check-model-answer:tasks-prev-intro}                        & The student checked the model answer for an introductory task in the previous week.                                           \\ \addlinespace[0.5em]
            \texttt{check-model-answer:tasks-prev-supplementary}                & The student checked the model answer for a supplementary task in the previous week.                                           \\ \addlinespace[0.5em]
            \texttt{{[}join|leave{]}-lecture:lecture-cur-wk}                    & The student joined/left the current week's lecture.                                                                           \\ \addlinespace[0.5em]
            \texttt{read:lecture-all}                                           & The student read general lecture materials.                                                                                   \\ \addlinespace[0.5em]
            \texttt{read:lecuture-cur}                                          & The student read the lecture materials during the lecture.                                                                    \\ \addlinespace[0.5em]
            \texttt{read:lecture-{[}prev|cur|next{]}-wk}                        & The student read the previous/current/next week's lecture materials.                                                          \\ \addlinespace[0.5em]
            \texttt{read:materials-book}                                        & The student read the book materials.                                                                                          \\ \addlinespace[0.5em]
            \texttt{read:tasks}                                                 & The student read general task materials.                                                                                      \\ \addlinespace[0.5em]
            \texttt{read:tasks-cur}                                             & The student read the current week's task materials.                                                                           \\ \addlinespace[0.5em]
            \texttt{read:tasks-extra}                                           & The student read extra task materials.                                                                                        \\ \addlinespace[0.5em]
            \texttt{read:tasks-next}                                            & The student read the next week's task materials.                                                                              \\ \addlinespace[0.5em]
            \texttt{read:tasks-prev}                                            & The student read the previous week's task materials.                                                                          \\ \addlinespace[0.5em]
            \texttt{read:tasks-pre}                                             & The student read pre-course task materials.                                                                                   \\ \addlinespace[0.5em]
            \texttt{read:tasks-review}                                          & The student read task review materials.                                                                                       \\ \addlinespace[0.5em]
            \texttt{read:tasks-supp}                                            & The student read supplementary task materials.                                                                                \\ \addlinespace[0.5em]
            \texttt{session-start:None}                                         & The student started a TA one-on-one session.                                                                                  \\ \addlinespace[0.5em]
            \texttt{session-end:None}                                           & The student ended a TA one-on-one session.                                                                                    \\ \addlinespace[0.5em]
            \texttt{watch-video:lecture-{[}prev|cur{]}-wk}                      & The student watched the previous/current week's lecture video.                                                                \\ \addlinespace[0.5em]
            \texttt{watch-video:tasks-review-{[}prev|cur|next{]}}               & The student watched the previous/current/next week's task review video.                                                       \\ \addlinespace[0.5em]
            \texttt{workshop-{[}start|end{]}:None}                              & The student joined/left a group study session.                                                                                \\ \bottomrule
        \end{tabular}
    }
    \caption{All event codes used in the trace logs analyzed in this study.
        The table is condensed by combining variations of the same code into the same row.
    }
    \label{tab:all_event_codes}
\end{table}

The trace logs used in the study were generated from raw logs obtained from the virtual learning environment.
The original logs contained rows of newline-separated entries in the following format: \begin{verbatim}
    2023-09-01 12:00:00,000 INFO: username [127.0.0.1]: ACTION
\end{verbatim} where \texttt{ACTION} was some action related to a students' interaction with the virtual learning environment.
The original logs were already limited to only the study participants and only to the duration of the CS1 course.

Original logs were further filtered and re-coded into trace logs as follows:

\begin{enumerate}
    \item If the action was not related to the studied course (e.g., not a course task, not a course document, general interaction with the system), the action was discarded.
    \item Actions were re-coded into initial event codes.
          Course week information was added to some codes (e.g., ``a student views tasks from week 1'' and ``a student views tasks from week 2'' were separate events).
    \item Events with week information were simplified into ``current week'' and ``previous week'' events (e.g., ``a student views tasks from week 2'' was simplified ``a student views current week's tasks'' if the event occurred during course week 2).
    \item Some consecutive events of the same type were merged.
          For example events of type ``a student views tasks'' were merged since the event would occur every time a student scrolls the task page.
    \item The final trace logs were generated in format \texttt{TIMESTAMP;STUDENT;EVENT\_CODE}.
\end{enumerate}
The final used event codes are listed in Table \ref{tab:all_event_codes}.

\section{The post-course questionnaire}\label{appendix:questionnaire}

[\emph{Note: Text within brackets are Authors' comments.}]

\vspace{1em}

\noindent
Reflect briefly on your learning style and on the course in general.
Answer the questions below with at least two sentences per question.
Please especially reflect on what kinds of course materials you tried and whether they were beneficial in studying the course.

\begin{enumerate}
    \item What was your usual study week in this course like? [\emph{Probing for general learning strategies.}]
    \item How did you schedule your studies during the course weeks? [\emph{Probing for time management.}]
    \item How did you feel about your ability to work towards the course? [\emph{Probing for effort regulation.}]
    \item What support channels and materials did you use? [\emph{Probing for peer learning \& help-seeking strategies.}]
    \item Did you change your study methods during the course?
          Why?
          How?
          [\emph{Probing for metacognitive self-regulation.}]
    \item How effective did you find your study methods? [\emph{Probing self-evaluation.}]
    \item Do you have any other thoughts on studying in the CS1? [\emph{Probing extra insights.}]
\end{enumerate}

\section{Learning tactics and event frequencies} \label{appendix:learning_tactic_clusters}

The table lists the most frequent event codes present in each learning tactic cluster.
Because student sessions can differ from week to week and from student to student, clusters may include some low-frequency events. We therefore focused only on the events within the 80th percentile of event frequencies in each cluster.
This threshold was chosen by plotting the frequencies of all event codes in the entire dataset on a scree plot. The elbow point in the relative frequencies was then visually selected, and the cumulative frequency of events up to that point was computed and used as the threshold.

\begin{longtable}{@{}lll@{}}
  \caption{Weekly learning tactics interpreted from student session clusters.
    For each tactic, short code, number of student sessions and the 80th percentile of
    events ($P_{80}$) are included.
  }\label{tab:learning_tactics} \\
  \toprule
  Code                                 & N   & Most frequent events ($P_{80}$) \\ \midrule
  \endfirsthead

  \multicolumn{3}{c}%
  {{\tablename\ \thetable{} -- continued from previous page}} \\
  \toprule
  Code                                 & N   & Most frequent events ($P_{80}$) \\ \midrule
  \endhead

  \bottomrule
  \endfoot

  \bottomrule
  \endlastfoot

  \texttt{F\_CourseMat\_Examples}      & 509 &
  \begin{tabular}[c]{@{}r@{\hspace{0.5em}}l@{}}
    45\% & \texttt{answer:materials-book-example} \\
    33\% & \texttt{read:materials-book} \\
    4\%  & \texttt{read:tasks-cur}
  \end{tabular}
  \\ \addlinespace[0.75em]
  \texttt{F\_Lec\_Engaged}             & 766 &
  \begin{tabular}[c]{@{}r@{\hspace{0.5em}}l@{}}
    36\% & \texttt{read:lecture-cur-wk} \\
    15\% & \texttt{read:lecture-all} \\
    12\% & \texttt{read:materials-book} \\
    11\% & \texttt{answer:lecture-cur-wk-example} \\
    6\%  & \texttt{join-lecture:lecture-cur-wk} \\
    4\%  & \texttt{read:tasks-cur}
  \end{tabular}
  \\ \addlinespace[0.75em]
  \texttt{F\_Lec\_Video}               & 469 &
  \begin{tabular}[c]{@{}r@{\hspace{0.5em}}l@{}}
    23\% & \texttt{read:lecture-cur-wk} \\
    20\% & \texttt{read:lecture-all} \\
    19\% & \texttt{watch-video:lecture-cur-wk} \\
    6\%  & \texttt{answer:lecture-cur-wk-example} \\
    6\%  & \texttt{read:lecture-prev-wk} \\
    5\%  & \texttt{read:tasks-cur} \\
    3\%  & \texttt{read:materials-book}
  \end{tabular}
  \\ \addlinespace[0.75em]
  \texttt{F\_CurTasks\_Intro}          & 350 &
  \begin{tabular}[c]{@{}r@{\hspace{0.5em}}l@{}}
    30\% & \texttt{answer-wrong:tasks-cur-intro} \\
    23\% & \texttt{read:tasks-cur} \\
    20\% & \texttt{answer:tasks-cur-intro} \\
    4\%  & \texttt{read:materials-book} \\
    3\%  & \texttt{read:tasks-prev} \\
    2\%  & \texttt{answer:tasks-cur-core}
  \end{tabular}
  \\ \addlinespace[0.75em]
  \texttt{F\_CurTasks\_Core\_Attempt}  & 356 &
  \begin{tabular}[c]{@{}r@{\hspace{0.5em}}l@{}}
    36\% & \texttt{answer-wrong:tasks-cur-core} \\
    18\% & \texttt{read:tasks-cur} \\
    10\% & \texttt{answer:tasks-cur-core} \\
    8\%  & \texttt{read:materials-book} \\
    4\%  & \texttt{answer-wrong:tasks-cur-basic} \\
    3\%  & \texttt{read:tasks-extra} \\
    3\%  & \texttt{read:tasks-prev}
  \end{tabular}
  \\ \addlinespace[0.75em]
  \texttt{F\_CurTasks\_Core\_Correct}  & 410 &
  \begin{tabular}[c]{@{}r@{\hspace{0.5em}}l@{}}
    36\% & \texttt{answer:tasks-cur-core} \\
    25\% & \texttt{read:tasks-cur} \\
    12\% & \texttt{answer-wrong:tasks-cur-core} \\
    5\%  & \texttt{read:materials-book} \\
    3\%  & \texttt{read:tasks-extra}
  \end{tabular}
  \\ \addlinespace[0.75em]
  \texttt{F\_CurTasks\_Basic\_Attempt} & 369 &
  \begin{tabular}[c]{@{}r@{\hspace{0.5em}}l@{}}
    37\% & \texttt{answer-wrong:tasks-cur-basic} \\
    18\% & \texttt{read:tasks-cur} \\
    10\% & \texttt{answer:tasks-cur-basic} \\
    8\%  & \texttt{read:materials-book} \\
    4\%  & \texttt{answer-wrong:tasks-cur-core} \\
    3\%  & \texttt{answer:tasks-cur-core}
  \end{tabular}
  \\ \addlinespace[0.75em]
  \texttt{F\_CurTasks\_Basic\_Correct} & 529 &
  \begin{tabular}[c]{@{}r@{\hspace{0.5em}}l@{}}
    33\% & \texttt{read:tasks-cur} \\
    32\% & \texttt{answer:tasks-cur-basic} \\
    7\%  & \texttt{answer-wrong:tasks-cur-basic} \\
    5\%  & \texttt{read:materials-book} \\
    2\%  & \texttt{read:tasks-prev} \\
    2\%  & \texttt{answer-wrong:tasks-cur-core}
  \end{tabular}
  \\ \addlinespace[0.75em]
  \texttt{F\_CurTasks\_Extra}          & 640 &
  \begin{tabular}[c]{@{}r@{\hspace{0.5em}}l@{}}
    24\% & \texttt{read:tasks-cur} \\
    18\% & \texttt{read:materials-book} \\
    13\% & \texttt{read:tasks-extra} \\
    9\%  & \texttt{answer-wrong:tasks-extra-supplementary} \\
    6\%  & \texttt{answer:tasks-extra-supplementary} \\
    3\%  & \texttt{answer:materials-book-example} \\
    3\%  & \texttt{answer:tasks-cur-basic} \\
    2\%  & \texttt{read:lecture-cur-wk} \\
    2\%  & \texttt{answer:tasks-cur-core}
  \end{tabular}
  \\ \addlinespace[0.75em]
  \texttt{TA\_Sess}                    & 57  &
  \begin{tabular}[c]{@{}r@{\hspace{0.5em}}l@{}}
    44\% & \texttt{session-end:None} \\
    44\% & \texttt{session-start:None}
  \end{tabular}
  \\ \addlinespace[0.75em]
  \texttt{F\_PrevTasks\_Basic}         & 361 &
  \begin{tabular}[c]{@{}r@{\hspace{0.5em}}l@{}}
    26\% & \texttt{read:tasks-prev} \\
    14\% & \texttt{answer:tasks-prev-basic} \\
    12\% & \texttt{answer-wrong:tasks-prev-basic} \\
    8\%  & \texttt{watch-video:tasks-review-prev} \\
    7\%  & \texttt{check-model-answer:tasks-prev-basic} \\
    5\%  & \texttt{answer:tasks-prev-core} \\
    3\%  & \texttt{answer:tasks-prev-intro} \\
    3\%  & \texttt{answer-wrong:tasks-prev-core} \\
    3\%  & \texttt{check-model-answer:tasks-prev-core}
  \end{tabular}
  \\ \addlinespace[0.75em]
  \texttt{F\_PrevTasks\_Deep}          & 629 &
  \begin{tabular}[c]{@{}r@{\hspace{0.5em}}l@{}}
    36\% & \texttt{read:tasks-prev} \\
    9\%  & \texttt{read:materials-book} \\
    8\%  & \texttt{read:tasks-cur} \\
    3\%  & \texttt{answer-wrong:tasks-prev-core} \\
    3\%  & \texttt{answer:tasks-prev-core} \\
    3\%  & \texttt{answer-wrong:tasks-prev-intro} \\
    3\%  & \texttt{read:lecture-prev-wk} \\
    2\%  & \texttt{answer:tasks-prev-intro} \\
    2\%  & \texttt{watch-video:tasks-review-prev} \\
    2\%  & \texttt{read:lecture-cur-wk} \\
    2\%  & \texttt{read:tasks-extra} \\
    2\%  & \texttt{answer-wrong:tasks-cur-bonus} \\
    1\%  & \texttt{read:lecture-all} \\
    1\%  & \texttt{answer:tasks-prev-basic} \\
    1\%  & \texttt{check-model-answer:tasks-prev-core}
  \end{tabular}
\end{longtable}

\section{Themes identified from questionnaires}\label{appendix:themes}

\begin{table}[H]
    \resizebox{0.875\textwidth}{!}{%
        \begin{tabular}{@{}lll@{}}
            \toprule
            Theme                                       & $N$ & Description                                                                               \\ \midrule
            Course Materials                            & 24  & The student indicates using the provided course materials.                                \\
            Weekly Assignments                          & 23  & The student reports completing weekly assignments.                                        \\
            Lecture Participation                       & 21  & The student reports attending lectures or viewing lecture materials.                      \\
            Emphasis on Regularity                      & 20  & The student highlights the importance of regular study habits.                            \\
            Group Sessions                              & 19  & The student reports attending computer lab sessions.                                      \\
            Positive Opinion of Own Effort              & 16  & The student gives an overall positive opinion of their own effort in the course.          \\
            Peer Support Groups                         & 15  & The student reports participating in a guided peer support group during the course.       \\
            Additional Online Materials                 & 15  & The student reports utilizing supplementary online resources.                             \\
            Importance of Acquaintances                 & 11  & The student emphasizes the role of acquaintances in completing the course.                \\
            Anticipation of Future Challenges           & 11  & The student exhibits self-regulation by preparing for upcoming topics or difficulties.    \\
            Perceived Course Workload                   & 11  & The student mentions that the course is labor-intensive.                                  \\
            Adaptation of Study Methods                 & 10  & The student adapts study methods to suit their individual needs.                          \\
            Neutral Opinion of Own Effort               & 8   & The student gives an overall neutral opinion of their own effort in the course.           \\
            Work-Induced Study Rhythm                   & 6   & The student indicates that employment provides structure to their studies.                \\
            Seeking External Help                       & 5   & The student reports seeking assistance from others to complete assignments.               \\
            Negative Opinion of Own Effort              & 5   & The student gives an overall negative opinion of their own effort in the course.          \\
            Constraints from Other Courses              & 5   & The student discusses how other courses limit their ability to focus on this one.         \\
            Externalizing Self-Regulation               & 4   & The student relies on others (e.g., teaching assistants) for guidance and regulation.     \\
            Personal Life Constraints                   & 4   & The student brings up additional personal life factors that impose constraints.           \\
            Emphasis on Group Work Benefits             & 4   & The student underscores the significance of group work in course completion.              \\
            Irregular Study Habits                      & 3   & The student mentions irregularity in their study activities.                              \\
            Synchronization of Lectures and Assignments & 3   & The student emphasizes the importance of timing between lectures and assignments.         \\
            Challenges in Studying                      & 3   & The student brings up challenges in studying the course material.                         \\
            Deviation from Established Practices        & 3   & The student indicates deviating from their usual study practices.                         \\
            Work-Related Limitations                    & 3   & The student mentions employment as a limiting factor in course participation.             \\
            Invested Effort                             & 3   & The student highlights having invested significant effort into the course.                \\
            Preference for Independent Work             & 2   & The student prefers to work alone rather than in groups.                                  \\
            Deadline-Driven Approach                    & 2   & The student operates primarily based on deadlines.                                        \\
            Lecturer-Induced Feelings of Inferiority    & 2   & The student recounts an instance where the lecturer caused feelings of inadequacy.        \\
            Random Study Practices                      & 2   & The student indicates randomness in their study methods.                                  \\
            Course Workload for Working Students        & 2   & The student highlights the demanding nature of the course workload for employed students. \\
            Dependence on Instructors                   & 1   & The student's progress depends heavily on instructor guidance and assistance.             \\
            Burnout Due to Workload                     & 1   & The student expresses burnout resulting from the course workload.                         \\
            AI Tools                                    & 1   & The student mentions using AI tools, such as ChatGPT, for assistance.                     \\
            Lack of Motivation                          & 1   & The student expresses a lack of motivation regarding the course.                          \\
            Language Barrier Challenges                 & 1   & The student identifies the language barrier as a challenge during the course.             \\
            Library Documentation                       & 1   & The student reports using documentation from the Jypeli library for project work.         \\
            Establishing Routine                        & 1   & The student demonstrates self-regulation by creating a routine for course completion.     \\
            Increasing Independence Over Time           & 1   & The student notes enhanced self-regulation as the course progresses.                      \\
            Expressed Motivation                        & 1   & The student conveys motivation toward the course.                                         \\
            Deliberate Note-Taking                      & 1   & The student practices self-regulation through conscious note-taking to enhance learning.  \\
            Low Goal Setting                            & 1   & The student sets low personal goals for the course.                                       \\
            Social Aspects of the Course                & 1   & The student brings up matters related to social interactions in the course.               \\
            Challenge of Transitioning to University    & 1   & The student mentions difficulties in transitioning to university-level studies.           \\
            Minimal Effort Invested                     & 1   & The student notes limited personal investment in the course.                              \\
            Concern Over Effort Sufficiency             & 1   & The student expresses concern about the adequacy of their effort in the course.           \\
            Adjusting Habits to Course Schedule         & 1   & The student adapts their habits to align with the course timetable.                       \\
            Awareness of Personal Goals                 & 1   & The student acknowledges and articulates personal goals for the course.                   \\ \bottomrule
        \end{tabular}%
    }
    \caption{
        Themes identified from content analysis of 29 student questionnaires to allow easier comparison and summary of student profiles. For each theme, theme name, number of student in which the theme emerged, and a brief description of the theme is provided.
    }
\end{table}

\end{document}